\documentclass{aa}  

\usepackage{graphicx}
\usepackage{txfonts}
\usepackage{natbib}
\bibpunct{(}{)}{;}{a}{}{,}
\begin{document}

   \title{Direct Imaging discovery of a second planet candidate
 \\ around the possibly transiting planet host CVSO 30
\thanks{Based on observations made with ESO Telescopes at the La Silla Paranal Observatory 
under programme IDs \mbox{090.C-0448(A)}, \mbox{290.C-5018(B)}, \mbox{092.C-0488(A)} and at the Centro Astron\'omico Hispano-Alem\'an in programme \mbox{H15-2.2-002}.}}
\titlerunning{Direct Imaging of a  second planet candidate in the transiting CVSO 30 system.}


   \author{T.~O.~B.~Schmidt\inst{1,2}
           \and
           R. Neuh\"auser\inst{2}
           \and
           C. Brice\~no\inst{3}
           \and
           N. Vogt\inst{4}
           \and
           St. Raetz\inst{5}
           \and
           A. Seifahrt\inst{6}
           \and
           C. Ginski\inst{7}
           \and
           M. Mugrauer\inst{2}           
           \and
\\
           S. Buder\inst{2,8}
           \and
           C. Adam\inst{2}
           \and
           P. Hauschildt\inst{1}
           \and
           S. Witte\inst{1}
           \and
           Ch. Helling\inst{9}
           \and
           J.~H.~M.~M.~Schmitt\inst{1}
}

   \institute{Hamburger Sternwarte, Gojenbergsweg 112, 21029 Hamburg, Germany, \email{tschmidt@hs.uni-hamburg.de}
              \and
              Astrophysikalisches Institut und Universit\"ats-Sternwarte, Universit\"at Jena, Schillerg\"a\ss chen 2-3, 07745 Jena, Germany
              \and
              Cerro Tololo Inter-American Observatory CTIO/AURA/NOAO, Colina El Pino s/n. Casilla 603, 1700000 La Serena, Chile
              \and
              Instituto de F\'isica y Astronom\'ia, Universidad de Valpara\'iso, Avenida Gran Breta\~na 1111, 2340000 Valpara\'iso, Chile
              \and
              European Space Agency ESA, ESTEC, SRE-S, Keplerlaan 1, NL-2201 AZ Noordwijk, the Netherlands
              \and
              Department of Astronomy and Astrophysics, University of Chicago, 5640 S. Ellis Ave., Chicago, IL 60637, USA
              \and              
              Sterrewacht Leiden, PO Box 9513, Niels Bohrweg 2, NL-2300RA Leiden, the Netherlands
              \and
              Max-Planck-Institute for Astronomy, Königstuhl 17, 69117 Heidelberg, Germany
              \and
              School of Physics and Astronomy SUPA, University of St. Andrews, North Haugh, St. Andrews, KY16 9SS, UK
            }

   \date{Received 2015; accepted}

 
  \abstract
   {Direct imaging has developed into a very successful technique for the detection of exoplanets in wide orbits, especially around young stars. Directly imaged planets can be both followed astrometrically on their orbits and observed spectroscopically and thus provide an essential tool for our understanding of the early solar system.}
   {We surveyed the 25 Ori association for direct-imaging companions.
This association has an age of only few million years. Among other targets, we observed CVSO 30, which has recently been identified as the first T Tauri star found to host a transiting planet candidate.}
   {We report on photometric and spectroscopic high-contrast observations with the Very Large Telescope, the Keck telescopes, and the Calar Alto observatory. They reveal a directly imaged planet candidate close to the young M3 star CVSO 30.} 
   {The JHK-band photometry of the newly identified candidate is at better than 1\,$\sigma$ consistent with late-type giants, early-T and early-M dwarfs, and free-floating planets. Other hypotheses such as~galaxies can be excluded at more than 3.5\,$\sigma$. A lucky imaging z$^{\prime}$ photometric detection limit z$^{\prime}$= 20.5 mag excludes early-M dwarfs and results in less than 10 M$_{\mathrm{Jup}}$ for CVSO 30 c if bound. We present spectroscopic observations of the wide companion that imply that the only remaining explanation for the object is that it is the first very young ($<$\,10 Myr) L\,--\,T-type planet bound to a star, meaning that it~appears bluer than expected as a result
of a decreasing cloud opacity at low effective temperatures. Only a planetary spectral model is consistent with the spectroscopy, and we deduce a best-fit mass of 4\,-\,5 Jupiter masses (total range 0.6 -- 10.2 Jupiter masses).}
   {This means that CVSO 30 is the first system in which both a close-in and a wide planet candidate are found to have a common host star. The orbits of the two possible planets could not be more different: they have orbital periods of 10.76 hours and about 27000 years. The two orbits may have formed during a mutual catastrophic event of planet-planet scattering.}


   \maketitle
%

\section{Introduction}

Since the first definite detection of a planet around another main-sequence star, 51 Peg \citep{1995Natur.378..355M}
made with high-precision radial velocity measurements,
various detection techniques have been applied to find a diverse population of exoplanets. The transit method, which was first
used for HD~209458
\citep{2000ApJ...529L..45C}, later allowed for a boost of exoplanet discoveries after the successful launch of two dedicated satellite missions, CoRoT \citep{2007AIPC..895..201B} and Kepler
\citep{2010ApJ...713L..79K,2010Sci...327..977B}. These two methods indirectly discern the presence of a planet by the influence the planet has on its host star. The methods are most sensitive to small and moderate planet-star-separations around old main-sequence stars that are fairly inactive because of their age. The sensitivity diminishes fast for separations beyond 5 au because transits become less likely and the radial velocity amplitude declines with increasing orbital period. 
In contrast, direct imaging allows discovering planets on wide orbits around nearby pre-main-sequence stars because such young planets 
are still bright at infrared wavelengths as a result of the gravitational contraction during their still-ongoing formation process.

\begin{figure*}
  \centering
  \includegraphics[width=0.38\textwidth]{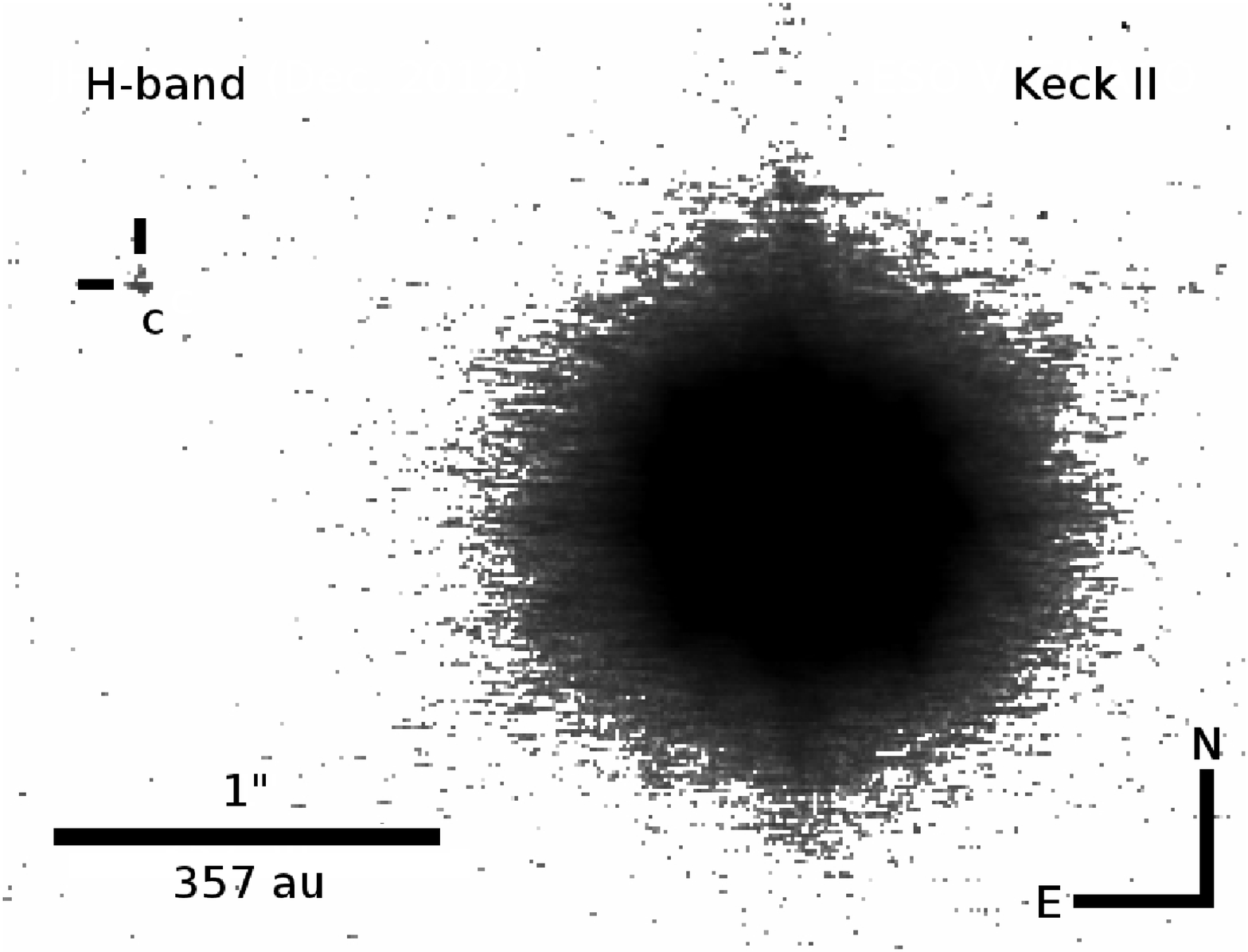}
  \includegraphics[width=0.4\textwidth]{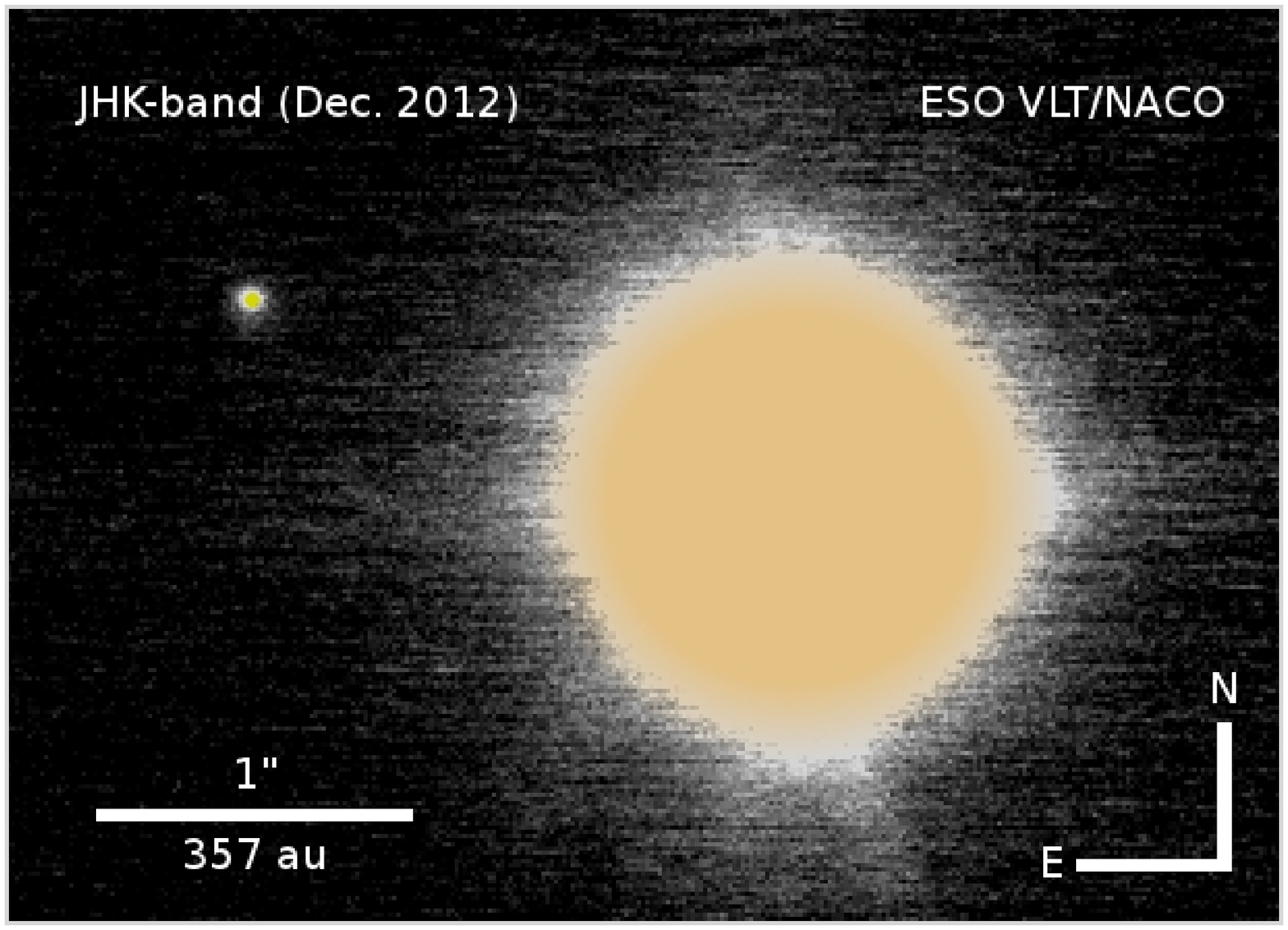}
  \caption{Direct images of CVSO 30 c. \textit{Left:} Keck image of data by \cite{2012ApJ...755...42V}, re-reduced. We note that the companion is north-east and not a contaminant south-east, as
  given in \cite{2012ApJ...755...42V}. \textit{Right:} Our new VLT epoch, clearly showing the planetary companion, which has similar colour as its host star (Fig.~\ref{FigPhot}). This excludes
that it is a false positive for the inner planet candidate CVSO 30 b.}
  \label{FigImage}
\end{figure*}

The total number of imaged planet candidates has increased
to about 50-60 objects today. The first detections were made
in 2005, when the first four co-moving planetary candidates were found around the solar-like stars DH Tau \citep{2005ApJ...620..984I}, GQ Lup \citep{2005AandA...435L..13N}, and AB Pic \citep{2005AandA...438L..29C}, all with masses near the threshold of 13 M$_{\mathrm{Jup}}$ that
divides brown dwarfs from planets according to the current IAU working definition, 
together with the planet candidate around the brown dwarf 2M1207 \citep{2005A&A...438L..25C}. A summary can be found in \citet{2012arXiv1201.3537N}, and the current status
is always available in several online encyclopaediae, such as the Extrasolar Planets Encyclopaedia at www.exoplanet.eu \citep{2011A&A...532A..79S}.
As in situ formation at $\sim$100 au to a few hundreds of au separation seems unlikely according to models, \citet{2006ApJ...637L.137B} argued that a third
body must exist that tossed these planets outward to their present distance from their young host stars. An alternative explanation might be a stellar encounter 
\citep{2001Icar..150..151A}. 

While early-type stars have less favourable planet-to-star contrast ratios, increasing evidence was found by millimeter-continuum measurements for larger
and more massive protoplanetary disks that are available for planet formation around these stars \citep{1997ApJ...490..792M,2013ApJ...771..129A}. These
conclusions were further strengthened when three of the most prominent planet candidates were found in 2008 and
2009 around the early F-type star
HR 8799 \citep{2008Sci...322.1348M}, which is the first system with multiple planets imaged around a star. The other two were around the A-type stars Fomalhaut \citep{2008Sci...322.1345K},
which is the first planet candidate discovered in the optical regime using the Hubble Space Telescope (HST), and $\beta$ Pic \citep{2009AandA...493L..21L,2010Sci...329...57L},
a planet within the large edge-on disk at only about twice the separation of Jupiter from the Sun. This had~previously been
predicted by \citet{2007A&A...466..389F}, for instance,
from the structural gaps in the disk.

Most of the direct-imaging surveys conducted so far have concentrated on AFGK stars. In 2012 a (proto-)planet candidate was discovered around the $\sim$2 Myr young
solar-like star LkCa 15 \citep{2012ApJ...745....5K}. This is a close ($\sim$11 AU) object found by single-dish interferometry, which is a technique that is also referred to as sparse aperture masking. Two companions of 4-5 M$_{\mathrm{Jup}}$ were recently
discovered around GJ 504 \citep{2013ApJ...774...11K}, which is a 160 Myr old solar-like star, and around HD 95086 \citep{2013ApJ...772L..15R}, an A-type star at about 10-17 Myr.
Additionally, first results from imaging surveys around
M dwarfs were published during the past two years, which increased our understanding of planetary systems around the most numerous stars in the Milky Way \citep{2013AandA...553L...5D,2015ApJS..216....7B}.

\begin{table}
\caption{Previously known CVSO 30 system data}
\label{TableCVSOsys}
\centering
\begin{scriptsize}
\begin{tabular}{lcc}
\hline
& \multicolumn{2}{c}{CVSO 30}\\
\hline\hline
Altern.~designations & \multicolumn{2}{c}{2MASS J05250755+0134243, PTF1 J052507.55+013424.3} \\
Location        & \multicolumn{2}{c}{25 Ori / Orion OB 1a [1,2]}\\
RA, Dec         & \multicolumn{2}{c}{05h 25m 07.57s, +01$^{\circ}$ 34$'$ 24.5$''$ [2]}\\
Spectral type   & \multicolumn{2}{c}{M3 (weak-line T-Tauri, WTTS) [2]}\\
Mass            & \multicolumn{2}{c}{0.34 / 0.44 M$_{\mathrm{\odot}}$ [2]}\\
Luminosity      & \multicolumn{2}{c}{0.25 L$_{\mathrm{\odot}}$ [2]}\\
Radius          & \multicolumn{2}{c}{1.39 R$_{\mathrm{\odot}}$ / 1.07 $\pm$ 0.10 R$_{\mathrm{\odot}}$ / [1.03 / 1.04 $\pm$ 0.01 R$_{\mathrm{\odot}}$] [2,3,4]}\\
Temperature     & \multicolumn{2}{c}{3470 K [2]}\\
Opt. extinction & \multicolumn{2}{c}{0.12 mag [2]}\\
\smallskip
Distance        & \multicolumn{2}{c}{[323$^{+233}_{-96}$, 322$^{+504}_{-122}$] pc / 357 $\pm$ 52 pc [2,5]}\\
Age             & \multicolumn{2}{c}{2.39$^{+3.41}_{-2.05}$ Myr [2,here]}\\
H$_{\alpha}$ equivalent width  & \multicolumn{2}{c}{-11.40 \AA \ [2]}\\
LiI equivalent width           & \multicolumn{2}{c}{0.40 \AA \ [2]}\\
v $\sin(i)_*$          & \multicolumn{2}{c}{80.6 $\pm$ 8.1 km s$^{-1}$ [3]}\\
Proper motion [E,N] & \multicolumn{2}{c}{[-0.1 $\pm$ 5.3, 0.9 $\pm$ 5.5] mas/yr [6]}\\
B, V, R photometry  & \multicolumn{2}{c}{[18.35, 16.26, 15.19] mag [7,2,3]}\\
J, H, K photometry  & \multicolumn{2}{c}{[12.232\,$\pm$\,0.028,\,11.559\,$\pm$\,0.026,\,11.357\,$\pm$\,0.021]\,mag [8]}\\
\hline
                    & \multicolumn{2}{c}{CVSO 30 b / PTFO 8-8695 b} \\
\hline\hline
(Projected) separation & \multicolumn{2}{c}{0.00838 $\pm$ 0.00072 au [3] } \\
Period (circular)      & \multicolumn{2}{c}{0.448413 $\pm$ 0.000040 d [3] } \\
Orbit.~inclination     & \multicolumn{2}{c}{61.8 $\pm$ 3.7 $^{\circ}$ [3] } \\
Orbit.~misalignment    & \multicolumn{2}{c}{69 $\pm$ 2 $^{\circ}$ / 73.1 $\pm$ 0.5 $^{\circ}$ [4] } \\
\hline
\end{tabular}
\begin{flushleft}
References: [1] \citet{2007ApJ...661.1119B}, [2] \citet{2005AJ....129..907B}, [3] \citet{2012ApJ...755...42V}, [4] \citet{2013ApJ...774...53B}, [5] \citet{2014MNRAS.444.1793D}
            [6] \citet{2013AJ....145...44Z}, \mbox{[7] \citet{2004AAS...205.4815Z}}, [8] \citet{2003tmc..book.....C,2006AJ....131.1163S}                                                                                         
\end{flushleft}
\end{scriptsize}
\end{table}

\begin{table}
\caption{CVSO 30 astrometry and photometry}
\label{TableCVSOAsPhot}
\centering
\begin{scriptsize}
\begin{tabular}{lcc}
\hline
                    & CVSO 30 b / & CVSO 30 c\\
                    & PTFO 8-8695 b \\
\hline\hline
                    & \multicolumn{2}{c}{Separation with respect
to~the host star [E,N]} \\
2010 September 25   & & [175.453,\,63.395] pixel \\
2012 December 3     & & [1.746 $\pm$ 0.006, \\
                    & & \ \ \ \ \ 0.621 $\pm$ 0.010] $''$ \\
\hline
(Projected) separation & 0.00838 $\pm$ 0.00072 au [1] & 662 $\pm$ 96 au \\
Period (circular)      & 0.448413 $\pm$ 0.000040 d [1] & $\sim$ 27250 years \\
Orbit.~inclination     & 61.8 $\pm$ 3.7 $^{\circ}$ [1] & \\
Orbit.~misalignment    & 69 $\pm$ 2 $^{\circ}$ / 73.1 $\pm$ 0.5 $^{\circ}$ [2] & \\
\hline
z$^{\prime}$ band (differential) & & > 6.8 mag \\
\hline
J band (differential)    & & 7.385 $\pm$ 0.045 mag \\
H band (differential)    & & 7.243 $\pm$ 0.014 mag \\
Ks band (differential)   & & 7.351 $\pm$ 0.022 mag \\
\hline
J band (differential)    & & 7.183 $\pm$ 0.035 mag \\
\hline
\end{tabular}
\begin{flushleft}
References: [1] \citet{2012ApJ...755...42V}, [2] \citet{2013ApJ...774...53B}                                                                                 
\end{flushleft}
\end{scriptsize}
\end{table}

In this article we describe for the first time the direct detection of a wide-separation (1.85$''$ or 662 au, see Fig.~\ref{FigImage}) directly imaged planet candidate around a star (\object{CVSO 30}) that also hosts a short-period transiting planet candidate; we refer to a more detailed discussion of this object in \citet{2012ApJ...755...42V}, \citet{2013ApJ...774...53B}, and \citet{2015ApJ...812...48Y}. A system that harbours two planets with such extreme orbits gives us for the first time the opportunity to study by observations
the possible outcome of planet-planet scattering theories, which have been used to explain the existence of close-in hot Jupiters in 1996 \citep{1996Sci...274..954R}. 


\section{25 Ori group and the CVSO 30 system properties}

\begin{table*}
\caption{VLT/NACO, VLT/SINFONI, archival KeckII/NIRC2, and Calar Alto/2.2m/AstraLux observation log}
\label{TableLog}
\centering
\begin{footnotesize}
\begin{tabular}{llccccccccccc}
\hline\hline
Instrument         & JD-2455000      & Date of         & DIT        & NDIT    & \#        & Airmass & DIMM$^a$   & $\tau_{0}^b$ & Strehl & S/N \\
                   & $[\mathrm{days}]$ & observation   & [s]        &         & images    &         & Seeing     & [ms]         & [\%]   & (brightest pixel) \\
\hline
NACO J             & 1264.69416       & 03 Dec 2012     & 15         & 4      & 15        & 1.13    & 0.8        & 3.7          & 3.2    & 5.9  \\
NACO H             & 1264.70764       & 03 Dec 2012     & 15         & 4      & 15        & 1.12    & 0.6        & 4.6          & 11.2   & 24.6 \\ 
NACO Ks            & 1264.72079       & 03 Dec 2012     & 15         & 4      & 15        & 1.11    & 0.7        & 4.6          & 23.7   & 11.1 \\
NACO J             & 1266.72899       & 05 Dec 2012     & 30         & 2      & 15        & 1.12    & 1.3        & 2.8          & 2.0    & 6.6  \\
SINFONI H+K        & 1592.82609       & 27 Oct 2013     & 300        & 2      & 3         & 1.12    & 0.5        & 5.0          &        & $\lessapprox$15 \\
\hline
NIRC2 H            & \, 465.05374     & 25 Sep 2010     & 3          & 10     & 12        & 1.25    & 0.4        &              & 7.0    & 7.8  \\
\hline
AstraLux z$^{\prime}$ & 2260.6696     & 26 Aug 2015     & 0.02945    & 1      & 70000     & 1.73    & 1.1        &              & no AO  & non-detection   \\
\hline
\end{tabular}
\begin{flushleft}
\textbf{Remarks}: (a) Differential image motion monitor (DIMM) seeing average of all images (b) coherence time of atmospheric fluctuations. 
\end{flushleft}                                                                                                     
\end{footnotesize}
\end{table*}

\begin{table}
\caption{CVSO 30 deduced planetary properties}
\label{TableCVSOPlanets}
\centering
\begin{scriptsize}
\begin{tabular}{lcc}
\hline
                    & CVSO 30 b / & CVSO 30 c\\
                    & PTFO 8-8695 b \\
\hline\hline
\smallskip
Opt. extinction                     & & 0.19$^{+2.51}_{-0.19}$ mag\\
\smallskip
Luminosity (vs.~${\mathrm{\odot}}$) & & -3.78$^{+0.33}_{-0.13}$ dex\\
\smallskip
Eff.~temperature T$_{\mathrm{eff}}$ & & 1600$^{+120}_{-300}$ K\\
Surface gravity $\log{g}$           & & 3.6$^{+1.4}_{-0.6}$ dex\\
Radius                              & 1.91 $\pm$ 0.21 R$_{\mathrm{Jup}}$ [1] & \\
\smallskip
                                    & 1.64 / 1.68 $\pm$ 0.07 R$_{\mathrm{Jup}}$ [2] & 1.63$^{+0.87}_{-0.34}$ R$_{\mathrm{Jup}}$\\
\smallskip
Mass                                & $<$ 5.5 $\pm$ 1.4 M$_{\mathrm{Jup}}$ [1] & 4.3$^{+4.9}_{-3.7}$ M$_{\mathrm{Jup}}$ ($\log{g}$ \& Roche)\\
\smallskip
                                    & 3.0 $\pm$ 0.2 M$_{\mathrm{Jup}}$     [2] & 4.7$^{+5.5}_{-2.0}$ M$_{\mathrm{Jup}}$ (L, age) \\
\smallskip
                                    & 3.6 $\pm$ 0.3 M$_{\mathrm{Jup}}$     [2] & 4.7$^{+3.6}_{-2.0}$ M$_{\mathrm{Jup}}$ (L, T$_{\mathrm{eff}}$, age) \\
\smallskip
                                    &                                          & < 10 M$_{\mathrm{Jup}}$ (z$^{\prime}$ imaging limit) \\
\hline
\end{tabular}
\begin{flushleft}
References: [1] \citet{2012ApJ...755...42V}, [2] \citet{2013ApJ...774...53B}                                                                                 
\end{flushleft}
\end{scriptsize}
\end{table}

Despite their importance for the evolution of protoplanetary disks and the early phases in the planet formation process, sufficiently large samples of 10 Myr old stars have been difficult to identify, mainly because the parent molecular clouds dissipate after a few Myr
and no longer serve as markers of these populations (see \citet{2007prpl.conf..345B} and references therein).
The 25 Ori cluster \citep[25 Ori,][]{2007ApJ...661.1119B} contains $> 200$ PMS stars in the mass range $0.1 < M/M_{\odot} < 3$. 
The Hipparcos OB and earlier A-type stars in 25 Ori are on the zero-age main sequence \citep[ZAMS,][]{2005AJ....129..856H}, implying a distance of $\sim$330 pc, with some of the A-type stars harbouring debris disks \citep{2006ApJ...652..472H}.
Isochrone fitting of the low-mass stars yields an age of 7-10 Myr \citep{2007prpl.conf..345B}. This is the most populous 10 Myr old sample within 500 pc,
which we consequently chose for a direct-imaging survey with ESO's VLT, the Very Large Telescope of the European Southern Observatory, to find young planetary and sub-stellar companions at or shortly after their formation. For this same reason, the 25 Ori cluster was also targeted in searches for transiting planets, like the Young Exoplanet Transit Initiative \citep[YETI,][]{2011AN....332..547N} and the Palomar Transient Factory \citep[PTF,][]{2012ApJ...755...42V}.

CVSO 30 (also 2MASS J05250755+0134243 and PTFO 8-8695) is a weak-line T Tauri star of spectral type M3 in 25 Ori at an average distance of 357$\,\pm\,$52 pc \citep{2014MNRAS.444.1793D}. 
It was confirmed as a T Tauri member of the 25 Ori cluster by the CIDA Variability Survey of Orion (CVSO), with properties shown in Table~\ref{TableCVSOsys}.
As shown in Fig.~1 in \citet{2012ApJ...755...42V}, CVSO 30 is one of the youngest objects within 25 Ori, its position in the colour-magnitude diagram corresponds to 2.39$^{+3.41}_{-2.05}$~Myr (if compared to evolutionary models of \citet{2000A&A...358..593S}). The object is highly variable, rotates fast, and has a mass of 0.34 -- 0.44 M$_{\odot}$ (depending on the evolutionary model), and an effective temperature of $\sim$3470 K. The rotation period of CVSO 30, which is possibly synchronised with the CVSO 30 b orbital period, is still debated \citep{2012ApJ...755...42V,2015MNRAS.450.3991K}. \citet{2015PASJ...67...94K} concluded that the stellar spin period is shorter than 0.671 d.

In 2012 the PTF team \citep{2012ApJ...755...42V} reported a young transiting planet candidate around CVSO 30, named PTFO 8-8695 b, with a fast co-rotating or near co-rotating 0.448413 day orbit.
The very same object, henceforth CVSO 30 b for simplicity, was independently detected with smaller telescopes within the YETI
\citep{2013prpl.conf2K047N,2014CoSka..43..513E},
confirming the presence of the transit events by quasi-simultaneous observations.

Keck and Hobby-Eberly Telescope (HET) spectra \citep{2012ApJ...755...42V} set an upper limit to the mass of the transiting companion of 5.5 $\pm$ 1.4 M$_{\mathrm{Jup}}$ from
the radial velocity variation, which exhibits a phase offset that is likely caused by spots on the surface of the star. This RV limit was already corrected for the derived orbital inclination 61.8 $\pm$ 3.7 $^{\circ}$ of the system. With an orbital radius of only about twice the stellar radius and a planetary radius of 1.91 $\pm$ 0.21 R$_{\mathrm{Jup}}$, the object appears to be at or within its Roche-limiting orbit, raising the possibility of past or ongoing mass loss. A false positive by a blended eclipsing binary is unlikely because the only present contaminant in Keck near-IR images (see Fig.~\ref{FigImage}) with 6.96 mag of contrast to the star would have to be very blue to be bright enough in the optical to mimic a transit; it is unlikely to be a star in that case. 

In 2013 \citet{2013ApJ...774...53B} fit the two separate light
curves observed in 2009 and 2010, which exhibited unusual differing shapes, simultaneously and
self-consistently with planetary masses of the companion of 3.0 -- 3.6 M$_{\mathrm{Jup}}$.
They assumed transits across an oblate, gravity-darkened stellar disk and precession of the planetary orbit ascending node. The fits show a high degree of spin-orbit misalignment of about 70$^{\circ}$, which leads
to the prediction that transits disappear for months at a time during the precession period of this system. The lower planet radius result of 
$\sim$1.65 R$_{\mathrm{Jup}}$ is consistent with a young, hydrogen-dominated planet that results from hot-start formation mechanisms \citep{2013ApJ...774...53B}.

\begin{table}
\caption{Astrometric calibration of VLT/NACO}
\label{TableAstroCal}
\centering
\begin{footnotesize}
\begin{tabular}{llcc@{\,$\pm$\,}l}
\hline\hline
Object & JD - 2456000 & Pixel scale &  \multicolumn{2}{c}{PA$^a$ }\\
       & $[\mathrm{days}]$ & [mas/pixel] & \multicolumn{2}{c}{[$\deg$]} \\
\hline
47 Tuc & 264.62525 & 13.265 $\pm$ 0.041 & +0.60 & 0.31 \\
\hline
\end{tabular}
\begin{flushleft}
\textbf{Remarks}: All data from Ks-band images. (a) PA is measured from N over E to S.
\end{flushleft}
\end{footnotesize}
\end{table}

\section{Astrometric and photometric analysis}

\begin{figure*}
\centering
  \includegraphics[width=0.49\textwidth]{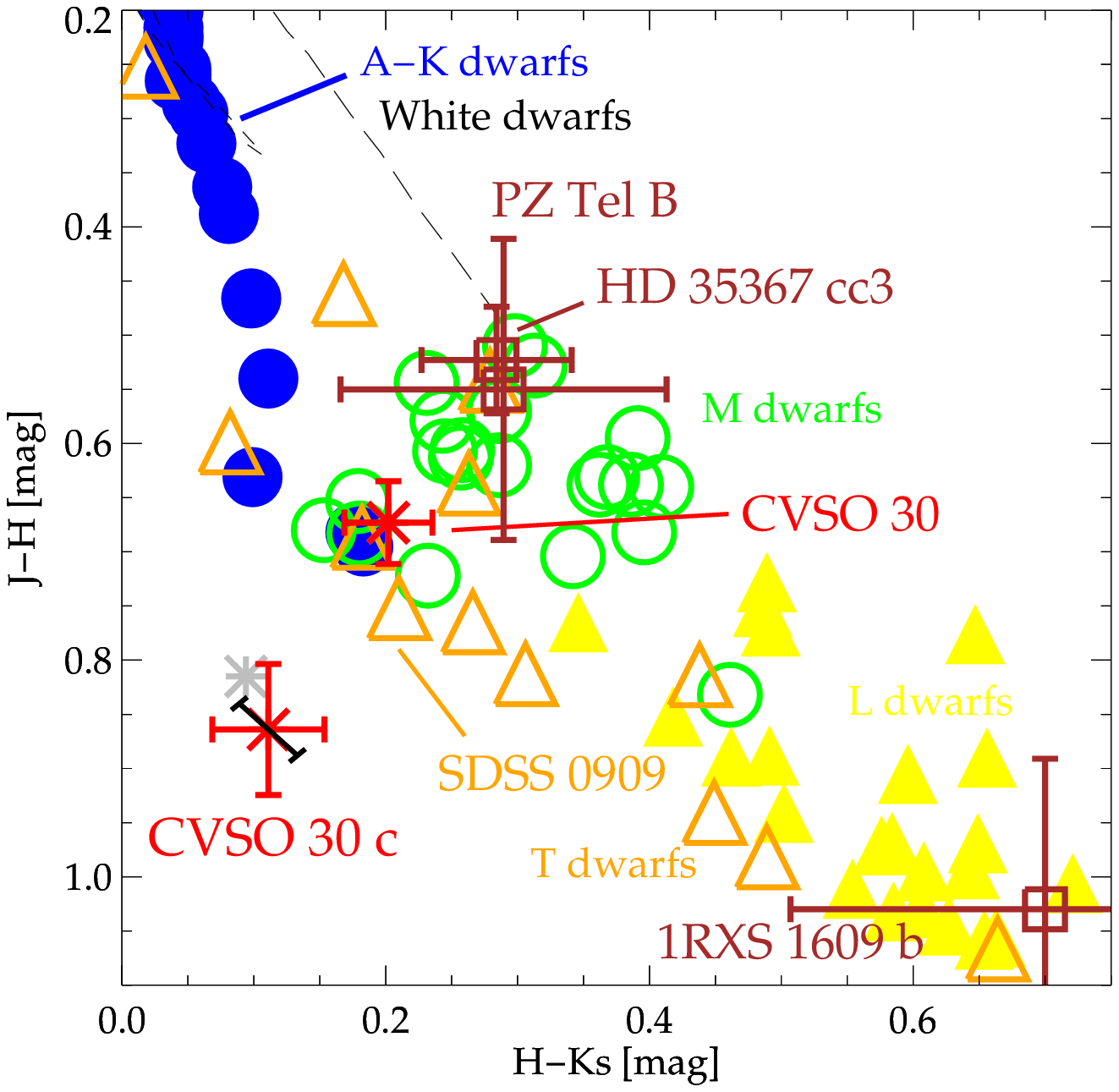}
  \includegraphics[width=0.49\textwidth]{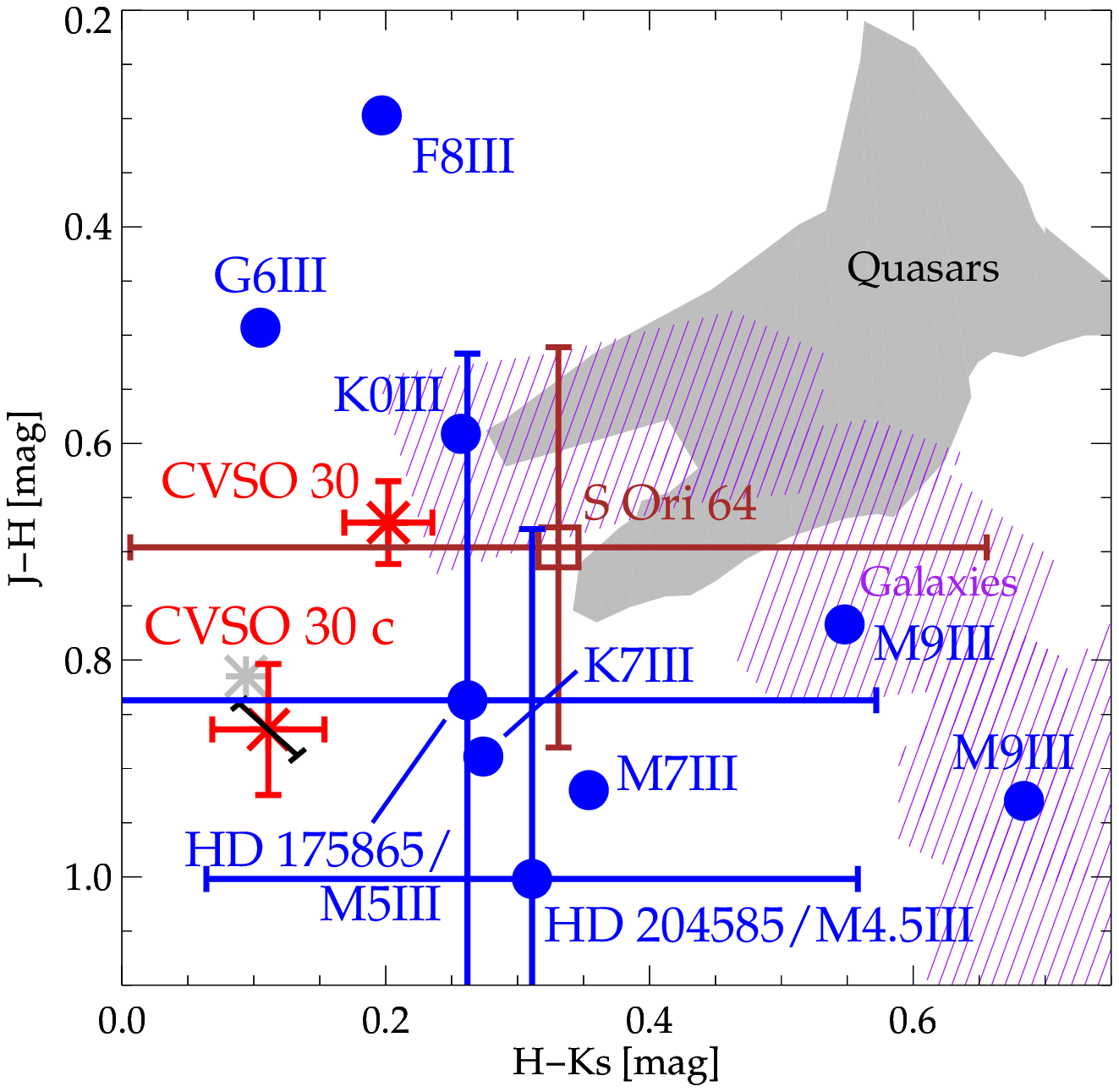}
  \caption{CVSO 30, CVSO 30 c, and comparison objects, superimposed onto the colour data from \citet{2006MNRAS.367..454H}. CVSO 30 c clearly stands out in the lower left corner,
  approximately consistent with colours of giants, early-M and early-T dwarfs, and free-floating planetary mass objects \citep{2000Sci...290..103Z,2012ApJ...754...30P}, e.g. consistent with absolute magnitude and J-Ks colour of S Ori 64.
  Its unusual blue colour can most likely be attributed to the youth of these objects \citep{2008ApJ...689.1327S}, leading to 
  L--T transition opacity drop at high brightnesses (see Fig.~\ref{FigColMag}). See Fig.~\ref{FigColMagAppend} for details.
  For CVSO 30 c we give the colours before (grey) and after (red) correcting for the NACO to the 2MASS filter set and we give the maximum possible systematic photometric offsets caused by the variability of the primary star that is used as reference (black).}
  \label{FigPhot}
\end{figure*}

After the discovery of the transiting planet candidate by \citet{2012ApJ...755...42V} and our independent detection of the transit signals with YETI, we included the system in our 25 Ori VLT/NACO direct-imaging survey with the intent to prove that the object labelled as a contaminant by \citet{2012ApJ...755...42V} is not able to produce the detected transiting signal and to confirm it as second planet. We performed our first high-resolution direct observations in December 2012 and obtained JHK-band photometry (Tables~\ref{TableCVSOAsPhot}\,and\,\ref{TableLog}
and Fig.~\ref{FigImage}). 

During the course of their study of the transiting planet CVSO 30 b, the PTF team used Keck II/NIRC2 H-band images obtained in 2010 to identify contaminants capable of creating a false-positive signal mimicking a planet.
We re-reduced these data and found that they already contain the planetary companion CVSO 30 c that we report here. In Fig.~\ref{FigImage} we show the companion, erroneously given to lie south-east in \citet{2012ApJ...755...42V}; it is located north-east of the host star CVSO 30.

We astrometrically calibrated the VLT/NACO detector epoch using a sub-field of 47 Tuc (Table~\ref{TableAstroCal}) to determine pixel scale and detector orientation in order to find precise values for the separation of CVSO 30 c with respect to CVSO 30 in right ascension and declination. From this,
we find the object to be $\sim$1.85$''$ NE of CVSO 30 at a position angle of $\sim$70$^{\circ}$ from north towards east, corresponding
to a projected separation of 662 $\pm$ 96 au at the distance of the star.
No astrometric calibrator could be found in the night of the Keck observations (hence the position of the object is given in pixels in Table~\ref{TableCVSOAsPhot}), but we note that the
position is consistent with the VLT data. We used the nominal pixel scale of NIRC2 of 0.009942 $''$/pixel ($\pm$ 0.00005$''$) and assumed 0$^{\circ}$ detector orientation for the Keck epoch, which results in 1.744 arcsec right ascension and 0.630 arcsec declination separation in the relative position of CVSO 30 c with respect to its host star. The consistency was
expected for a companion because the proper motion of CVSO 30 is too small to distinguish a background source from a sub-stellar companion based on common proper motion (Table~\ref{TableCVSOsys}).

CVSO 30 is in general currently not suitable for a common proper motion analysis because the errors in proper motion exceed the proper motion values (Table~\ref{TableCVSOsys}).
As orbital motion around the host star might be detectable, we performed a dedicated orbit estimation for the wide companion.
The analysis shows that even after two to three years of epoch difference, no significant orbital motion is expected for the wide companion (Fig.~\ref{FigOrbit}).

Using the Two Micron All Sky Survey (2MASS) \citep{2003tmc..book.....C,2006AJ....131.1163S}
photometry for the primary and our NACO images for differential brightness measurements, we find CVSO 30 c to exhibit an unusually blue H-Ks colour, while its J-H colour indicates the companion candidate to be redder than the primary. This implies that the companion is too red to be an eclipsing background binary mimicking the transiting signal of CVSO 30 as a false-positive signal, which is further indication for the planetary nature of CVSO 30 b.

The differential photometry (Table~\ref{TableCVSOAsPhot}) of CVSO 30 c was achieved using psf fitting with the \textit{Starfinder} package of IDL \citep{2000SPIE.4007..879D} using the primary star
CVSO 30 as psf reference. First the noise of the final jittered image was computed, taking the photon noise, the gain, RON, and the number of combined images into account, and then it was transferred to the starfinder routine for psf fitting. This resulted in the values given in Table \ref{TableCVSOAsPhot}. The values were checked with aperture photometry.

\begin{table}
\caption{Photometric rejection significance, spectroscopic reduced $\chi^2$ results, and corresponding formal significance without systematics for different comparison objects.}
\label{TableChi}
\centering
\begin{scriptsize}
\begin{tabular}{ccccccc}
\hline
Object & SpT & \multicolumn{2}{c}{Photometry} & add. & \multicolumn{2}{c}{Spectroscopy}                  \\
       &     & J-H           & H-Ks           & ref. & H-band                  & K-band                  \\
       &     & [$\sigma$]    & [$\sigma$]     &      & [$\sigma$\,/\,$\chi^2_r$] & [$\sigma$\,/\,$\chi^2_r$] \\
\hline\hline
HD 237903 & K7V        & 3.4 & 0.5 & [1]       & >6\,/\,2.66 &  >6\,/\,1.60 \\
Gl 846    & M0V        & 2.8 & 1.8 & [1]       & >6\,/\,2.38 & 5.4\,/\,1.51 \\
Gl 229    & M1V        & 0.6 & 0.3 & [1]       & >6\,/\,2.37 & 5.3\,/\,1.50 \\
Gl 806    & M2V        & 4.5 & 2.5 & [1]       & >6\,/\,2.73 & 4.3\,/\,1.40 \\
Gl 388    & M3V        & 3.7 & 2.8 & [2],[1]   & >6\,/\,2.57 & 3.7\,/\,1.33 \\
Gl 213    & M4V        & 5.5 & 2.6 & [2],[1]   & >6\,/\,2.80 & 2.5\,/\,1.21 \\
Gl 51     & M5V        & 3.8 & 3.5 & [2],[1]   & >6\,/\,2.47 & 2.6\,/\,1.21 \\
Gl 406    & M6V        & 3.4 & 4.6 & [2],[1]   & >6\,/\,2.50 & 2.5\,/\,1.20 \\
Gl 644C   & M7V        & 4.1 & 5.1 & [2],[1]   & >6\,/\,2.87 & 2.2\,/\,1.17 \\
Gl 752B   & M8V        & 2.6 & 6.4 & [2],[1]   & >6\,/\,2.76 & 2.3\,/\,1.18 \\
LHS 2065  & M9V        & 1.7 & 7.5 & [1]       & >6\,/\,2.45 & 2.2\,/\,1.17 \\
LHS 2924  & L0         & 1.4 & 6.5 & [2],[1]   & >6\,/\,2.77 & 2.1\,/\,1.16 \\
2MUCD 20581 & L1       & 2.2 & 7.5 & [2]       & >6\,/\,3.96 & 3.7\,/\,1.33 \\
Kelu-1AB  & L2+L3.5    & 2.2 & 9.8 & [2]       & >6\,/\,3.68 & 3.6\,/\,1.32 \\
2MUCD 11291 & L3       & 1.8 & >10 & [2]       & >6\,/\,3.66 & 3.8\,/\,1.34 \\
2MUCD 12128 & L4.5     & 5.5 & >12 & [2]       & >6\,/\,3.09 & 3.4\,/\,1.29 \\
2MUCD 11296 & L5.5     & 1.3 & >10 & [2]       & >6\,/\,4.60 & 5.5\,/\,1.52 \\
2MUCD 11314 & L6       & 2.0 & 8.4 & [2]       & >6\,/\,3.64 &  >6\,/\,1.66 \\
2MUCD 10721 & L7.5     & 5.8 & >11 & [2]       & >6\,/\,3.49 & 3.4\,/\,1.29 \\
2MUCD 10158 & L8.5     & 2.5 & 9.8 & [2]       & >6\,/\,4.87 & 5.0\,/\,1.47 \\
SDSS 1520+354 & T0     & 1.0 & 5.4 & [3]       & >6\,/\,4.63 &  >6\,/\,2.15 \\
SDSS 0909+652 & T1.5   & 0.3 & 0.4 & [4]       & >6\,/\,8.04 &  >6\,/\,3.64 \\
SDSS 1254-012 & T2     & 0.8 & 2.0 & [2]       & >6\,/\,7.97 &  >6\,/\,2.90 \\
2MASS 055-140 & T4     & 9.4 & 0.1 & [2]       & >6\,/\,16.2 &  >6\,/\,19.1 \\ 
\hline
HD 204585 & M4.5III    & 0.4 & 0.8 & [1]       & >6\,/\,1.86 &  >6\,/\,1.88 \\
HD 175865 & M5III      & 0.1 & 0.5 & [1]       & >6\,/\,1.91 &  >6\,/\,1.78 \\
BK Vir    & M7III      & 0.4 & 0.9 & [1]       & >6\,/\,2.47 &  >6\,/\,1.72 \\
HY Aqr    & M8-9III    & 0.6 & 8.5 & [1]       & >6\,/\,5.78 &  >6\,/\,1.61 \\
Galaxies  & various    & 4.2 & 3.0 & [5],[6]   & >6\,/\,2.28 &  >6\,/\,1.61 \\
Quasars   & ---        & 4.4 & 3.9 & [5]       & ---         & ---          \\
White Dwarfs & various & 6.4 & 3.9 & [5]       & ---         & ---          \\      
\hline
CVSO 30   & M3         & 2.7 & 1.7 &           & >6\,/\,3.31 & 6.0\,/\,1.57 \\
PZ Tel B  & M7         & 2.1 & 1.4 & [7]       & >6\,/\,3.29 & 3.0\,/\,1.25 \\
CT Cha b  & M9         & 0.4 & 1.3 & [8],[9]   & >6\,/\,2.27 & 1.8\,/\,1.13 \\
2M0441 Bb & L1         & 0.5 & 2.4 & [10]      & >6\,/\,3.13 & 1.9\,/\,1.13 \\
1RXS 1609 b   & L4     & 1.1 & 3.0 & [11]      & >6\,/\,2.70 & 2.3\,/\,1.18 \\
$\beta$ Pic b & L4     & 1.0 & 3.5 & [12]      & >6\,/\,2.06 & ---          \\
2M1207 b      & L7     & 3.5 & 4.4 & [13]      & >6\,/\,2.66 & 2.5\,/\,1.20 \\
S Ori 64      & L/T    & 0.9 & 0.7 & [14]      & ---         & ---          \\
DP (Fig.~\ref{FigSpec2}) & --- & --- & --- & [15] & 2.2\,/\,1.16 & 2.0\,/\,1.14 \\
\hline
\end{tabular}                                                                                                    
\begin{flushleft}
References: [1] \citet{2009ApJS..185..289R}, [2] \citet{2005ApJ...623.1115C}, [3] \citet{2010ApJ...710.1142B}, [4] \citet{2006AJ....131.2722C}, [5] \citet{2006MNRAS.367..454H}, [6] \citet{2001MNRAS.326..745M}, [7] \citet{2014A&A...566A..85S}, [8] \citet{2009AIPC.1094..852S}, [9] \citet{2008AandA...491..311S}, [10] \citet{2015ApJ...811L..30B},
[11] \citet{2008ApJ...689L.153L}, [12] \citet{2015ApJ...798L...3C}, [13] \citet{2010A&A...517A..76P},
[14] \citet{2012ApJ...754...30P}, from VISTA to 2MASS magnitudes
using colour equations from http://casu.ast.cam.ac.uk/surveys-projects/vista/technical/photometric-properties,
[15] \citet{2008ApJ...675L.105H}                                                                       
\end{flushleft}
\end{scriptsize}
\end{table}

As given in \citet{2012ApJ...755...42V}, our psf reference CVSO 30 varies by 0.17 mag (min to max) in the R band, consistent with our estimates within YETI.
The present steep wavelength dependence of the variability amplitudes is best described by hot star-spots \citep{2015MNRAS.450.3991K}, therefore we can extrapolate from measurements of the very similar T Tauri GQ Lup \citep{2007A&A...468.1039B}. According to this, 0.17 mag in R correspond to about 0.1 mag and 0.055 mag variability in J and Ks band, respectively. As the hot spots change the bands simultaneously,
this gives rise to a maximum systematic offset of 0.045 mag in J-Ks colour. We give an estimate of this variability as black error bars for a possible additional systematic offset of CVSO 30 c in Fig.~\ref{FigPhot}.

The colours of CVSO 30 and CVSO 30 c are very similar (Table \ref{TableCVSOAsPhot}\,and\,Fig.~\ref{FigPhot}). We currently lack a spectrum of CVSO 30 c in J band, therefore we used the M3V star Gl 388 \citep{2005ApJ...623.1115C,2009ApJS..185..289R} and the L3/L4 brown dwarf 2MASS J11463449+2230527 \citep{2005ApJ...623.1115C} to derive a preliminary filter correction between 2MASS and NACO for CVSO 30 and CVSO 30 c. The colours of CVSO 30 are well known from 2MASS (Table \ref{TableCVSOsys}), the differential brightnesses to CVSO 30 c vary from NACO to 2MASS by 28 mmag in J, -21 mmag in H, and -38 mmag in Ks. Thus CVSO 30 c is 49 mmag redder in J-H and 17 mmag redder in H-Ks in 2MASS (red in Fig.~\ref{FigPhot}) than in the NACO results (grey in Fig.~\ref{FigPhot}).

In Fig.~\ref{FigPhot} and Table \ref{TableChi} we compare CVSO 30 c to the colours of several possible sources. We find that background stars of spectral types OBAFGK are too blue in J-H, late-M dwarfs are too blue in J-H and too red in H-K, while foreground L- and late T-dwarfs are either too red in H-K or too blue in J-H. In addition, background galaxies, quasars, and H/He white dwarfs are also inconsistent with the values of CVSO 30 c. Only late-type giants, early-M and early-T dwarfs, and planetary mass free-floating objects such as are found in the $\sigma$ Orionis star cluster have comparable colours \citep{2000Sci...290..103Z,2012ApJ...754...30P}.


\section{CVSO 30 c spectroscopic analysis}

\begin{figure}
  \centering
  \includegraphics[width=0.16\textwidth]{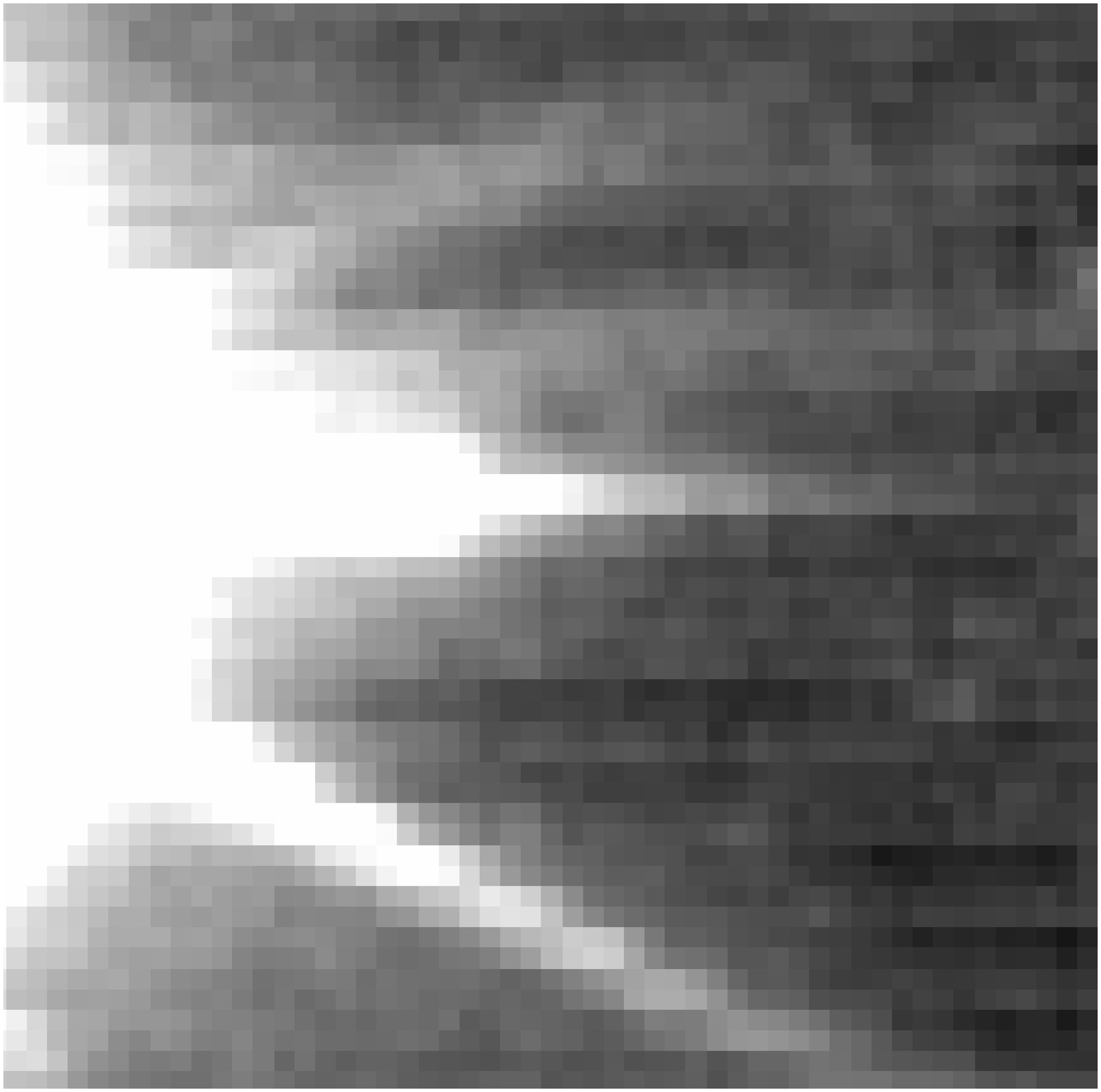}
  \includegraphics[width=0.16\textwidth]{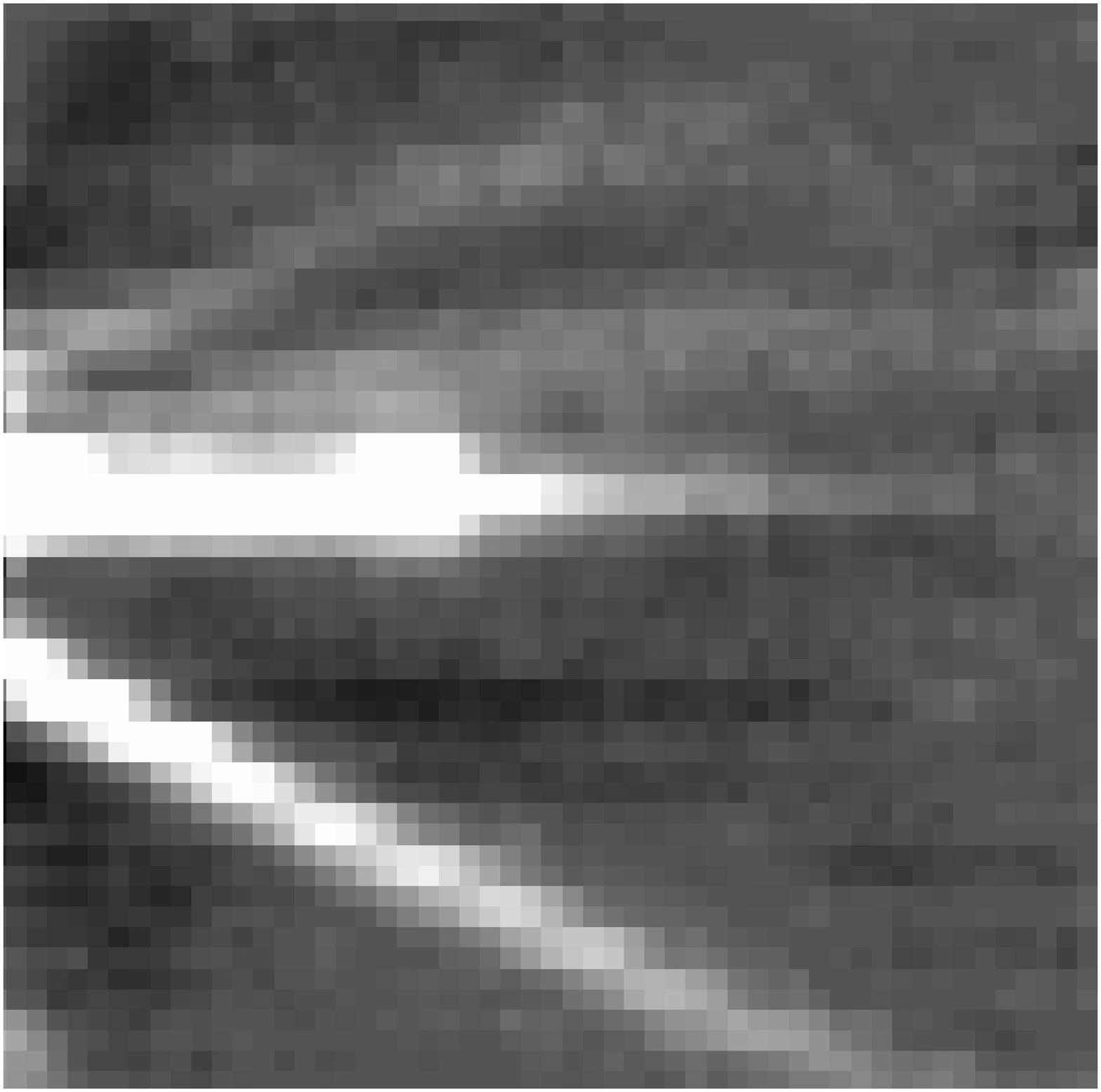}
  \includegraphics[width=0.16\textwidth]{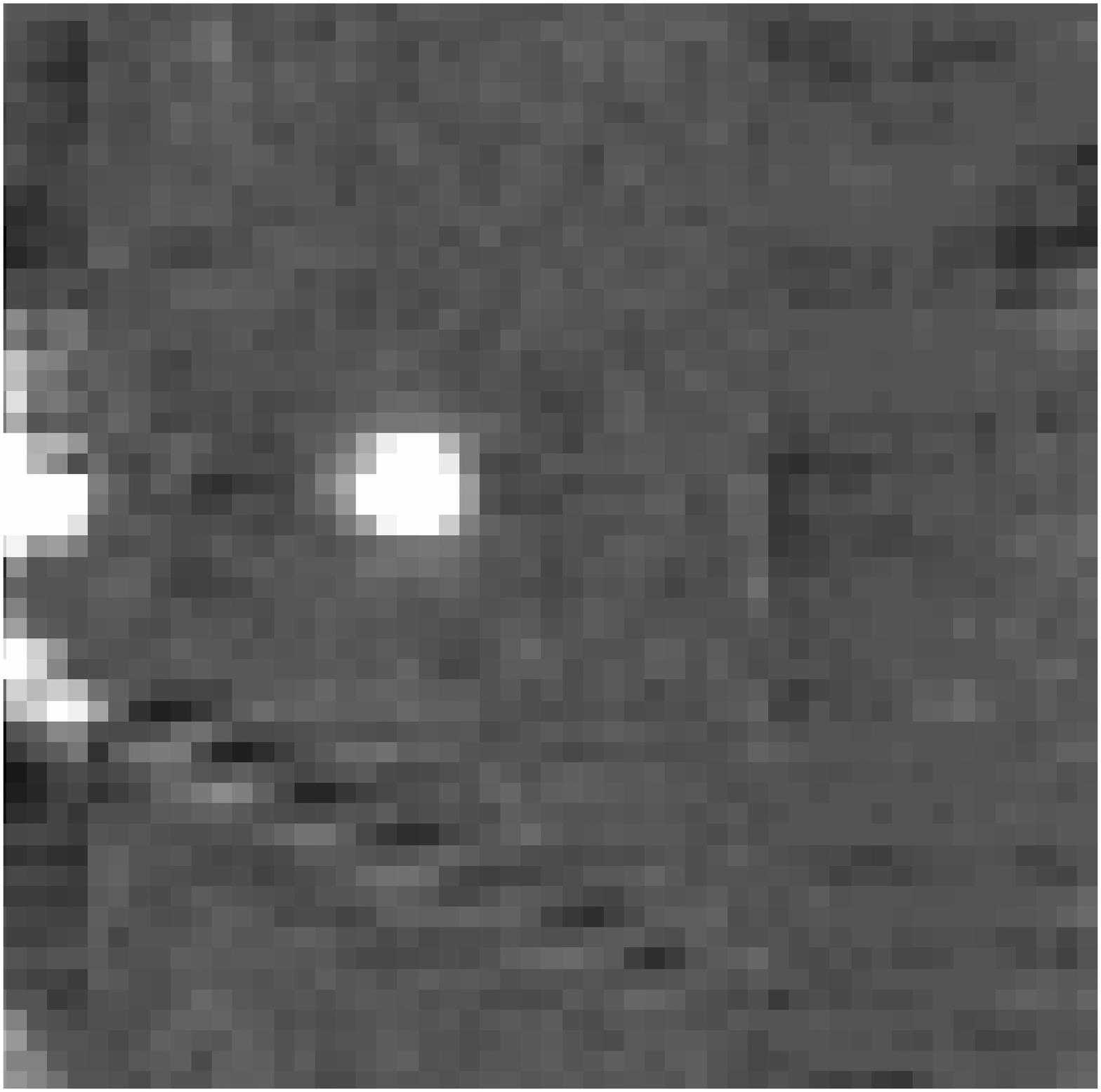}
  \caption{Median in wavelength direction of the reduced VLT/SINFONI integral field cubes. \textit{Left:} Cube after reduction. \textit{Centre:} Cube after removing the primary halo, assumed to be centred at the separation of 1.85$``$, as measured in the VLT/NACO images. North is about 70$^{\circ}$ from the right-hand side towards the bottom of the plots. \textit{Right:} Cube after removing
the primary halo, spectral deconvolution, and polynomial flattening of the resulting background, used for the extraction of the final spectrum.}
  \label{FigCubes}
\end{figure}

\begin{figure*}
  \centering
22  \includegraphics[width=0.68\textwidth]{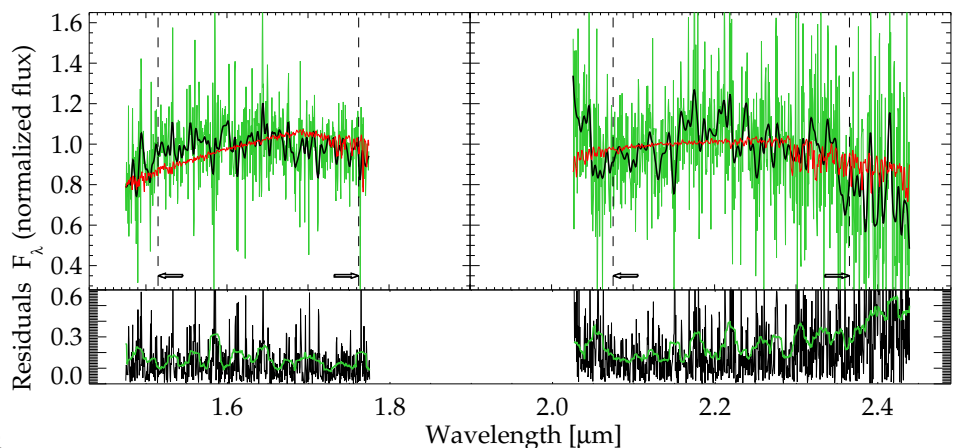}
  \caption{Spectrum of CVSO 30 c as extracted from the spectral deconvolution-corrected cube in the right panel of Fig.~\ref{FigCubes}. \textit{Top:} The spectrum in resolution 700 (black) is
  shown after binning of the original extracted spectrum in resolution 1500 (green). The best-fitting Drift-Phoenix model of \cite{2008ApJ...675L.105H} is shown in red, fitting both the
  individually normalised H and K spectra. This type of normalisation was necessary because the redder colour of the models, in comparison to the unusually blue nature of CVSO 30 c, would steer
  the best-fitting model to higher temperatures, which would
prevent fitting the individual features in H and K band. The best-fitting model (red) corresponds to 1600 K, surface gravity log g 3.6 dex,
  metallicity [M/H] 0.3 dex, and 0.19 mag of visual extinction. \textit{Bottom:} Absolute value of the difference between spectrum and model from the top panel (black) versus noise floor at the
  corresponding position (green).
  }
  \label{FigSpec2}
\end{figure*}

\begin{figure}
  \centering
  \includegraphics[width=0.35\textwidth]{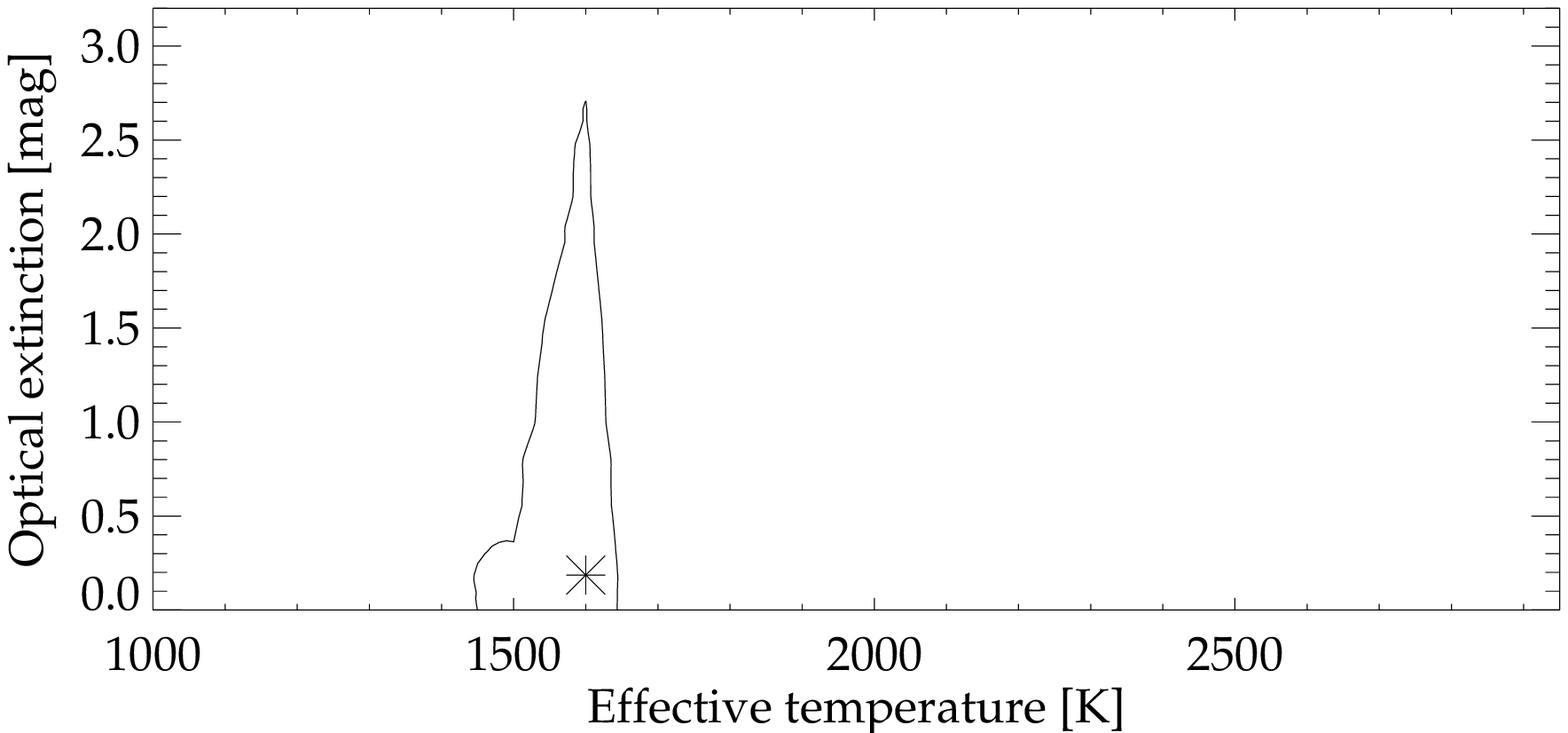}
  \includegraphics[width=0.35\textwidth]{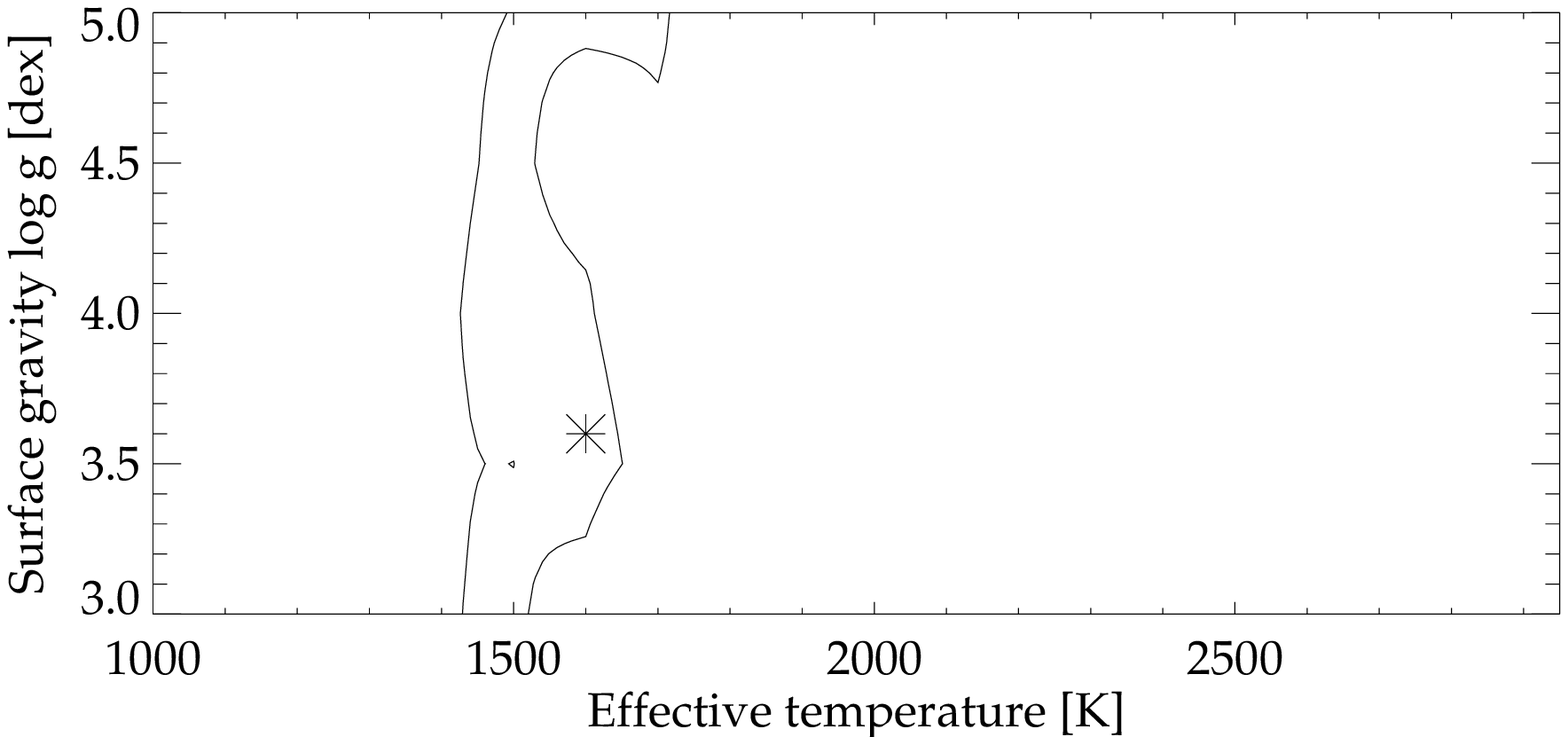}
  \includegraphics[width=0.35\textwidth]{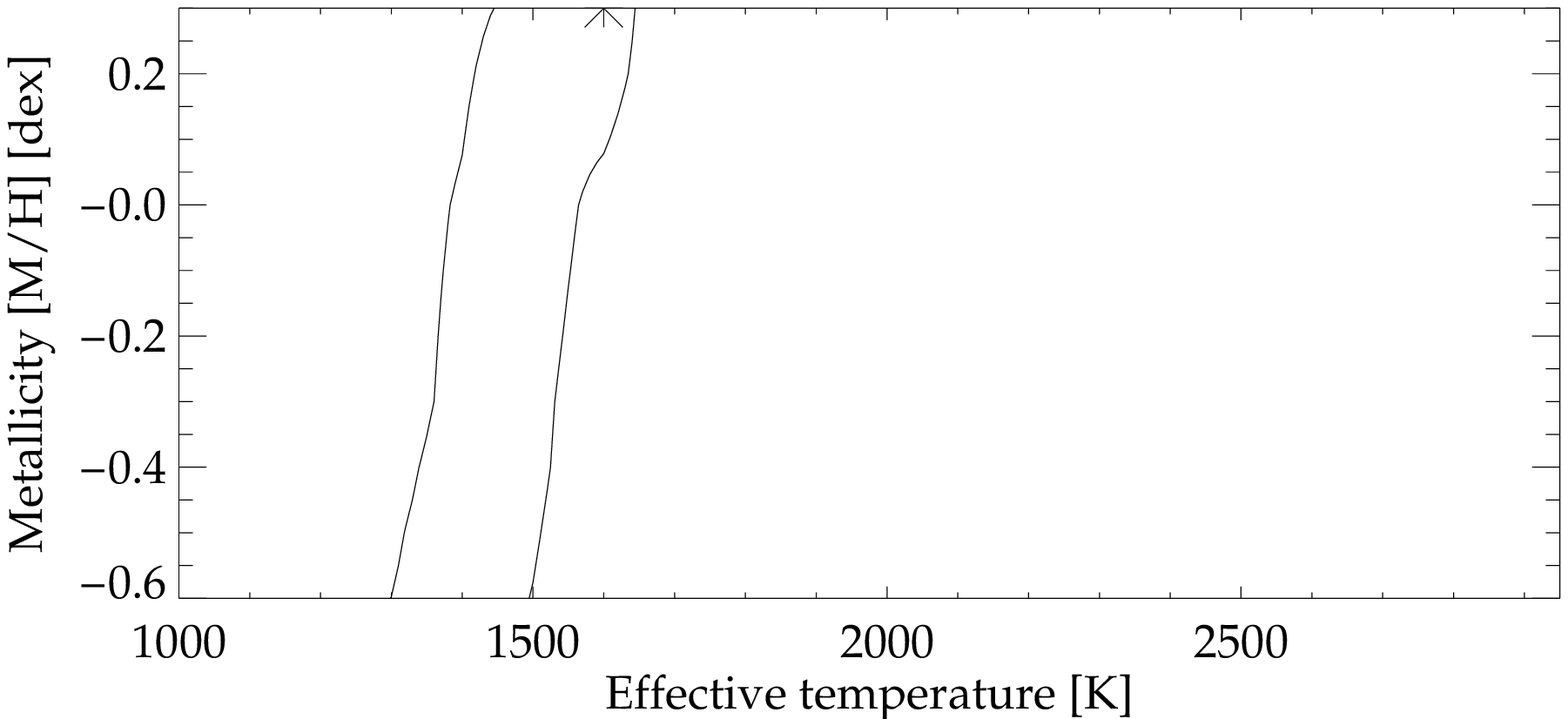}
  \caption{1 $\sigma$ contour plots of the $\chi^2$ Drift-Phoenix model fit to the spectrum shown in Fig.~\ref{FigSpec2}. Contour plot in extinction vs.~effective temperature (top), surface gravity log g vs.~effective temperature (centre) and metallicity [M/H] vs.~effective temperature (bottom). The fit shows a best fit at 1600 K, low extinction of 0.19 mag, higher values becoming
increasingly less likely, and a best fit at log g 3.6. While all surface gravities seem to be almost of equal probability, a high surface gravity foreground brown dwarf can be excluded from the shape of the H band in Fig.~\ref{FigSpec}. Although the young planetary models differ in photometric colours, this could be because of a not yet fully understood change in the cloud properties at the L-T transition that is indicated by the change in brightness of the L-T transition with age of the system, which we show in Fig.~\ref{FigColMag}.}
  \label{FigContSpec}
\end{figure}

A common proper motion analysis is not feasible because of the low proper motion of the host star (Table~\ref{TableCVSOsys}), therefore we carried out spectroscopic follow-up observations at the end of 2013, using the ESO VLT integral field unit SINFONI. The observations were made in H+K band with 100 mas/spaxel scale (FoV: 3 arcsec x 3 arcsec). The instrument provides information in the two spatial directions of the sky in addition to the simultaneous H- and K-band spectra. An unfortunate timing of the observations led to a parallactic angle at which a spike, 
probably of the telescope secondary mounting, was superimposed onto the well-separated spectrum of the companion candidate CVSO 30 c.

After correcting, the resulting spectrum can be compared to model atmospheres to determine its basic properties and to other sub-stellar companions to assess its youth and the reliability of the models at this low age, surface gravity, and temperature regime.

In an attempt to optimally subtract the spike of the host star, we performed several standard and customised reduction steps. After dark subtraction, flat-fielding, wavelength calibration, and
cube reconstruction, we found that the spike was superimposed onto the companion in every one of the three individual exposures, but at slightly different orientation angles
(Fig.~\ref{FigCubes}, left panel). As a first step, we used the NACO astrometry to determine the central position of the primary, which is itself outside the observed field of view of the integral field observations. The orientation of the SINFONI observations was intentionally chosen to leave the connection line of primary and companion exactly in x direction. The primary is
about 1.85 arcsec exactly to the left of CVSO 30 c in the data because the x direction offers a twice as good sampling regarding the number of pixels for the separation. 
We were thus able to subtract the radial symmetric halo of the host star from the data cube (Fig.~\ref{FigCubes}, central panel) using the nominal spatial scale.
This is necessary because the halo of the primary star is determined by the AO performance at the different wavelength.
At this stage, we extracted a first spectrum by subtracting an average spectrum of the spike, left and right of the companion psf from the superposition of companion and spike.
We find the results in Fig.~\ref{FigSpecAppend} before (red spectrum) and after (blue spectrum) spike subtraction, which also removes the still-present OH lines.
The horizontal spike in Fig.~\ref{FigCubes} appears too narrow to the right. This is a projection effect because the rotation of the spike within the three median-combined cubes
leads to less overlap on the right-hand side of the cube than on the left-hand side. For this reason, the continuum in Fig.~\ref{FigSpecAppend} is not trustworthy because the flux of the spike
below the companion candidate is not the average of the spike flux to the left and right of the object.

We tried several methods to remove the spike and decided to follow the spectral deconvolution technique \citep{2002ApJ...578..543S,2007MNRAS.378.1229T}.
This method is able to distinguish both the wavelength-dependent airy rings and speckles and the spike from the light of the wavelength-independent companion position by using the long wavelength coverage of the observed data cube.
As given in \citet{2007MNRAS.378.1229T} for the same instrument, the bifurcation radius for SINFONI H+K is for $\epsilon$=1.1 r=246 mas, and for $\epsilon$=1.2 r=268 mas, which means that parts of the
data without contamination of the companion could be found at the much higher separation of about 1.85 arcsec.
The reduction was then completed by applying a polynomial background correction around CVSO 30 c because the previous reduction steps left a low spatial frequency remnant around it (Fig.~\ref{FigCubes}, right panel). Finally, the optimal extraction algorithm \citep{1986PASP...98..609H} was performed around the companion and subtracted by the corresponding background flux from the close, well-corrected vicinity, and the telluric atmosphere correction using HD 61957, a B3V spectroscopic standard observed in the same night.   

\begin{figure*}
  \centering
  \includegraphics[width=\textwidth]{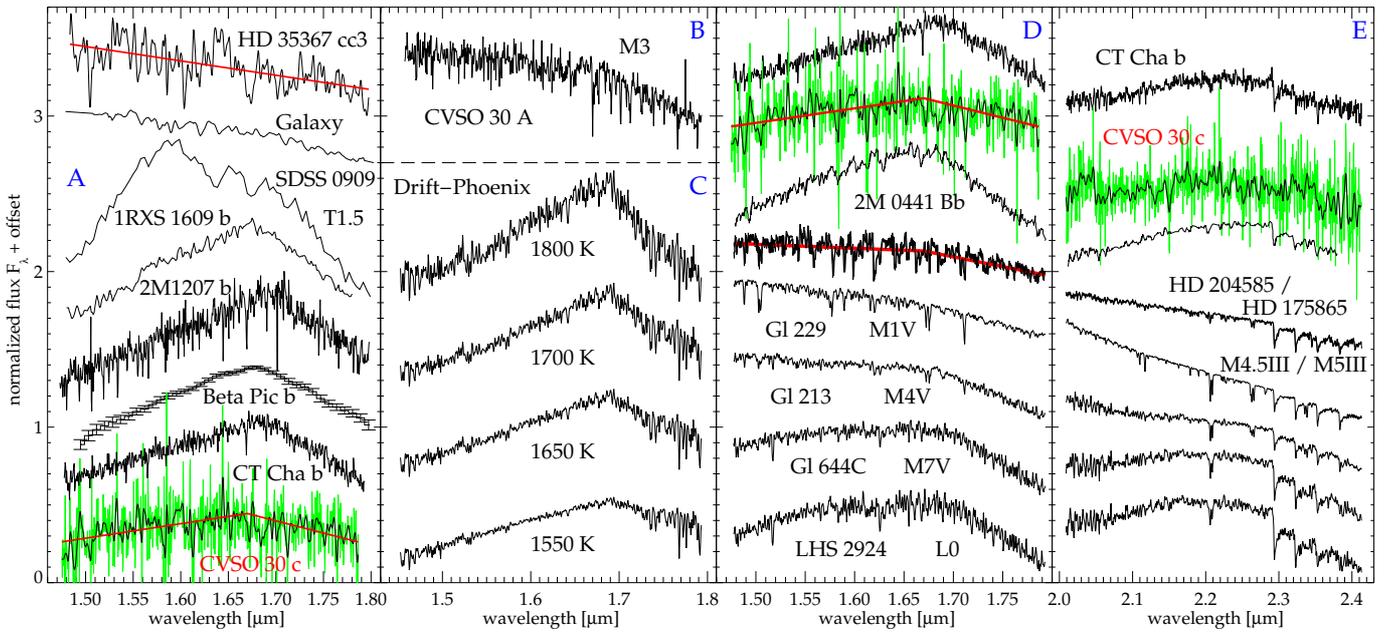}
  \caption{
  H-band spectrum of CVSO 30 c (lower left) compared to several known planetary candidates 
  and background objects (subplots A, D, E). 
  The triangular shape of the H band (A), with red linear fits guiding the eye, indicates that it is not a background galaxy, but a sub-stellar companion. Beta Pic b has approximately
  the same luminosity and temperature \citep{2015ApJ...798L...3C} but a different surface gravity, hence about twice the mass of CVSO 30 c. As shown (C), the Drift-Phoenix models indicate that the H band becomes less steep with temperature. This means that CVSO 30 c is even slightly lower in temperature than $\beta$ Pic b.
  In the upper left panel another candidate is shown, detected at 4.3'' from the A1 star HD 35367, which is about 0.5 mag brighter in the K band than CVSO 30 c, but is obviously located in the background.
  In addition, the H band (D) and K band (E) of CT Cha b and 2M 0441 Bb, the best-fitting comparison objects, are given in K band. These two and CVSO 30 c are given in (D, E) with identical offsets in H band and K band. Additionally, the best-fitting giants and a sample of late-type dwarfs is shown for comparison.
  References and individual reduced $\chi^2_r$ comparison values are given in Table \ref{TableChi}. Low-resolution spectra of free-floating planetary candidates are not shown, but can be found in \citet{2001ApJ...558L.117M}.
  }
  \label{FigSpec}
\end{figure*}

We first compare the spectrum of CVSO 30 c to spectra derived from Drift-Phoenix atmosphere simulations. These are dedicated radiative transfer models that take the strong continuum altering influence of dust cloud formation in the detectable parts of planetary atmospheres into account \citep{2008ApJ...675L.105H}. From a $\chi^2$ 
comparison of the H- and K-band spectra to the model grid, we find an effective temperature of about 1800 - 1900 K, while the individual fit of the H- and  K-band spectrum give a lower T$_{eff}$ of about 1600 K. In addition, the slope of the blue part of the triangular H band is too steep in the atmosphere models of about 1800 K and does not fit the continuum well. The higher T$_{eff}$ is only needed to fit the unusually blue H-Ks colour of the object, as already discussed in the previous photometry section and visible in Fig.~\ref{FigPhot}, since the models do yet not include a good description of the dust opacity drop at the L-T transition. We thus decided to fit the H and K band simultaneously, but normalising them individually, to cope with the unusual colours, while using all the present information for the fit. In this way, we find a best-fitting T$_{\mathrm{eff}}$=~1600$^{+120}_{-300}$ K, an extinction A$_V$=~0.19$^{+2.51}_{-0.19}$ mag, a surface gravity $\log{g}$\,[cm/s$^2$]=~3.6$^{+1.4}_{-0.6}$ dex, and a metallicity log[(M/H)/(M/H)$_{\odot}$]=~0.3$_{-0.9}$ dex at the upper supersolar edge of the grid.
The 1 $\sigma$ fitting contours are shown in Fig.~\ref{FigContSpec},
where they delineate the full regime for the error bars, and the best fit itself is shown in Fig.~\ref{FigSpec2}.

In Fig.~\ref{FigSpec} we compare the spectrum of CVSO 30 c to the triangular shaped H-band spectrum of the $\beta$ Pic b planet
that was obtained with the Gemini Planet Imager \citep[GPI,][]{2015ApJ...798L...3C}.
We also compare this to other planetary mass objects. $\beta$ Pic b is particularly suited as comparison object because it is young (10--20 Myr) and has about the same luminosity and effective temperature (1600--1700 K) while being of higher mass (10--12 M$_{\mathrm{Jup}}$).
We show linear fits to the blue and red part of the H band and the triangular shape of the chosen Drift-Phoenix models. In contrast to M5~--~L5 companions, 
for which the H$_2$O index in \citet{2007ApJ...657..511A} shows an increase in water absorption, the absorption becomes shallower for later spectral types. This means that even though the formal $\chi^2$ fit finds a best temperature of 1600 K for CVSO 30 c, the temperature is likely to be lower than for $\beta$ Pic b, exhibiting a steeper H-band spectrum.
The object's spectrum is not consistent with a giant of any spectral type. The best-fitting giants with consistent photometry (Fig.~\ref{FigPhot}\,and\,Table \ref{TableChi}) are shown as comparison in Fig.~\ref{FigSpec} and would be at a distance of about 200 Mpc. To improve the fit in the K band, the spectral type would have to be later than M7III, while the H band does not fit for these objects.
Finally, CVSO 30 c, which is comparable but younger, must have a lower surface gravity than $\beta$ Pic b, which is determined to have a 1$\sigma$ upper limit of $\log{g}$\,[cm/s$^2$]=~4.3 dex according to the linear prior orbit fit in \citet{2014AandA...567L...9B}.
This corrects the surface gravity of CVSO 30 c to $\log{g}$\,[cm/s$^2$]=~3.6$^{+0.7}_{-0.6}$ dex.

\section{AstraLux lucky imaging follow-up observations}

\begin{figure}
  \centering
  \includegraphics[width=0.30\textwidth]{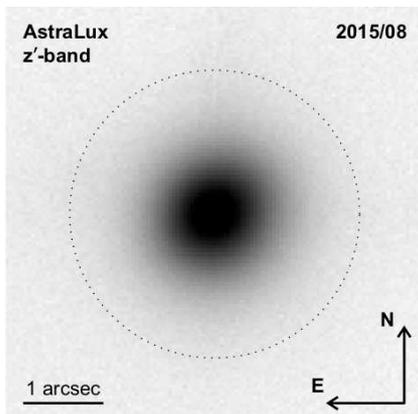}
  \caption{AstraLux z$^{\prime}$ band image of CVSO 30, taken on Aug 27, 2015. The dotted circle indicates an angular separation of 1.8 arcsec to CVSO 30 (see Fig.~\ref{AstraLuxLimit}). No other
objects are detected except for the star, which is located in the centre of the AstraLux image.
  }
  \label{AstraLux}
\end{figure}

\begin{figure}
  \centering
  \includegraphics[width=0.35\textwidth]{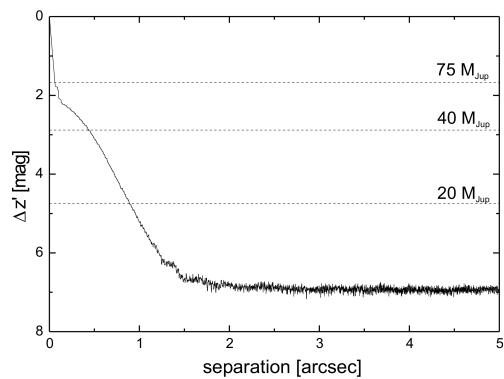}
  \caption{S/N=\,5 detection limit of our AstraLux observation of CVSO 30 (Fig.~\ref{AstraLux}). The reached magnitude difference that is dependent on the angular separation to the star is shown. The horizontal dashed lines indicate the expected magnitude differences of sub-stellar companions of the star at an age of 3 Myr. Beyond about 1.8 arcsec (or $\sim$640 au of projected separation), all companions with masses down to 10 M$_{\mathrm{Jup}}$ can be excluded around CVSO 30.
  }
  \label{AstraLuxLimit}
\end{figure}

We performed a follow-up of CVSO 30 with 2000\,s of AstraLux integration in z$^{\prime}$. The individual AstraLux images were combined using our own pipeline for the reduction of lucky imaging data. The fully reduced AstraLux image is shown in Fig.~\ref{AstraLux}. z$^{\prime}$ photometry of CVSO 30 was not measured so far, but can be derived from its magnitudes in other photometric bands using the colour transformation equations\footnote{$r - R = 0.77 \cdot (V - R) - 0.37$ and $r - z^{\prime} = 1.584 \cdot (R - I) - 0.386$} from \citet{2006A&A...460..339J}. The V- and R-band photometry of CVSO 30, as given by \citet{2005AJ....129..907B} and \citet{2012ApJ...755...42V} (V = 16.26 $\pm$ 0.19 mag, and R = 15.19 $\pm$ 0.085 mag), and the I-band photometry of the star, listed in the 2005 DENIS database (I = 13.695 $\pm$ 0.030 mag), yield z$^{\prime}$ = 13.66 mag.

The (S/N= 5) detection limit reached in the AstraLux observation is given in Fig.~\ref{AstraLuxLimit}. At an angular separation of about 1.8 arcsec from CVSO 30 (or $\sim$640 au of projected separation), companions that are $\Delta$z$^{\prime}$= 6.8 mag fainter than the star are still detectable at S/N = 5. The reached detection limit at this angular separation is z$^{\prime}$= 20.5 mag, which is just a tenth of magnitude above the limiting magnitude in the background-noise-limited region around the star at angular separations larger than 2 arcsec. This results in a limiting absolute magnitude of M$_{\mathrm{z}^{\prime}}$= 12.7 mag, allowing the detection of sub-stellar companions of the star with masses down to 10 M$_{\mathrm{Jup}}$ according to the evolutionary models
of \citet{2015A&A...577A..42B}. 

Furthermore, the AstraLux observations also exclude all young (3 Myr) stellar objects (mass higher than 75 M$_{\mathrm{Jup}}$) that are unrelated to CVSO 30, which are located in the AstraLux field of view at distances closer than about 3410 pc. All young M dwarfs with an age of 3 Myr and masses above 15 M$_{\mathrm{Jup}}$ (T$_{\mathrm{eff}}$ > 2400 K) can be ruled out up to 530 pc. All old stellar objects (mass higher than 75 M$_{\mathrm{Jup}}$) with an age of 5 Gyr can be excluded when they are located closer than about 130 pc.

The AstraLux upper limit results in z$^{\prime}$\,-\,Ks\,$\gtrsim$ 1.75 mag, which corresponds to excluding $\gtrsim$ 0.2 M$_{\odot}$ or $\gtrsim$ 3300 K \citep{2015A&A...577A..42B} as possible sources or about earlier than M4.5V in spectal type \citep{1995ApJS..101..117K}. Because any object later than $\sim$\,M2V\,/\,M3V can be excluded by $\gtrsim$ 4 $\sigma$ from JHKs photometry (Table \ref{TableChi}), no M dwarf can be a false positive of the new companion candidate CVSO 30 c.

\section{Mass determination and conclusions}

With the object brightness determined from the direct near-IR imaging and the information provided by the spectroscopic analysis, we can directly estimate the basic parameters of CVSO 30 c.  
To determine the luminosity, we considered the extinction law by \citet{1985ApJ...288..618R}, a bolometric correction of B.C.$_K$= 3.3$^{+0.0}_{-0.7}$ mag for spectral type L5-T4 \citep{2004AJ....127.3516G}, and a distance of 357$\,\pm\,$52 pc to the 25 Orionis cluster.
From the 2MASS brightness of the primary and the differential brightness measured in our VLT NACO data (Table~\ref{TableCVSOAsPhot}) and the extinction value towards the companion derived from spectroscopy, we find $\log{L_{bol}/L_{\odot}}= -3.78^{+0.33}_{-0.13}$~dex. From the luminosity and effective temperature, we calculate the radius to be R=~1.63$^{+0.87}_{-0.34}$ R$_{\mathrm{Jup}}$. In combination with the derived surface gravity, this would correspond to a mass of M=~4.3
M$_{\mathrm{Jup}}$, dominated in its errors by high distance and surface gravity uncertainties. 
While the latter value and the photometry (Fig.~\ref{FigPhot}) would be consistent with a high surface gravity, thus old foreground T-type brown dwarf, but inconsistent with an L-type brown dwarf, the available spectroscopy excludes an old T-type brown dwarf (Fig.~\ref{FigSpec}\, and \,Table \ref{TableChi}).
While the photometry is also consistent with early-M dwarfs, the K-band spectroscopy and z$^{\prime}$ upper limit show the opposite behaviour, being only consistent with late-M dwarfs, excluding all types with 
high significance. Similarly, the remaining H-band spectroscopy excludes all comparison objects.
Only the best-fitting Drift-Phoenix model (Fig.~\ref{FigSpec2}) shows 
low deviation in H band, consistent with the fact that the only available very young directly imaged planet candidates exhibit higher temperatures, thus a steeper H band (Fig.~\ref{FigSpec}).

Although recent observations by \citet{2015ApJ...812...48Y} cast doubts on the existence of the inner transiting planet candidate CVSO 30 b or PTFO 8-8695 b, we assume its existence throughout the remaining discussion because all five hypotheses have difficulties
in reproducing the observations presented in \citet{2015ApJ...812...48Y}, including for example~different types of starspots. The inner planet hypothesis gives another constraint, namely that the system has to be stable with both its planets. As described in \citet{2012ApJ...755...42V}, CVSO 30 b is very close to its Roche radius, the radius of stability. 
Assuming the values for mass of CVSO 30 b, its radius and orbital period (Tables~\ref{TableCVSOAsPhot} and \ref{TableCVSOPlanets}), we find from the Roche limit an upper limit for the mass of CVSO 30 of $\leq$ 0.92 M$_{\odot}$ for a stable inner system comprised of CVSO 30 A and b. 
This mass limit for CVSO 30 is fulfilled
at 1 Myr for masses of CVSO 30 c of $\leq$ 6.9 M$_{\mathrm{Jup}}$ at $\leq$ 760 pc up to 5.8 Myr with masses of CVSO 30 c of $\leq$ 9.2 M$_{\mathrm{Jup}}$ at $\leq$ 455 pc, according to BT-Settl evolutionary models \citep{2014IAUS..299..271A,2015A&A...577A..42B}. Higher ages are not consistent with the age estimate of the primary, but even at 20 Myr we find a mass of CVSO 30 c of $\leq$ 12.1 M$_{\mathrm{Jup}}$ at $\leq$ 340 pc. With the Roche stability criterion for CVSO 30 b, the previous calculations result in a mass estimate of M=~4.3$^{+4.9}_{-3.7}$ M$_{\mathrm{J up}}$ for CVSO 30 c.

For the approximate age of CVSO 30, 2--3 Myr BT-Settl evolutionary models \citep{2014IAUS..299..271A,2015A&A...577A..42B} predict an apparent brightness of m$_K\sim$ 18.5 mag (assuming the distance to 25 Ori), effective temperature $\sim$1575 K, mass 4\,--\,5 M$_{\mathrm{Jup}}$, and $\log{L_{bol}/L_{\odot}}\sim$~-3.8 dex. These expected values are very close to the best-fit atmospheric model spectra fits above, and even the derived visual extinction of about 0.19 mag is very close to the value of the primary $\sim$0.12 mag \citep{2005AJ....129..907B}.

\begin{figure}
  \centering
  \includegraphics[width=0.5\textwidth]{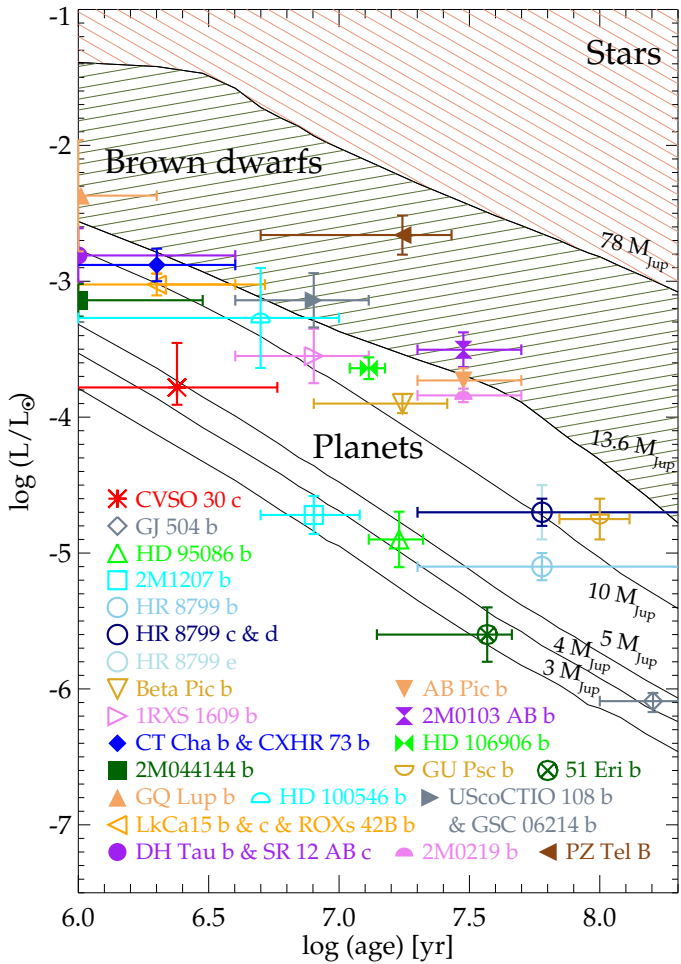}
  \caption{Evolution of young stars, brown dwarfs, and planets with BT-Settl evolutionary tracks \citep{2014IAUS..299..271A,2015A&A...577A..42B}. Shown are a few of the planet candidates known so far in comparison to the new sub-stellar companion candidate CVSO 30 c (see Table~\ref{TableRefEvol}).}
  \label{FigEvol}
\end{figure}

Of course, these evolutionary models can also be used to determine the resulting mass from the luminosity and age of the companion candidate and system, respectively. To put CVSO 30 c into context, we show the models and several of the currently known directly imaged planet candidates in Fig.~\ref{FigEvol}. The new companion is one of the youngest and lowest mass companions, and we find a mass of
4.7$^{+5.5}_{-2.0}$ M$_{\mathrm{Jup}}$ because the luminosity is not very precise as a result of the rather scarce knowledge of the distance of the system.
However, if we take temperature additionally into account, we find a more precise mass determination of 4.7$^{+3.6}_{-2.0}$ M$_{\mathrm{Jup}}$, which places CVSO 30 c well within the planetary regime and would mean that it is very close in mass to the probable inner companion of the system CVSO 30 b with about 2.8 -- 6.9 M$_{\mathrm{Jup}}$ \citep{2012ApJ...755...42V,2013ApJ...774...53B}.

\begin{figure}
  \centering
  \includegraphics[angle=90,width=0.5\textwidth]{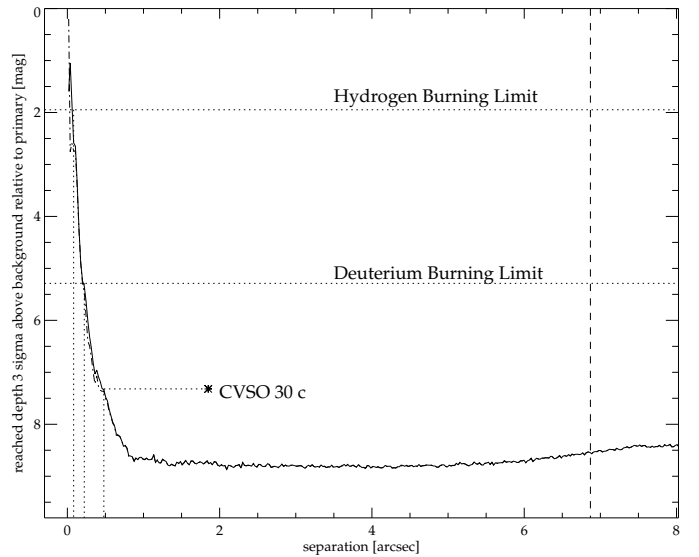}
  \caption{Dynamic range per pixel achieved in our VLT / NACO Ks-band observations, given as 3 $\sigma$ contrast to the primary star. The companion would have been detectable until 0.48 arc seconds or 171 au separation. A depth of 20.2 mag was reached at maximum, corresponding to 2.8 M$_{\mathrm{Jup}}$.}
  \label{FigDynRan}
\end{figure}

In Fig.~\ref{FigDynRan} we show the depth reached per pixel in the Ks-band epoch of 20.2 mag, corresponding to 2.8 M$_{\mathrm{Jup}}$ at the age of CVSO 30, using the same models as above. Brown dwarfs could be found from 30 au outwards, planets from 79 au outwards, and CVSO 30 c could have been found from 171 au outwards.

The core-accretion model 
\citep{1969Icar...10..109S,1973ApJ...183.1051G,1996Icar..124...62P},
one of the much debated planet formation scenarios, is unlikely to form an object in situ at
$\geq$662 au because the timescale would be prohibitively long at such separations. 
In principle, the object could also have formed in a star-like fashion by turbulent core fragmentation as in the case of a binary star system, since the opacity limit for fragmentation is a few Jupiter masses \citep{2009MNRAS.392..590B}, but the large separation and high mass ratio argue against this hypothesis.

The even more obvious possibility would be planet-planet scattering
because an inner planet candidate CVSO 30 b of comparable mass is present that could have been scattered inward at the very same scattering event.
Several authors simulated such events and found mostly highly eccentric orbits for the outer scattered planets of up to 100s or 1000s of au \citep{2009MNRAS.392..413S,2011ApJ...742...72N}, which is similar to the minimum separation of our outer planet candidate of 662 au. 
The closest match to CVSO 30 bc of a model simulation was presented by \citet{2011ApJ...742...72N} with an object at $\sim$300 au, which has an inner hot planet with
which it was scattered. Scattering or gravitational interaction might not be that uncommon as $72\%\pm16\%$ 
of hot Jupiters are part of multi-planet and/or multi-star systems \citep{2015ApJ...800..138N}.

The luminosity of CVSO 30 c is only consistent with hot-start models that usually represent the objects formed by gravitational disk-instability, not with cold-start models that are attributed to 
core-accretion-formed planets \citep{2007ApJ...655..541M}. However, as stated in \citet{2012ApJ...745..174S}, first-principle calculations cannot yet specify the initial (post-formation) entropies of objects with certainty in the different formation scenarios, hence CVSO 30 c could have formed through a gravitational disk-instability or core accretion and might have been scattered with CVSO 30 b afterwards. 

\begin{figure}
  \centering
  \includegraphics[width=0.5\textwidth]{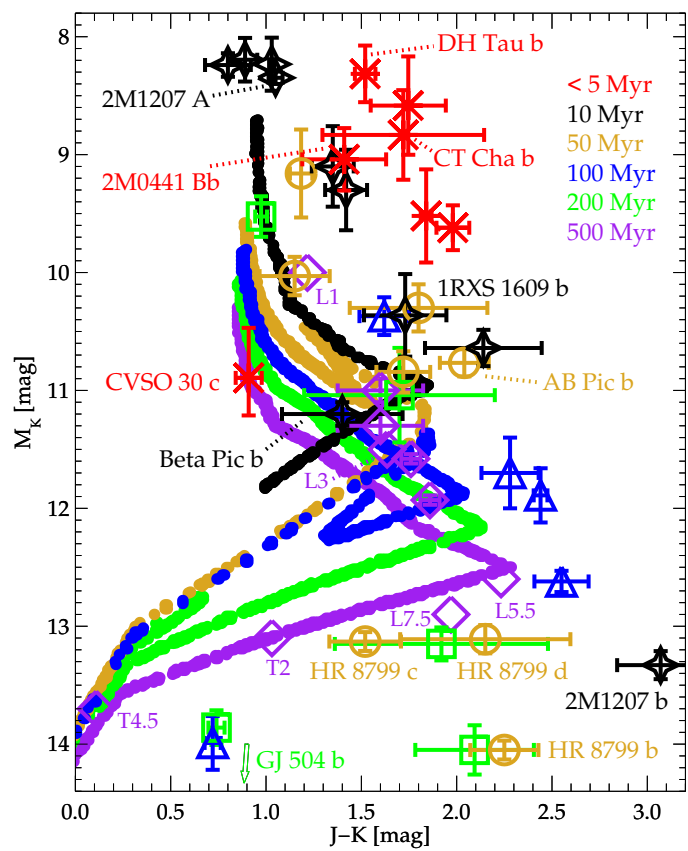}
  \caption{Colour-magnitude diagram of a simulated cluster brown dwarf population from \citet{2008ApJ...689.1327S}. Each sequence corresponds to a different age as given in the legend.
  Superimposed we show the positions of several planet candidates and CVSO 30 c. Its unusual blue colour can most likely be attributed to its youth; it is about 2.4 Myr old. The younger the objects, the brighter they are because of not-yet-occurred contraction. Hence they reach the L- and T-dwarf regime at higher brightnesses. If this extrapolation is correct, CVSO 30 c is at the L-T transition, which is roughly consistent with its low effective temperature results. See the discussion and Table~\ref{TableRefColMag} and Fig.~\ref{FigColMagAppend} for details.
  }
  \label{FigColMag}
\end{figure}

In this context, it would also be important to clarify the nature of the unusually blue H-Ks colour of CVSO 30 c. It is consistent with colours of free-floating planets (Fig.~\ref{FigPhot}) and might
be caused by its youth, allowing the companion to be very bright, still already being at the L-T transition, which would be consistent with simulations of cluster brown dwarfs at very young ages and
their colours in \citet{2008ApJ...689.1327S} (Fig.~\ref{FigColMag}). This would imply a temperature at the lower end of the 1$\sigma$ errors found for CVSO 30 c, $\leq$ 1400 K, which is consistent with the less steep H band in comparison to $\beta$ Pic b of about 1600--1700K \citep{2015ApJ...798L...3C}, however, as shown in Fig.~\ref{FigSpec}. 
For old brown dwarfs the L-T transition occurs at T$_{eff}$ 1200--1400 K, when methane absorption bands start to be ubiquitously seen. However, in the $\sim$30 Myr old planet candidates around HR 8799 no strong methane is found, while the spectrum of the $\sim$90 Myr old object around GU Psc shows strong methane absorption \citep{2014ApJ...787....5N} all at temperatures of about 1000--1100 K. Thus the L-T transition might be gravity dependent \citep{2012ApJ...754..135M}.
Binarity of CVSO 30 c cannot be excluded either, which would also explain the unusual blue H-Ks colour.  

Since we cannot confirm that CVSO 30 c is co-moving with its host star from our proper motion analysis, we cannot exclude the possibility that CVSO 30 c is a free-floating young planet belonging to the 25 Ori cluster, which is not gravitationally bound to CVSO 30. However, such a coincidence is highly improbable. \citet{2000Sci...290..103Z} searched 847 arcmin$^2$ of the $\sigma$ Orionis star cluster for free-floating planets and found only six candidates in the survey with similar colours as CVSO 30 c. This means that the probability to find a free-floating planet by chance within a radius of 1.85'' around the transiting planet host star CVSO 30 is about 2$\cdot$10$^{-5}$.

With a mass ratio of planet candidate to star q= 0.0115 $\pm$ 0.0015, CVSO 30 c (and CVSO 30 b) is among the imaged planets with the lowest mass ratio
\citep[see e.g.][]{2014MNRAS.445.3694D}.

In summary, CVSO 30 b and c for the first time allow a comprehensive study of both a transiting and a directly imaged 
planet candidate within the same system, hence at the same age and even similar masses, using RV, transit photometry, direct imaging, and spectroscopy.
Within a few years, the GAIA satellite mission
\citep{2005ASPC..338....3P}
will provide the distance to the system to a precision of about 10 pc, which will additionally restrict the 
mass of CVSO 30 c. 
Simulations of a possible scattering event will profit from the current (end) conditions found for the system. Considering that the inner planet is very close to the Roche stability limit and the outer planet is far away from its host star, the future evolution and stability of the system is also very interesting for dedicated modelling. 
To investigate how often 
these scattering events occur, inner planets need also be searched for around other stars with directly imaged wide planets.

\begin{acknowledgements}
We thank the ESO and CAHA staff for support, especially during service-mode observations. Moreover, we would like to thank Jeff Chilcote and David Lafreni\`ere for kindly providing electronic versions of comparison spectra from their publications and the anonymous referee, the editor 
and our language editor 
for helpful comments that improved this manuscript.

TOBS and JHMMS acknowledge support by the DFG Graduiertenkolleg 1351 ``Extrasolar Planets and their Host Stars``.
RN and SR would like to thank the DFG for support in the Priority Programme SPP 1385 on the
“First Ten Million Years of the Solar system” in project NE 515/33-1.
SR is currently a Research Fellow at ESA/ESTEC.

This publication makes use of data products from the Two Micron All Sky Survey, which is a joint project of the University of Massachusetts and the Infrared Processing and Analysis Center/California Institute of Technology, funded by the National Aeronautics and Space Administration and the National Science Foundation. This research has made use of the VizieR catalog access tool and the Simbad database, both operated at the Observatoire Strasbourg.
This research has made use of NASA's Astrophysics Data System.

\end{acknowledgements}

\listofobjects



\begin{thebibliography}{161}
\expandafter\ifx\csname natexlab\endcsname\relax\def\natexlab#1{#1}\fi

\bibitem[{{Adams} \& {Laughlin}(2001)}]{2001Icar..150..151A}
{Adams}, F.~C. \& {Laughlin}, G. 2001, \icarus, 150, 151

\bibitem[{{Allard}(2014)}]{2014IAUS..299..271A}
{Allard}, F. 2014, in IAU Symposium, Vol. 299, IAU Symposium, ed. M.~{Booth},
  B.~C. {Matthews}, \& J.~R. {Graham}, 271--272

\bibitem[{{Aller} {et~al.}(2013){Aller}, {Kraus}, {Liu}, {Burgett}, {Chambers},
  {Hodapp}, {Kaiser}, {Magnier}, \& {Price}}]{2013ApJ...773...63A}
{Aller}, K.~M., {Kraus}, A.~L., {Liu}, M.~C., {et~al.} 2013, \apj, 773, 63

\bibitem[{{Allers} {et~al.}(2007){Allers}, {Jaffe}, {Luhman}, {Liu}, {Wilson},
  {Skrutskie}, {Nelson}, {Peterson}, {Smith}, \&
  {Cushing}}]{2007ApJ...657..511A}
{Allers}, K.~N., {Jaffe}, D.~T., {Luhman}, K.~L., {et~al.} 2007, \apj, 657, 511

\bibitem[{{Andrews} {et~al.}(2013){Andrews}, {Rosenfeld}, {Kraus}, \&
  {Wilner}}]{2013ApJ...771..129A}
{Andrews}, S.~M., {Rosenfeld}, K.~A., {Kraus}, A.~L., \& {Wilner}, D.~J. 2013,
  \apj, 771, 129

\bibitem[{{Artigau} {et~al.}(2015){Artigau}, {Gagn{\'e}}, {Faherty}, {Malo},
  {Naud}, {Doyon}, {Lafreni{\`e}re}, \& {Beletsky}}]{2015ApJ...806..254A}
{Artigau}, {\'E}., {Gagn{\'e}}, J., {Faherty}, J., {et~al.} 2015, \apj, 806,
  254

\bibitem[{{Baglin} {et~al.}(2007){Baglin}, {Auvergne}, {Barge}, {Michel},
  {Catala}, {Deleuil}, \& {Weiss}}]{2007AIPC..895..201B}
{Baglin}, A., {Auvergne}, M., {Barge}, P., {et~al.} 2007, in American Institute
  of Physics Conference Series, Vol. 895, Fifty Years of Romanian Astrophysics,
  ed. C.~{Dumitrache}, N.~A. {Popescu}, M.~D. {Suran}, \& V.~{Mioc}, 201--209

\bibitem[{{Bailey} {et~al.}(2014){Bailey}, {Meshkat}, {Reiter}, {Morzinski},
  {Males}, {Su}, {Hinz}, {Kenworthy}, {Stark}, {Mamajek}, {Briguglio}, {Close},
  {Follette}, {Puglisi}, {Rodigas}, {Weinberger}, \&
  {Xompero}}]{2014ApJ...780L...4B}
{Bailey}, V., {Meshkat}, T., {Reiter}, M., {et~al.} 2014, \apjl, 780, L4

\bibitem[{{Baraffe} {et~al.}(2015){Baraffe}, {Homeier}, {Allard}, \&
  {Chabrier}}]{2015A&A...577A..42B}
{Baraffe}, I., {Homeier}, D., {Allard}, F., \& {Chabrier}, G. 2015, \aap, 577,
  A42

\bibitem[{{Barnes} {et~al.}(2013){Barnes}, {van Eyken}, {Jackson}, {Ciardi}, \&
  {Fortney}}]{2013ApJ...774...53B}
{Barnes}, J.~W., {van Eyken}, J.~C., {Jackson}, B.~K., {Ciardi}, D.~R., \&
  {Fortney}, J.~J. 2013, \apj, 774, 53

\bibitem[{{Bate}(2009)}]{2009MNRAS.392..590B}
{Bate}, M.~R. 2009, \mnras, 392, 590

\bibitem[{{B{\'e}jar} {et~al.}(2008){B{\'e}jar}, {Zapatero Osorio},
  {P{\'e}rez-Garrido}, {{\'A}lvarez}, {Mart{\'{\i}}n}, {Rebolo},
  {Vill{\'o}-P{\'e}rez}, \& {D{\'{\i}}az-S{\'a}nchez}}]{2008ApJ...673L.185B}
{B{\'e}jar}, V.~J.~S., {Zapatero Osorio}, M.~R., {P{\'e}rez-Garrido}, A.,
  {et~al.} 2008, \apjl, 673, L185

\bibitem[{{Biller} {et~al.}(2010){Biller}, {Liu}, {Wahhaj}, {Nielsen}, {Close},
  {Dupuy}, {Hayward}, {Burrows}, {Chun}, {Ftaclas}, {Clarke}, {Hartung},
  {Males}, {Reid}, {Shkolnik}, {Skemer}, {Tecza}, {Thatte}, {Alencar},
  {Artymowicz}, {Boss}, {de Gouveia Dal Pino}, {Gregorio-Hetem}, {Ida},
  {Kuchner}, {Lin}, \& {Toomey}}]{2010ApJ...720L..82B}
{Biller}, B.~A., {Liu}, M.~C., {Wahhaj}, Z., {et~al.} 2010, \apjl, 720, L82

\bibitem[{{Binks} \& {Jeffries}(2014)}]{2014MNRAS.438L..11B}
{Binks}, A.~S. \& {Jeffries}, R.~D. 2014, \mnras, 438, L11

\bibitem[{{Bonavita} {et~al.}(2014){Bonavita}, {Daemgen}, {Desidera},
  {Jayawardhana}, {Janson}, \& {Lafreni{\`e}re}}]{2014ApJ...791L..40B}
{Bonavita}, M., {Daemgen}, S., {Desidera}, S., {et~al.} 2014, \apjl, 791, L40

\bibitem[{{Bonnefoy} {et~al.}(2013){Bonnefoy}, {Boccaletti}, {Lagrange},
  {Allard}, {Mordasini}, {Beust}, {Chauvin}, {Girard}, {Homeier}, {Apai},
  {Lacour}, \& {Rouan}}]{2013AandA...555A.107B}
{Bonnefoy}, M., {Boccaletti}, A., {Lagrange}, A.-M., {et~al.} 2013, \aap, 555,
  A107

\bibitem[{{Bonnefoy} {et~al.}(2014{\natexlab{a}}){Bonnefoy}, {Chauvin},
  {Lagrange}, {Rojo}, {Allard}, {Pinte}, {Dumas}, \&
  {Homeier}}]{2014AandA...562A.127B}
{Bonnefoy}, M., {Chauvin}, G., {Lagrange}, A.-M., {et~al.} 2014{\natexlab{a}},
  \aap, 562, A127

\bibitem[{{Bonnefoy} {et~al.}(2014{\natexlab{b}}){Bonnefoy}, {Marleau},
  {Galicher}, {Beust}, {Lagrange}, {Baudino}, {Chauvin}, {Borgniet}, {Meunier},
  {Rameau}, {Boccaletti}, {Cumming}, {Helling}, {Homeier}, {Allard}, \&
  {Delorme}}]{2014AandA...567L...9B}
{Bonnefoy}, M., {Marleau}, G.-D., {Galicher}, R., {et~al.} 2014{\natexlab{b}},
  \aap, 567, L9

\bibitem[{{Borucki} {et~al.}(2010){Borucki}, {Koch}, {Basri}, {Batalha},
  {Brown}, {Caldwell}, {Caldwell}, {Christensen-Dalsgaard}, {Cochran},
  {DeVore}, {Dunham}, {Dupree}, {Gautier}, {Geary}, {Gilliland}, {Gould},
  {Howell}, {Jenkins}, {Kondo}, {Latham}, {Marcy}, {Meibom}, {Kjeldsen},
  {Lissauer}, {Monet}, {Morrison}, {Sasselov}, {Tarter}, {Boss}, {Brownlee},
  {Owen}, {Buzasi}, {Charbonneau}, {Doyle}, {Fortney}, {Ford}, {Holman},
  {Seager}, {Steffen}, {Welsh}, {Rowe}, {Anderson}, {Buchhave}, {Ciardi},
  {Walkowicz}, {Sherry}, {Horch}, {Isaacson}, {Everett}, {Fischer}, {Torres},
  {Johnson}, {Endl}, {MacQueen}, {Bryson}, {Dotson}, {Haas}, {Kolodziejczak},
  {Van Cleve}, {Chandrasekaran}, {Twicken}, {Quintana}, {Clarke}, {Allen},
  {Li}, {Wu}, {Tenenbaum}, {Verner}, {Bruhweiler}, {Barnes}, \&
  {Prsa}}]{2010Sci...327..977B}
{Borucki}, W.~J., {Koch}, D., {Basri}, G., {et~al.} 2010, Science, 327, 977

\bibitem[{{Boss}(1997)}]{1997Sci...276.1836B}
{Boss}, A.~P. 1997, Science, 276, 1836

\bibitem[{{Boss}(2006)}]{2006ApJ...637L.137B}
{Boss}, A.~P. 2006, \apjl, 637, L137

\bibitem[{{Bowler} \& {Hillenbrand}(2015)}]{2015ApJ...811L..30B}
{Bowler}, B.~P. \& {Hillenbrand}, L.~A. 2015, \apjl, 811, L30

\bibitem[{{Bowler} {et~al.}(2013){Bowler}, {Liu}, {Shkolnik}, \&
  {Dupuy}}]{2013ApJ...774...55B}
{Bowler}, B.~P., {Liu}, M.~C., {Shkolnik}, E.~L., \& {Dupuy}, T.~J. 2013, \apj,
  774, 55

\bibitem[{{Bowler} {et~al.}(2015){Bowler}, {Liu}, {Shkolnik}, \&
  {Tamura}}]{2015ApJS..216....7B}
{Bowler}, B.~P., {Liu}, M.~C., {Shkolnik}, E.~L., \& {Tamura}, M. 2015, \apjs,
  216, 7

\bibitem[{{Brice{\~n}o} {et~al.}(2005){Brice{\~n}o}, {Calvet}, {Hern{\'a}ndez},
  {Vivas}, {Hartmann}, {Downes}, \& {Berlind}}]{2005AJ....129..907B}
{Brice{\~n}o}, C., {Calvet}, N., {Hern{\'a}ndez}, J., {et~al.} 2005, \aj, 129,
  907

\bibitem[{{Brice{\~n}o} {et~al.}(2007{\natexlab{a}}){Brice{\~n}o}, {Hartmann},
  {Hern{\'a}ndez}, {Calvet}, {Vivas}, {Furesz}, \&
  {Szentgyorgyi}}]{2007ApJ...661.1119B}
{Brice{\~n}o}, C., {Hartmann}, L., {Hern{\'a}ndez}, J., {et~al.}
  2007{\natexlab{a}}, \apj, 661, 1119

\bibitem[{{Brice{\~n}o} {et~al.}(2007{\natexlab{b}}){Brice{\~n}o}, {Preibisch},
  {Sherry}, {Mamajek}, {Mathieu}, {Walter}, \&
  {Zinnecker}}]{2007prpl.conf..345B}
{Brice{\~n}o}, C., {Preibisch}, T., {Sherry}, W.~H., {et~al.}
  2007{\natexlab{b}}, Protostars and Planets V, 345

\bibitem[{{Broeg} {et~al.}(2007){Broeg}, {Schmidt}, {Guenther}, {Gaedke},
  {Bedalov}, {Neuh{\"a}user}, \& {Walter}}]{2007A&A...468.1039B}
{Broeg}, C., {Schmidt}, T.~O.~B., {Guenther}, E., {et~al.} 2007, \aap, 468,
  1039

\bibitem[{{Burgasser} {et~al.}(2010{\natexlab{a}}){Burgasser}, {Cruz},
  {Cushing}, {Gelino}, {Looper}, {Faherty}, {Kirkpatrick}, \&
  {Reid}}]{2010ApJ...710.1142B}
{Burgasser}, A.~J., {Cruz}, K.~L., {Cushing}, M., {et~al.} 2010{\natexlab{a}},
  \apj, 710, 1142

\bibitem[{{Burgasser} {et~al.}(2010{\natexlab{b}}){Burgasser}, {Simcoe},
  {Bochanski}, {Saumon}, {Mamajek}, {Cushing}, {Marley}, {McMurtry}, {Pipher},
  \& {Forrest}}]{2010ApJ...725.1405B}
{Burgasser}, A.~J., {Simcoe}, R.~A., {Bochanski}, J.~J., {et~al.}
  2010{\natexlab{b}}, \apj, 725, 1405

\bibitem[{{Burningham} {et~al.}(2011){Burningham}, {Leggett}, {Homeier},
  {Saumon}, {Lucas}, {Pinfield}, {Tinney}, {Allard}, {Marley}, {Jones},
  {Murray}, {Ishii}, {Day-Jones}, {Gomes}, \& {Zhang}}]{2011MNRAS.414.3590B}
{Burningham}, B., {Leggett}, S.~K., {Homeier}, D., {et~al.} 2011, \mnras, 414,
  3590

\bibitem[{{Cameron}(1978)}]{1978M&P....18....5C}
{Cameron}, A.~G.~W. 1978, Moon and Planets, 18, 5

\bibitem[{{Carson} {et~al.}(2013){Carson}, {Thalmann}, {Janson}, {Kozakis},
  {Bonnefoy}, {Biller}, {Schlieder}, {Currie}, {McElwain}, {Goto}, {Henning},
  {Brandner}, {Feldt}, {Kandori}, {Kuzuhara}, {Stevens}, {Wong}, {Gainey},
  {Fukagawa}, {Kuwada}, {Brandt}, {Kwon}, {Abe}, {Egner}, {Grady}, {Guyon},
  {Hashimoto}, {Hayano}, {Hayashi}, {Hayashi}, {Hodapp}, {Ishii}, {Iye},
  {Knapp}, {Kudo}, {Kusakabe}, {Matsuo}, {Miyama}, {Morino}, {Moro-Martin},
  {Nishimura}, {Pyo}, {Serabyn}, {Suto}, {Suzuki}, {Takami}, {Takato},
  {Terada}, {Tomono}, {Turner}, {Watanabe}, {Wisniewski}, {Yamada}, {Takami},
  {Usuda}, \& {Tamura}}]{2013ApJ...763L..32C}
{Carson}, J., {Thalmann}, C., {Janson}, M., {et~al.} 2013, \apjl, 763, L32

\bibitem[{{Charbonneau} {et~al.}(2000){Charbonneau}, {Brown}, {Latham}, \&
  {Mayor}}]{2000ApJ...529L..45C}
{Charbonneau}, D., {Brown}, T.~M., {Latham}, D.~W., \& {Mayor}, M. 2000, \apjl,
  529, L45

\bibitem[{{Chauvin} {et~al.}(2004){Chauvin}, {Lagrange}, {Dumas}, {Zuckerman},
  {Mouillet}, {Song}, {Beuzit}, \& {Lowrance}}]{2004AandA...425L..29C}
{Chauvin}, G., {Lagrange}, A.-M., {Dumas}, C., {et~al.} 2004, \aap, 425, L29

\bibitem[{{Chauvin} {et~al.}(2005{\natexlab{a}}){Chauvin}, {Lagrange}, {Dumas},
  {Zuckerman}, {Mouillet}, {Song}, {Beuzit}, \&
  {Lowrance}}]{2005A&A...438L..25C}
{Chauvin}, G., {Lagrange}, A.-M., {Dumas}, C., {et~al.} 2005{\natexlab{a}},
  \aap, 438, L25

\bibitem[{{Chauvin} {et~al.}(2005{\natexlab{b}}){Chauvin}, {Lagrange},
  {Lacombe}, {Dumas}, {Mouillet}, {Zuckerman}, {Gendron}, {Song}, {Beuzit},
  {Lowrance}, \& {Fusco}}]{2005AandA...430.1027C}
{Chauvin}, G., {Lagrange}, A.-M., {Lacombe}, F., {et~al.} 2005{\natexlab{b}},
  \aap, 430, 1027

\bibitem[{{Chauvin} {et~al.}(2005{\natexlab{c}}){Chauvin}, {Lagrange},
  {Zuckerman}, {Dumas}, {Mouillet}, {Song}, {Beuzit}, {Lowrance}, \&
  {Bessell}}]{2005AandA...438L..29C}
{Chauvin}, G., {Lagrange}, A.-M., {Zuckerman}, B., {et~al.} 2005{\natexlab{c}},
  \aap, 438, L29

\bibitem[{{Chilcote} {et~al.}(2015){Chilcote}, {Barman}, {Fitzgerald},
  {Graham}, {Larkin}, {Macintosh}, {Bauman}, {Burrows}, {Cardwell}, {De Rosa},
  {Dillon}, {Doyon}, {Dunn}, {Erikson}, {Gavel}, {Goodsell}, {Hartung},
  {Hibon}, {Ingraham}, {Kalas}, {Konopacky}, {Maire}, {Marchis}, {Marley},
  {Marois}, {Millar-Blanchaer}, {Morzinski}, {Norton}, {Oppenheimer}, {Palmer},
  {Patience}, {Perrin}, {Poyneer}, {Pueyo}, {Rantakyr{\"o}}, {Sadakuni},
  {Saddlemyer}, {Savransky}, {Serio}, {Sivaramakrishnan}, {Song}, {Soummer},
  {Thomas}, {Wallace}, {Wiktorowicz}, \& {Wolff}}]{2015ApJ...798L...3C}
{Chilcote}, J., {Barman}, T., {Fitzgerald}, M.~P., {et~al.} 2015, \apjl, 798,
  L3

\bibitem[{{Chiu} {et~al.}(2006){Chiu}, {Fan}, {Leggett}, {Golimowski}, {Zheng},
  {Geballe}, {Schneider}, \& {Brinkmann}}]{2006AJ....131.2722C}
{Chiu}, K., {Fan}, X., {Leggett}, S.~K., {et~al.} 2006, \aj, 131, 2722

\bibitem[{{Close}(2010)}]{2010Natur.468.1048C}
{Close}, L. 2010, \nat, 468, 1048

\bibitem[{{Close} {et~al.}(2003){Close}, {Siegler}, {Freed}, \&
  {Biller}}]{2003ApJ...587..407C}
{Close}, L.~M., {Siegler}, N., {Freed}, M., \& {Biller}, B. 2003, \apj, 587,
  407

\bibitem[{{Close} {et~al.}(2007){Close}, {Zuckerman}, {Song}, {Barman},
  {Marois}, {Rice}, {Siegler}, {Macintosh}, {Becklin}, {Campbell}, {Lyke},
  {Conrad}, \& {Le Mignant}}]{2007ApJ...660.1492C}
{Close}, L.~M., {Zuckerman}, B., {Song}, I., {et~al.} 2007, \apj, 660, 1492

\bibitem[{{Currie} {et~al.}(2014){Currie}, {Burrows}, \&
  {Daemgen}}]{2014ApJ...787..104C}
{Currie}, T., {Burrows}, A., \& {Daemgen}, S. 2014, \apj, 787, 104

\bibitem[{{Cushing} {et~al.}(2005){Cushing}, {Rayner}, \&
  {Vacca}}]{2005ApJ...623.1115C}
{Cushing}, M.~C., {Rayner}, J.~T., \& {Vacca}, W.~D. 2005, \apj, 623, 1115

\bibitem[{{Cutri} {et~al.}(2003){Cutri}, {Skrutskie}, {van Dyk}, {Beichman},
  {Carpenter}, {Chester}, {Cambresy}, {Evans}, {Fowler}, {Gizis}, {Howard},
  {Huchra}, {Jarrett}, {Kopan}, {Kirkpatrick}, {Light}, {Marsh}, {McCallon},
  {Schneider}, {Stiening}, {Sykes}, {Weinberg}, {Wheaton}, {Wheelock}, \&
  {Zacarias}}]{2003tmc..book.....C}
{Cutri}, R.~M., {Skrutskie}, M.~F., {van Dyk}, S., {et~al.} 2003, {2MASS All
  Sky Catalog of point sources.}

\bibitem[{{De Rosa} {et~al.}(2014){De Rosa}, {Patience}, {Ward-Duong}, {Vigan},
  {Marois}, {Song}, {Macintosh}, {Graham}, {Doyon}, {Bessell}, {Lai},
  {McCarthy}, \& {Kulesa}}]{2014MNRAS.445.3694D}
{De Rosa}, R.~J., {Patience}, J., {Ward-Duong}, K., {et~al.} 2014, \mnras, 445,
  3694

\bibitem[{{Delorme} {et~al.}(2013){Delorme}, {Gagn{\'e}}, {Girard}, {Lagrange},
  {Chauvin}, {Naud}, {Lafreni{\`e}re}, {Doyon}, {Riedel}, {Bonnefoy}, \&
  {Malo}}]{2013AandA...553L...5D}
{Delorme}, P., {Gagn{\'e}}, J., {Girard}, J.~H., {et~al.} 2013, \aap, 553, L5

\bibitem[{{Diolaiti} {et~al.}(2000){Diolaiti}, {Bendinelli}, {Bonaccini},
  {Close}, {Currie}, \& {Parmeggiani}}]{2000SPIE.4007..879D}
{Diolaiti}, E., {Bendinelli}, O., {Bonaccini}, D., {et~al.} 2000, in Society of
  Photo-Optical Instrumentation Engineers (SPIE) Conference Series, Vol. 4007,
  Society of Photo-Optical Instrumentation Engineers (SPIE) Conference Series,
  ed. P.~L. {Wizinowich}, 879--888

\bibitem[{{Downes} {et~al.}(2014){Downes}, {Brice{\~n}o}, {Mateu},
  {Hern{\'a}ndez}, {Vivas}, {Calvet}, {Hartmann}, {Petr-Gotzens}, \&
  {Allen}}]{2014MNRAS.444.1793D}
{Downes}, J.~J., {Brice{\~n}o}, C., {Mateu}, C., {et~al.} 2014, \mnras, 444,
  1793

\bibitem[{{Ducourant} {et~al.}(2008){Ducourant}, {Teixeira}, {Chauvin},
  {Daigne}, {Le Campion}, {Song}, \& {Zuckerman}}]{2008AandA...477L...1D}
{Ducourant}, C., {Teixeira}, R., {Chauvin}, G., {et~al.} 2008, \aap, 477, L1

\bibitem[{{Dupuy} {et~al.}(2014){Dupuy}, {Liu}, \&
  {Ireland}}]{2014ApJ...790..133D}
{Dupuy}, T.~J., {Liu}, M.~C., \& {Ireland}, M.~J. 2014, \apj, 790, 133

\bibitem[{{Errmann} {et~al.}(2014){Errmann}, {Raetz}, {Kitze}, {Neuh{\"a}user},
  \& {YETI Team}}]{2014CoSka..43..513E}
{Errmann}, R., {Raetz}, S., {Kitze}, M., {Neuh{\"a}user}, R., \& {YETI Team}.
  2014, Contributions of the Astronomical Observatory Skalnate Pleso, 43, 513

\bibitem[{{Faherty} {et~al.}(2013){Faherty}, {Rice}, {Cruz}, {Mamajek}, \&
  {N{\'u}{\~n}ez}}]{2013AJ....145....2F}
{Faherty}, J.~K., {Rice}, E.~L., {Cruz}, K.~L., {Mamajek}, E.~E., \&
  {N{\'u}{\~n}ez}, A. 2013, \aj, 145, 2

\bibitem[{{Ford} \& {Rasio}(2008)}]{2008ApJ...686..621F}
{Ford}, E.~B. \& {Rasio}, F.~A. 2008, \apj, 686, 621

\bibitem[{{Freistetter} {et~al.}(2007){Freistetter}, {Krivov}, \&
  {L{\"o}hne}}]{2007A&A...466..389F}
{Freistetter}, F., {Krivov}, A.~V., \& {L{\"o}hne}, T. 2007, \aap, 466, 389

\bibitem[{{Galicher} {et~al.}(2014){Galicher}, {Rameau}, {Bonnefoy}, {Baudino},
  {Currie}, {Boccaletti}, {Chauvin}, {Lagrange}, \&
  {Marois}}]{2014AandA...565L...4G}
{Galicher}, R., {Rameau}, J., {Bonnefoy}, M., {et~al.} 2014, \aap, 565, L4

\bibitem[{{Gauza} {et~al.}(2015){Gauza}, {B{\'e}jar}, {P{\'e}rez-Garrido},
  {Rosa Zapatero Osorio}, {Lodieu}, {Rebolo}, {Pall{\'e}}, \&
  {Nowak}}]{2015ApJ...804...96G}
{Gauza}, B., {B{\'e}jar}, V.~J.~S., {P{\'e}rez-Garrido}, A., {et~al.} 2015,
  \apj, 804, 96

\bibitem[{{Gizis} {et~al.}(2015){Gizis}, {Allers}, {Liu}, {Harris}, {Faherty},
  {Burgasser}, \& {Kirkpatrick}}]{2015ApJ...799..203G}
{Gizis}, J.~E., {Allers}, K.~N., {Liu}, M.~C., {et~al.} 2015, \apj, 799, 203

\bibitem[{{Goldreich} \& {Ward}(1973)}]{1973ApJ...183.1051G}
{Goldreich}, P. \& {Ward}, W.~R. 1973, \apj, 183, 1051

\bibitem[{{Golimowski} {et~al.}(2004){Golimowski}, {Leggett}, {Marley}, {Fan},
  {Geballe}, {Knapp}, {Vrba}, {Henden}, {Luginbuhl}, {Guetter}, {Munn},
  {Canzian}, {Zheng}, {Tsvetanov}, {Chiu}, {Glazebrook}, {Hoversten},
  {Schneider}, \& {Brinkmann}}]{2004AJ....127.3516G}
{Golimowski}, D.~A., {Leggett}, S.~K., {Marley}, M.~S., {et~al.} 2004, \aj,
  127, 3516

\bibitem[{{Helling} {et~al.}(2008){Helling}, {Dehn}, {Woitke}, \&
  {Hauschildt}}]{2008ApJ...675L.105H}
{Helling}, C., {Dehn}, M., {Woitke}, P., \& {Hauschildt}, P.~H. 2008, \apjl,
  675, L105

\bibitem[{{Hern{\'a}ndez} {et~al.}(2006){Hern{\'a}ndez}, {Brice{\~n}o},
  {Calvet}, {Hartmann}, {Muzerolle}, \& {Quintero}}]{2006ApJ...652..472H}
{Hern{\'a}ndez}, J., {Brice{\~n}o}, C., {Calvet}, N., {et~al.} 2006, \apj, 652,
  472

\bibitem[{{Hern{\'a}ndez} {et~al.}(2005){Hern{\'a}ndez}, {Calvet}, {Hartmann},
  {Brice{\~n}o}, {Sicilia-Aguilar}, \& {Berlind}}]{2005AJ....129..856H}
{Hern{\'a}ndez}, J., {Calvet}, N., {Hartmann}, L., {et~al.} 2005, \aj, 129, 856

\bibitem[{{Hewett} {et~al.}(2006){Hewett}, {Warren}, {Leggett}, \&
  {Hodgkin}}]{2006MNRAS.367..454H}
{Hewett}, P.~C., {Warren}, S.~J., {Leggett}, S.~K., \& {Hodgkin}, S.~T. 2006,
  \mnras, 367, 454

\bibitem[{{Hinkley} {et~al.}(2013){Hinkley}, {Pueyo}, {Faherty}, {Oppenheimer},
  {Mamajek}, {Kraus}, {Rice}, {Ireland}, {David}, {Hillenbrand}, {Vasisht},
  {Cady}, {Brenner}, {Veicht}, {Nilsson}, {Zimmerman}, {Parry}, {Beichman},
  {Dekany}, {Roberts}, {Roberts}, {Baranec}, {Crepp}, {Burruss}, {Wallace},
  {King}, {Zhai}, {Lockhart}, {Shao}, {Soummer}, {Sivaramakrishnan}, \&
  {Wilson}}]{2013ApJ...779..153H}
{Hinkley}, S., {Pueyo}, L., {Faherty}, J.~K., {et~al.} 2013, \apj, 779, 153

\bibitem[{{Horne}(1986)}]{1986PASP...98..609H}
{Horne}, K. 1986, \pasp, 98, 609

\bibitem[{{Ireland} {et~al.}(2011){Ireland}, {Kraus}, {Martinache}, {Law}, \&
  {Hillenbrand}}]{2011ApJ...726..113I}
{Ireland}, M.~J., {Kraus}, A., {Martinache}, F., {Law}, N., \& {Hillenbrand},
  L.~A. 2011, \apj, 726, 113

\bibitem[{{Itoh} {et~al.}(2005){Itoh}, {Hayashi}, {Tamura}, {Tsuji}, {Oasa},
  {Fukagawa}, {Hayashi}, {Naoi}, {Ishii}, {Mayama}, {Morino}, {Yamashita},
  {Pyo}, {Nishikawa}, {Usuda}, {Murakawa}, {Suto}, {Oya}, {Takato}, {Ando},
  {Miyama}, {Kobayashi}, \& {Kaifu}}]{2005ApJ...620..984I}
{Itoh}, Y., {Hayashi}, M., {Tamura}, M., {et~al.} 2005, \apj, 620, 984

\bibitem[{{Jayawardhana} \& {Ivanov}(2006)}]{2006Sci...313.1279J}
{Jayawardhana}, R. \& {Ivanov}, V.~D. 2006, Science, 313, 1279

\bibitem[{{Jenkins} {et~al.}(2012){Jenkins}, {Pavlenko}, {Ivanyuk}, {Gallardo},
  {Jones}, {Day-Jones}, {Jones}, {Ruiz}, {Pinfield}, \&
  {Yakovina}}]{2012MNRAS.420.3587J}
{Jenkins}, J.~S., {Pavlenko}, Y.~V., {Ivanyuk}, O., {et~al.} 2012, \mnras, 420,
  3587

\bibitem[{{Jordi} {et~al.}(2006){Jordi}, {Grebel}, \&
  {Ammon}}]{2006A&A...460..339J}
{Jordi}, K., {Grebel}, E.~K., \& {Ammon}, K. 2006, \aap, 460, 339

\bibitem[{{Kalas} {et~al.}(2008){Kalas}, {Graham}, {Chiang}, {Fitzgerald},
  {Clampin}, {Kite}, {Stapelfeldt}, {Marois}, \& {Krist}}]{2008Sci...322.1345K}
{Kalas}, P., {Graham}, J.~R., {Chiang}, E., {et~al.} 2008, Science, 322, 1345

\bibitem[{{Kamiaka} {et~al.}(2015){Kamiaka}, {Masuda}, {Xue}, {Suto},
  {Nishioka}, {Murakami}, {Inayama}, {Saitoh}, {Tanaka}, \&
  {Yonehara}}]{2015PASJ...67...94K}
{Kamiaka}, S., {Masuda}, K., {Xue}, Y., {et~al.} 2015, \pasj, 67, 94

\bibitem[{{Kenyon} \& {Hartmann}(1995)}]{1995ApJS..101..117K}
{Kenyon}, S.~J. \& {Hartmann}, L. 1995, \apjs, 101, 117

\bibitem[{{Kirkpatrick} {et~al.}(2001){Kirkpatrick}, {Dahn}, {Monet}, {Reid},
  {Gizis}, {Liebert}, \& {Burgasser}}]{2001AJ....121.3235K}
{Kirkpatrick}, J.~D., {Dahn}, C.~C., {Monet}, D.~G., {et~al.} 2001, \aj, 121,
  3235

\bibitem[{{Kirkpatrick} {et~al.}(2000){Kirkpatrick}, {Reid}, {Liebert},
  {Gizis}, {Burgasser}, {Monet}, {Dahn}, {Nelson}, \&
  {Williams}}]{2000AJ....120..447K}
{Kirkpatrick}, J.~D., {Reid}, I.~N., {Liebert}, J., {et~al.} 2000, \aj, 120,
  447

\bibitem[{{Koch} {et~al.}(2010){Koch}, {Borucki}, {Basri}, {Batalha}, {Brown},
  {Caldwell}, {Christensen-Dalsgaard}, {Cochran}, {DeVore}, {Dunham},
  {Gautier}, {Geary}, {Gilliland}, {Gould}, {Jenkins}, {Kondo}, {Latham},
  {Lissauer}, {Marcy}, {Monet}, {Sasselov}, {Boss}, {Brownlee}, {Caldwell},
  {Dupree}, {Howell}, {Kjeldsen}, {Meibom}, {Morrison}, {Owen}, {Reitsema},
  {Tarter}, {Bryson}, {Dotson}, {Gazis}, {Haas}, {Kolodziejczak}, {Rowe}, {Van
  Cleve}, {Allen}, {Chandrasekaran}, {Clarke}, {Li}, {Quintana}, {Tenenbaum},
  {Twicken}, \& {Wu}}]{2010ApJ...713L..79K}
{Koch}, D.~G., {Borucki}, W.~J., {Basri}, G., {et~al.} 2010, \apjl, 713, L79

\bibitem[{{Koen}(2015)}]{2015MNRAS.450.3991K}
{Koen}, C. 2015, \mnras, 450, 3991

\bibitem[{{Konopacky} {et~al.}(2013){Konopacky}, {Barman}, {Macintosh}, \&
  {Marois}}]{2013Sci...339.1398K}
{Konopacky}, Q.~M., {Barman}, T.~S., {Macintosh}, B.~A., \& {Marois}, C. 2013,
  Science, 339, 1398

\bibitem[{{Kraus} \& {Ireland}(2012)}]{2012ApJ...745....5K}
{Kraus}, A.~L. \& {Ireland}, M.~J. 2012, \apj, 745, 5

\bibitem[{{Kraus} {et~al.}(2014){Kraus}, {Ireland}, {Cieza}, {Hinkley},
  {Dupuy}, {Bowler}, \& {Liu}}]{2014ApJ...781...20K}
{Kraus}, A.~L., {Ireland}, M.~J., {Cieza}, L.~A., {et~al.} 2014, \apj, 781, 20

\bibitem[{{Kuzuhara} {et~al.}(2011){Kuzuhara}, {Tamura}, {Ishii}, {Kudo},
  {Nishiyama}, \& {Kandori}}]{2011AJ....141..119K}
{Kuzuhara}, M., {Tamura}, M., {Ishii}, M., {et~al.} 2011, \aj, 141, 119

\bibitem[{{Kuzuhara} {et~al.}(2013){Kuzuhara}, {Tamura}, {Kudo}, {Janson},
  {Kandori}, {Brandt}, {Thalmann}, {Spiegel}, {Biller}, {Carson}, {Hori},
  {Suzuki}, {Burrows}, {Henning}, {Turner}, {McElwain}, {Moro-Mart{\'{\i}}n},
  {Suenaga}, {Takahashi}, {Kwon}, {Lucas}, {Abe}, {Brandner}, {Egner}, {Feldt},
  {Fujiwara}, {Goto}, {Grady}, {Guyon}, {Hashimoto}, {Hayano}, {Hayashi},
  {Hayashi}, {Hodapp}, {Ishii}, {Iye}, {Knapp}, {Matsuo}, {Mayama}, {Miyama},
  {Morino}, {Nishikawa}, {Nishimura}, {Kotani}, {Kusakabe}, {Pyo}, {Serabyn},
  {Suto}, {Takami}, {Takato}, {Terada}, {Tomono}, {Watanabe}, {Wisniewski},
  {Yamada}, {Takami}, \& {Usuda}}]{2013ApJ...774...11K}
{Kuzuhara}, M., {Tamura}, M., {Kudo}, T., {et~al.} 2013, \apj, 774, 11

\bibitem[{{Lafreni{\`e}re} {et~al.}(2011){Lafreni{\`e}re}, {Jayawardhana},
  {Janson}, {Helling}, {Witte}, \& {Hauschildt}}]{2011ApJ...730...42L}
{Lafreni{\`e}re}, D., {Jayawardhana}, R., {Janson}, M., {et~al.} 2011, \apj,
  730, 42

\bibitem[{{Lafreni{\`e}re} {et~al.}(2008){Lafreni{\`e}re}, {Jayawardhana}, \&
  {van Kerkwijk}}]{2008ApJ...689L.153L}
{Lafreni{\`e}re}, D., {Jayawardhana}, R., \& {van Kerkwijk}, M.~H. 2008, \apjl,
  689, L153

\bibitem[{{Lagrange} {et~al.}(2010){Lagrange}, {Bonnefoy}, {Chauvin}, {Apai},
  {Ehrenreich}, {Boccaletti}, {Gratadour}, {Rouan}, {Mouillet}, {Lacour}, \&
  {Kasper}}]{2010Sci...329...57L}
{Lagrange}, A.-M., {Bonnefoy}, M., {Chauvin}, G., {et~al.} 2010, Science, 329,
  57

\bibitem[{{Lagrange} {et~al.}(2009){Lagrange}, {Gratadour}, {Chauvin}, {Fusco},
  {Ehrenreich}, {Mouillet}, {Rousset}, {Rouan}, {Allard}, {Gendron}, {Charton},
  {Mugnier}, {Rabou}, {Montri}, \& {Lacombe}}]{2009AandA...493L..21L}
{Lagrange}, A.-M., {Gratadour}, D., {Chauvin}, G., {et~al.} 2009, \aap, 493,
  L21

\bibitem[{{Latham} {et~al.}(2011){Latham}, {Rowe}, {Quinn}, {Batalha},
  {Borucki}, {Brown}, {Bryson}, {Buchhave}, {Caldwell}, {Carter},
  {Christiansen}, {Ciardi}, {Cochran}, {Dunham}, {Fabrycky}, {Ford}, {Gautier},
  {Gilliland}, {Holman}, {Howell}, {Ibrahim}, {Isaacson}, {Jenkins}, {Koch},
  {Lissauer}, {Marcy}, {Quintana}, {Ragozzine}, {Sasselov}, {Shporer},
  {Steffen}, {Welsh}, \& {Wohler}}]{2011ApJ...732L..24L}
{Latham}, D.~W., {Rowe}, J.~F., {Quinn}, S.~N., {et~al.} 2011, \apjl, 732, L24

\bibitem[{{Lissauer} {et~al.}(2013){Lissauer}, {Jontof-Hutter}, {Rowe},
  {Fabrycky}, {Lopez}, {Agol}, {Marcy}, {Deck}, {Fischer}, {Fortney}, {Howell},
  {Isaacson}, {Jenkins}, {Kolbl}, {Sasselov}, {Short}, \&
  {Welsh}}]{2013ApJ...770..131L}
{Lissauer}, J.~J., {Jontof-Hutter}, D., {Rowe}, J.~F., {et~al.} 2013, \apj,
  770, 131

\bibitem[{{Liu} {et~al.}(2013){Liu}, {Magnier}, {Deacon}, {Allers}, {Dupuy},
  {Kotson}, {Aller}, {Burgett}, {Chambers}, {Draper}, {Hodapp}, {Jedicke},
  {Kaiser}, {Kudritzki}, {Metcalfe}, {Morgan}, {Price}, {Tonry}, \&
  {Wainscoat}}]{2013ApJ...777L..20L}
{Liu}, M.~C., {Magnier}, E.~A., {Deacon}, N.~R., {et~al.} 2013, \apjl, 777, L20

\bibitem[{{Luhman} {et~al.}(2005{\natexlab{a}}){Luhman}, {Adame}, {D'Alessio},
  {Calvet}, {Hartmann}, {Megeath}, \& {Fazio}}]{2005ApJ...635L..93L}
{Luhman}, K.~L., {Adame}, L., {D'Alessio}, P., {et~al.} 2005{\natexlab{a}},
  \apjl, 635, L93

\bibitem[{{Luhman} {et~al.}(2005{\natexlab{b}}){Luhman}, {D'Alessio}, {Calvet},
  {Allen}, {Hartmann}, {Megeath}, {Myers}, \& {Fazio}}]{2005ApJ...620L..51L}
{Luhman}, K.~L., {D'Alessio}, P., {Calvet}, N., {et~al.} 2005{\natexlab{b}},
  \apjl, 620, L51

\bibitem[{{Luhman} {et~al.}(2007){Luhman}, {Patten}, {Marengo}, {Schuster},
  {Hora}, {Ellis}, {Stauffer}, {Sonnett}, {Winston}, {Gutermuth}, {Megeath},
  {Backman}, {Henry}, {Werner}, \& {Fazio}}]{2007ApJ...654..570L}
{Luhman}, K.~L., {Patten}, B.~M., {Marengo}, M., {et~al.} 2007, \apj, 654, 570

\bibitem[{{Luhman} {et~al.}(2006){Luhman}, {Wilson}, {Brandner}, {Skrutskie},
  {Nelson}, {Smith}, {Peterson}, {Cushing}, \& {Young}}]{2006ApJ...649..894L}
{Luhman}, K.~L., {Wilson}, J.~C., {Brandner}, W., {et~al.} 2006, \apj, 649, 894

\bibitem[{{Macintosh} {et~al.}(2015){Macintosh}, {Graham}, {Barman}, {De Rosa},
  {Konopacky}, {Marley}, {Marois}, {Nielsen}, {Pueyo}, {Rajan}, {Rameau},
  {Saumon}, {Wang}, {Patience}, {Ammons}, {Arriaga}, {Artigau}, {Beckwith},
  {Brewster}, {Bruzzone}, {Bulger}, {Burningham}, {Burrows}, {Chen}, {Chiang},
  {Chilcote}, {Dawson}, {Dong}, {Doyon}, {Draper}, {Duch{\^e}ne}, {Esposito},
  {Fabrycky}, {Fitzgerald}, {Follette}, {Fortney}, {Gerard}, {Goodsell},
  {Greenbaum}, {Hibon}, {Hinkley}, {Cotten}, {Hung}, {Ingraham},
  {Johnson-Groh}, {Kalas}, {Lafreniere}, {Larkin}, {Lee}, {Line}, {Long},
  {Maire}, {Marchis}, {Matthews}, {Max}, {Metchev}, {Millar-Blanchaer},
  {Mittal}, {Morley}, {Morzinski}, {Murray-Clay}, {Oppenheimer}, {Palmer},
  {Patel}, {Perrin}, {Poyneer}, {Rafikov}, {Rantakyr{\"o}}, {Rice}, {Rojo},
  {Rudy}, {Ruffio}, {Ruiz}, {Sadakuni}, {Saddlemyer}, {Salama}, {Savransky},
  {Schneider}, {Sivaramakrishnan}, {Song}, {Soummer}, {Thomas}, {Vasisht},
  {Wallace}, {Ward-Duong}, {Wiktorowicz}, {Wolff}, \&
  {Zuckerman}}]{2015Sci...350...64M}
{Macintosh}, B., {Graham}, J.~R., {Barman}, T., {et~al.} 2015, Science, 350, 64

\bibitem[{{Mamajek} \& {Bell}(2014)}]{2014MNRAS.445.2169M}
{Mamajek}, E.~E. \& {Bell}, C.~P.~M. 2014, \mnras, 445, 2169

\bibitem[{{Mannings} \& {Sargent}(1997)}]{1997ApJ...490..792M}
{Mannings}, V. \& {Sargent}, A.~I. 1997, \apj, 490, 792

\bibitem[{{Mannucci} {et~al.}(2001){Mannucci}, {Basile}, {Poggianti},
  {Cimatti}, {Daddi}, {Pozzetti}, \& {Vanzi}}]{2001MNRAS.326..745M}
{Mannucci}, F., {Basile}, F., {Poggianti}, B.~M., {et~al.} 2001, \mnras, 326,
  745

\bibitem[{{Marley}(2013)}]{2013Sci...339.1393M}
{Marley}, M.~S. 2013, Science, 339, 1393

\bibitem[{{Marley} {et~al.}(2007){Marley}, {Fortney}, {Hubickyj},
  {Bodenheimer}, \& {Lissauer}}]{2007ApJ...655..541M}
{Marley}, M.~S., {Fortney}, J.~J., {Hubickyj}, O., {Bodenheimer}, P., \&
  {Lissauer}, J.~J. 2007, \apj, 655, 541

\bibitem[{{Marley} {et~al.}(2012){Marley}, {Saumon}, {Cushing}, {Ackerman},
  {Fortney}, \& {Freedman}}]{2012ApJ...754..135M}
{Marley}, M.~S., {Saumon}, D., {Cushing}, M., {et~al.} 2012, \apj, 754, 135

\bibitem[{{Marois} {et~al.}(2008){Marois}, {Macintosh}, {Barman}, {Zuckerman},
  {Song}, {Patience}, {Lafreni{\`e}re}, \& {Doyon}}]{2008Sci...322.1348M}
{Marois}, C., {Macintosh}, B., {Barman}, T., {et~al.} 2008, Science, 322, 1348

\bibitem[{{Marois} {et~al.}(2010){Marois}, {Zuckerman}, {Konopacky},
  {Macintosh}, \& {Barman}}]{2010Natur.468.1080M}
{Marois}, C., {Zuckerman}, B., {Konopacky}, Q.~M., {Macintosh}, B., \&
  {Barman}, T. 2010, \nat, 468, 1080

\bibitem[{{Mart{\'{\i}}n} {et~al.}(2001){Mart{\'{\i}}n}, {Zapatero Osorio},
  {Barrado y Navascu{\'e}s}, {B{\'e}jar}, \& {Rebolo}}]{2001ApJ...558L.117M}
{Mart{\'{\i}}n}, E.~L., {Zapatero Osorio}, M.~R., {Barrado y Navascu{\'e}s},
  D., {B{\'e}jar}, V.~J.~S., \& {Rebolo}, R. 2001, \apjl, 558, L117

\bibitem[{{Mayor} \& {Queloz}(1995)}]{1995Natur.378..355M}
{Mayor}, M. \& {Queloz}, D. 1995, \nat, 378, 355

\bibitem[{{Metchev} \& {Hillenbrand}(2006)}]{2006ApJ...651.1166M}
{Metchev}, S.~A. \& {Hillenbrand}, L.~A. 2006, \apj, 651, 1166

\bibitem[{{Mohanty} {et~al.}(2007){Mohanty}, {Jayawardhana}, {Hu{\'e}lamo}, \&
  {Mamajek}}]{2007ApJ...657.1064M}
{Mohanty}, S., {Jayawardhana}, R., {Hu{\'e}lamo}, N., \& {Mamajek}, E. 2007,
  \apj, 657, 1064

\bibitem[{{Montet} {et~al.}(2015){Montet}, {Bowler}, {Shkolnik}, {Deck},
  {Wang}, {Horch}, {Liu}, {Hillenbrand}, {Kraus}, \&
  {Charbonneau}}]{2015ApJ...813L..11M}
{Montet}, B.~T., {Bowler}, B.~P., {Shkolnik}, E.~L., {et~al.} 2015, \apjl, 813,
  L11

\bibitem[{{Moya} {et~al.}(2010){Moya}, {Amado}, {Barrado}, {Garc{\'{\i}}a
  Hern{\'a}ndez}, {Aberasturi}, {Montesinos}, \&
  {Aceituno}}]{2010MNRAS.405L..81M}
{Moya}, A., {Amado}, P.~J., {Barrado}, D., {et~al.} 2010, \mnras, 405, L81

\bibitem[{{Mu{\~n}oz} {et~al.}(2015){Mu{\~n}oz}, {Kratter}, {Vogelsberger},
  {Hernquist}, \& {Springel}}]{2015MNRAS.446.2010M}
{Mu{\~n}oz}, D.~J., {Kratter}, K., {Vogelsberger}, M., {Hernquist}, L., \&
  {Springel}, V. 2015, \mnras, 446, 2010

\bibitem[{{Mugrauer} \& {Neuh{\"a}user}(2005)}]{2005AN....326..701M}
{Mugrauer}, M. \& {Neuh{\"a}user}, R. 2005, Astronomische Nachrichten, 326, 701

\bibitem[{{Mugrauer} {et~al.}(2010){Mugrauer}, {Vogt}, {Neuh{\"a}user}, \&
  {Schmidt}}]{2010AandA...523L...1M}
{Mugrauer}, M., {Vogt}, N., {Neuh{\"a}user}, R., \& {Schmidt}, T.~O.~B. 2010,
  \aap, 523, L1

\bibitem[{{Nagasawa} \& {Ida}(2011)}]{2011ApJ...742...72N}
{Nagasawa}, M. \& {Ida}, S. 2011, \apj, 742, 72

\bibitem[{{Naud} {et~al.}(2014){Naud}, {Artigau}, {Malo}, {Albert}, {Doyon},
  {Lafreni{\`e}re}, {Gagn{\'e}}, {Saumon}, {Morley}, {Allard}, {Homeier},
  {Beichman}, {Gelino}, \& {Boucher}}]{2014ApJ...787....5N}
{Naud}, M.-E., {Artigau}, {\'E}., {Malo}, L., {et~al.} 2014, \apj, 787, 5

\bibitem[{{Neuh{\"a}user} {et~al.}(2011){Neuh{\"a}user}, {Errmann}, {Berndt},
  {Maciejewski}, {Takahashi}, {Chen}, {Dimitrov}, {Pribulla}, {Nikogossian},
  {Jensen}, {Marschall}, {Wu}, {Kellerer}, {Walter}, {Brice{\~n}o}, {Chini},
  {Fernandez}, {Raetz}, {Torres}, {Latham}, {Quinn}, {Niedzielski},
  {Bukowiecki}, {Nowak}, {Tomov}, {Tachihara}, {Hu}, {Hung}, {Kjurkchieva},
  {Radeva}, {Mihov}, {Slavcheva-Mihova}, {Bozhinova}, {Budaj}, {Va{\v n}ko},
  {Kundra}, {Hamb{\'a}lek}, {Krushevska}, {Movsessian}, {Harutyunyan},
  {Downes}, {Hernandez}, {Hoffmeister}, {Cohen}, {Abel}, {Ahmad}, {Chapman},
  {Eckert}, {Goodman}, {Guerard}, {Kim}, {Koontharana}, {Sokol}, {Trinh},
  {Wang}, {Zhou}, {Redmer}, {Kramm}, {Nettelmann}, {Mugrauer}, {Schmidt},
  {Moualla}, {Ginski}, {Marka}, {Adam}, {Seeliger}, {Baar}, {Roell}, {Schmidt},
  {Trepl}, {Eisenbei{\ss}}, {Fiedler}, {Tetzlaff}, {Schmidt}, {Hohle}, {Kitze},
  {Chakrova}, {Gr{\"a}fe}, {Schreyer}, {Hambaryan}, {Broeg}, {Koppenhoefer}, \&
  {Pandey}}]{2011AN....332..547N}
{Neuh{\"a}user}, R., {Errmann}, R., {Berndt}, A., {et~al.} 2011, Astronomische
  Nachrichten, 332, 547

\bibitem[{{Neuh{\"a}user} {et~al.}(2013){Neuh{\"a}user}, {Errmann}, {Raetz},
  {Chen}, {Hu}, {Torres}, {Kellerer}, {Kitze}, \& {YETI
  Team}}]{2013prpl.conf2K047N}
{Neuh{\"a}user}, R., {Errmann}, R., {Raetz}, S., {et~al.} 2013, in Protostars
  and Planets VI Posters, 47

\bibitem[{{Neuh{\"a}user} {et~al.}(2005){Neuh{\"a}user}, {Guenther},
  {Wuchterl}, {Mugrauer}, {Bedalov}, \& {Hauschildt}}]{2005AandA...435L..13N}
{Neuh{\"a}user}, R., {Guenther}, E.~W., {Wuchterl}, G., {et~al.} 2005, \aap,
  435, L13

\bibitem[{{Neuh{\"a}user} \& {Schmidt}(2012)}]{2012arXiv1201.3537N}
{Neuh{\"a}user}, R. \& {Schmidt}, T.~O.~B. 2012, Topics in Adaptive Optics
  (InTech)

\bibitem[{{Ngo} {et~al.}(2015){Ngo}, {Knutson}, {Hinkley}, {Crepp}, {Bechter},
  {Batygin}, {Howard}, {Johnson}, {Morton}, \&
  {Muirhead}}]{2015ApJ...800..138N}
{Ngo}, H., {Knutson}, H.~A., {Hinkley}, S., {et~al.} 2015, \apj, 800, 138

\bibitem[{{Pani{\'c}} {et~al.}(2009){Pani{\'c}}, {Hogerheijde}, {Wilner}, \&
  {Qi}}]{2009A&A...501..269P}
{Pani{\'c}}, O., {Hogerheijde}, M.~R., {Wilner}, D., \& {Qi}, C. 2009, \aap,
  501, 269

\bibitem[{{Patience} {et~al.}(2010){Patience}, {King}, {de Rosa}, \&
  {Marois}}]{2010A&A...517A..76P}
{Patience}, J., {King}, R.~R., {de Rosa}, R.~J., \& {Marois}, C. 2010, \aap,
  517, A76

\bibitem[{{Pe{\~n}a Ram{\'{\i}}rez} {et~al.}(2012){Pe{\~n}a Ram{\'{\i}}rez},
  {B{\'e}jar}, {Zapatero Osorio}, {Petr-Gotzens}, \&
  {Mart{\'{\i}}n}}]{2012ApJ...754...30P}
{Pe{\~n}a Ram{\'{\i}}rez}, K., {B{\'e}jar}, V.~J.~S., {Zapatero Osorio}, M.~R.,
  {Petr-Gotzens}, M.~G., \& {Mart{\'{\i}}n}, E.~L. 2012, \apj, 754, 30

\bibitem[{{Pecaut} {et~al.}(2012){Pecaut}, {Mamajek}, \&
  {Bubar}}]{2012ApJ...746..154P}
{Pecaut}, M.~J., {Mamajek}, E.~E., \& {Bubar}, E.~J. 2012, \apj, 746, 154

\bibitem[{{Perryman}(2005)}]{2005ASPC..338....3P}
{Perryman}, M.~A.~C. 2005, in Astronomical Society of the Pacific Conference
  Series, Vol. 338, Astrometry in the Age of the Next Generation of Large
  Telescopes, ed. P.~K. {Seidelmann} \& A.~K.~B. {Monet}, 3

\bibitem[{{Pollack} {et~al.}(1996){Pollack}, {Hubickyj}, {Bodenheimer},
  {Lissauer}, {Podolak}, \& {Greenzweig}}]{1996Icar..124...62P}
{Pollack}, J.~B., {Hubickyj}, O., {Bodenheimer}, P., {et~al.} 1996, \icarus,
  124, 62

\bibitem[{{Potter} {et~al.}(2002){Potter}, {Mart{\'{\i}}n}, {Cushing},
  {Baudoz}, {Brandner}, {Guyon}, \& {Neuh{\"a}user}}]{2002ApJ...567L.133P}
{Potter}, D., {Mart{\'{\i}}n}, E.~L., {Cushing}, M.~C., {et~al.} 2002, \apjl,
  567, L133

\bibitem[{{Preibisch} {et~al.}(2002){Preibisch}, {Brown}, {Bridges},
  {Guenther}, \& {Zinnecker}}]{2002AJ....124..404P}
{Preibisch}, T., {Brown}, A.~G.~A., {Bridges}, T., {Guenther}, E., \&
  {Zinnecker}, H. 2002, \aj, 124, 404

\bibitem[{{Quanz} {et~al.}(2015){Quanz}, {Amara}, {Meyer}, {Girard},
  {Kenworthy}, \& {Kasper}}]{2015ApJ...807...64Q}
{Quanz}, S.~P., {Amara}, A., {Meyer}, M.~R., {et~al.} 2015, \apj, 807, 64

\bibitem[{{Quanz} {et~al.}(2013){Quanz}, {Amara}, {Meyer}, {Kenworthy},
  {Kasper}, \& {Girard}}]{2013ApJ...766L...1Q}
{Quanz}, S.~P., {Amara}, A., {Meyer}, M.~R., {et~al.} 2013, \apjl, 766, L1

\bibitem[{{Rameau} {et~al.}(2013){Rameau}, {Chauvin}, {Lagrange}, {Boccaletti},
  {Quanz}, {Bonnefoy}, {Girard}, {Delorme}, {Desidera}, {Klahr}, {Mordasini},
  {Dumas}, \& {Bonavita}}]{2013ApJ...772L..15R}
{Rameau}, J., {Chauvin}, G., {Lagrange}, A.-M., {et~al.} 2013, \apjl, 772, L15

\bibitem[{{Rasio} \& {Ford}(1996)}]{1996Sci...274..954R}
{Rasio}, F.~A. \& {Ford}, E.~B. 1996, Science, 274, 954

\bibitem[{{Rayner} {et~al.}(2009){Rayner}, {Cushing}, \&
  {Vacca}}]{2009ApJS..185..289R}
{Rayner}, J.~T., {Cushing}, M.~C., \& {Vacca}, W.~D. 2009, \apjs, 185, 289

\bibitem[{{Rebolo} {et~al.}(1998){Rebolo}, {Zapatero Osorio}, {Madruga},
  {Bejar}, {Arribas}, \& {Licandro}}]{1998Sci...282.1309R}
{Rebolo}, R., {Zapatero Osorio}, M.~R., {Madruga}, S., {et~al.} 1998, Science,
  282, 1309

\bibitem[{{Reid} \& {Walkowicz}(2006)}]{2006PASP..118..671R}
{Reid}, I.~N. \& {Walkowicz}, L.~M. 2006, \pasp, 118, 671

\bibitem[{{Reipurth} \& {Clarke}(2001)}]{2001AJ....122..432R}
{Reipurth}, B. \& {Clarke}, C. 2001, \aj, 122, 432

\bibitem[{{Rieke} \& {Lebofsky}(1985)}]{1985ApJ...288..618R}
{Rieke}, G.~H. \& {Lebofsky}, M.~J. 1985, \apj, 288, 618

\bibitem[{{Safronov} \& {Zvjagina}(1969)}]{1969Icar...10..109S}
{Safronov}, V.~S. \& {Zvjagina}, E.~V. 1969, \icarus, 10, 109

\bibitem[{{Sallum} {et~al.}(2015){Sallum}, {Follette}, {Eisner}, {Close},
  {Hinz}, {Kratter}, {Males}, {Skemer}, {Macintosh}, {Tuthill}, {Bailey},
  {Defr{\`e}re}, {Morzinski}, {Rodigas}, {Spalding}, {Vaz}, \&
  {Weinberger}}]{2015Natur.527..342S}
{Sallum}, S., {Follette}, K.~B., {Eisner}, J.~A., {et~al.} 2015, \nat, 527, 342

\bibitem[{{Saumon} \& {Marley}(2008)}]{2008ApJ...689.1327S}
{Saumon}, D. \& {Marley}, M.~S. 2008, \apj, 689, 1327

\bibitem[{{Schmidt} {et~al.}(2014){Schmidt}, {Mugrauer}, {Neuh{\"a}user},
  {Vogt}, {Witte}, {Hauschildt}, {Helling}, \&
  {Seifahrt}}]{2014A&A...566A..85S}
{Schmidt}, T.~O.~B., {Mugrauer}, M., {Neuh{\"a}user}, R., {et~al.} 2014, \aap,
  566, A85

\bibitem[{{Schmidt} {et~al.}(2009){Schmidt}, {Neuh{\"a}user}, {Mugrauer},
  {Bedalov}, \& {Vogt}}]{2009AIPC.1094..852S}
{Schmidt}, T.~O.~B., {Neuh{\"a}user}, R., {Mugrauer}, M., {Bedalov}, A., \&
  {Vogt}, N. 2009, in American Institute of Physics Conference Series, Vol.
  1094, 15th Cambridge Workshop on Cool Stars, Stellar Systems, and the Sun,
  ed. E.~{Stempels}, 852--855

\bibitem[{{Schmidt} {et~al.}(2008){Schmidt}, {Neuh{\"a}user}, {Seifahrt},
  {Vogt}, {Bedalov}, {Helling}, {Witte}, \&
  {Hauschildt}}]{2008AandA...491..311S}
{Schmidt}, T.~O.~B., {Neuh{\"a}user}, R., {Seifahrt}, A., {et~al.} 2008, \aap,
  491, 311

\bibitem[{{Schneider} {et~al.}(2011){Schneider}, {Dedieu}, {Le Sidaner},
  {Savalle}, \& {Zolotukhin}}]{2011A&A...532A..79S}
{Schneider}, J., {Dedieu}, C., {Le Sidaner}, P., {Savalle}, R., \&
  {Zolotukhin}, I. 2011, \aap, 532, A79

\bibitem[{{Siess} {et~al.}(2000){Siess}, {Dufour}, \&
  {Forestini}}]{2000A&A...358..593S}
{Siess}, L., {Dufour}, E., \& {Forestini}, M. 2000, \aap, 358, 593

\bibitem[{{Skrutskie} {et~al.}(2006){Skrutskie}, {Cutri}, {Stiening},
  {Weinberg}, {Schneider}, {Carpenter}, {Beichman}, {Capps}, {Chester},
  {Elias}, {Huchra}, {Liebert}, {Lonsdale}, {Monet}, {Price}, {Seitzer},
  {Jarrett}, {Kirkpatrick}, {Gizis}, {Howard}, {Evans}, {Fowler}, {Fullmer},
  {Hurt}, {Light}, {Kopan}, {Marsh}, {McCallon}, {Tam}, {Van Dyk}, \&
  {Wheelock}}]{2006AJ....131.1163S}
{Skrutskie}, M.~F., {Cutri}, R.~M., {Stiening}, R., {et~al.} 2006, \aj, 131,
  1163

\bibitem[{{Sparks} \& {Ford}(2002)}]{2002ApJ...578..543S}
{Sparks}, W.~B. \& {Ford}, H.~C. 2002, \apj, 578, 543

\bibitem[{{Spiegel} \& {Burrows}(2012)}]{2012ApJ...745..174S}
{Spiegel}, D.~S. \& {Burrows}, A. 2012, \apj, 745, 174

\bibitem[{{Stamatellos} \& {Whitworth}(2009)}]{2009MNRAS.392..413S}
{Stamatellos}, D. \& {Whitworth}, A.~P. 2009, \mnras, 392, 413

\bibitem[{{Thatte} {et~al.}(2007){Thatte}, {Abuter}, {Tecza}, {Nielsen},
  {Clarke}, \& {Close}}]{2007MNRAS.378.1229T}
{Thatte}, N., {Abuter}, R., {Tecza}, M., {et~al.} 2007, \mnras, 378, 1229

\bibitem[{{Todorov} {et~al.}(2010){Todorov}, {Luhman}, \&
  {McLeod}}]{2010ApJ...714L..84T}
{Todorov}, K., {Luhman}, K.~L., \& {McLeod}, K.~K. 2010, \apjl, 714, L84

\bibitem[{{van Eyken} {et~al.}(2012){van Eyken}, {Ciardi}, {von Braun}, {Kane},
  {Plavchan}, {Bender}, {Brown}, {Crepp}, {Fulton}, {Howard}, {Howell},
  {Mahadevan}, {Marcy}, {Shporer}, {Szkody}, {Akeson}, {Beichman}, {Boden},
  {Gelino}, {Hoard}, {Ram{\'{\i}}rez}, {Rebull}, {Stauffer}, {Bloom}, {Cenko},
  {Kasliwal}, {Kulkarni}, {Law}, {Nugent}, {Ofek}, {Poznanski}, {Quimby},
  {Walters}, {Grillmair}, {Laher}, {Levitan}, {Sesar}, \&
  {Surace}}]{2012ApJ...755...42V}
{van Eyken}, J.~C., {Ciardi}, D.~R., {von Braun}, K., {et~al.} 2012, \apj, 755,
  42

\bibitem[{{Vorobyov}(2013)}]{2013A&A...552A.129V}
{Vorobyov}, E.~I. 2013, \aap, 552, A129

\bibitem[{{Wahhaj} {et~al.}(2011){Wahhaj}, {Liu}, {Biller}, {Clarke},
  {Nielsen}, {Close}, {Hayward}, {Mamajek}, {Cushing}, {Dupuy}, {Tecza},
  {Thatte}, {Chun}, {Ftaclas}, {Hartung}, {Reid}, {Shkolnik}, {Alencar},
  {Artymowicz}, {Boss}, {de Gouveia Dal Pino}, {Gregorio-Hetem}, {Ida},
  {Kuchner}, {Lin}, \& {Toomey}}]{2011ApJ...729..139W}
{Wahhaj}, Z., {Liu}, M.~C., {Biller}, B.~A., {et~al.} 2011, \apj, 729, 139

\bibitem[{{Weinberg} {et~al.}(1987){Weinberg}, {Shapiro}, \&
  {Wasserman}}]{1987ApJ...312..367W}
{Weinberg}, M.~D., {Shapiro}, S.~L., \& {Wasserman}, I. 1987, \apj, 312, 367

\bibitem[{{Wright} {et~al.}(2009){Wright}, {Upadhyay}, {Marcy}, {Fischer},
  {Ford}, \& {Johnson}}]{2009ApJ...693.1084W}
{Wright}, J.~T., {Upadhyay}, S., {Marcy}, G.~W., {et~al.} 2009, \apj, 693, 1084

\bibitem[{{Yu} {et~al.}(2015){Yu}, {Winn}, {Gillon}, {Albrecht}, {Rappaport},
  {Bieryla}, {Dai}, {Delrez}, {Hillenbrand}, {Holman}, {Howard}, {Huang},
  {Isaacson}, {Jehin}, {Lendl}, {Montet}, {Muirhead}, {Sanchis-Ojeda}, \&
  {Triaud}}]{2015ApJ...812...48Y}
{Yu}, L., {Winn}, J.~N., {Gillon}, M., {et~al.} 2015, \apj, 812, 48

\bibitem[{{Zacharias} {et~al.}(2013){Zacharias}, {Finch}, {Girard}, {Henden},
  {Bartlett}, {Monet}, \& {Zacharias}}]{2013AJ....145...44Z}
{Zacharias}, N., {Finch}, C.~T., {Girard}, T.~M., {et~al.} 2013, \aj, 145, 44

\bibitem[{{Zacharias} {et~al.}(2004){Zacharias}, {Monet}, {Levine}, {Urban},
  {Gaume}, \& {Wycoff}}]{2004AAS...205.4815Z}
{Zacharias}, N., {Monet}, D.~G., {Levine}, S.~E., {et~al.} 2004, in Bulletin of
  the American Astronomical Society, Vol.~36, American Astronomical Society
  Meeting Abstracts, 1418

\bibitem[{{Zapatero Osorio} {et~al.}(2000){Zapatero Osorio}, {B{\'e}jar},
  {Mart{\'{\i}}n}, {Rebolo}, {Barrado y Navascu{\'e}s}, {Bailer-Jones}, \&
  {Mundt}}]{2000Sci...290..103Z}
{Zapatero Osorio}, M.~R., {B{\'e}jar}, V.~J.~S., {Mart{\'{\i}}n}, E.~L.,
  {et~al.} 2000, Science, 290, 103

\bibitem[{{Zuckerman} {et~al.}(2011){Zuckerman}, {Rhee}, {Song}, \&
  {Bessell}}]{2011ApJ...732...61Z}
{Zuckerman}, B., {Rhee}, J.~H., {Song}, I., \& {Bessell}, M.~S. 2011, \apj,
  732, 61

\end{thebibliography}


\Online

\begin{appendix} 
\section{Supplementary material}

\begin{figure}
  \centering
  \includegraphics[width=0.43\textwidth]{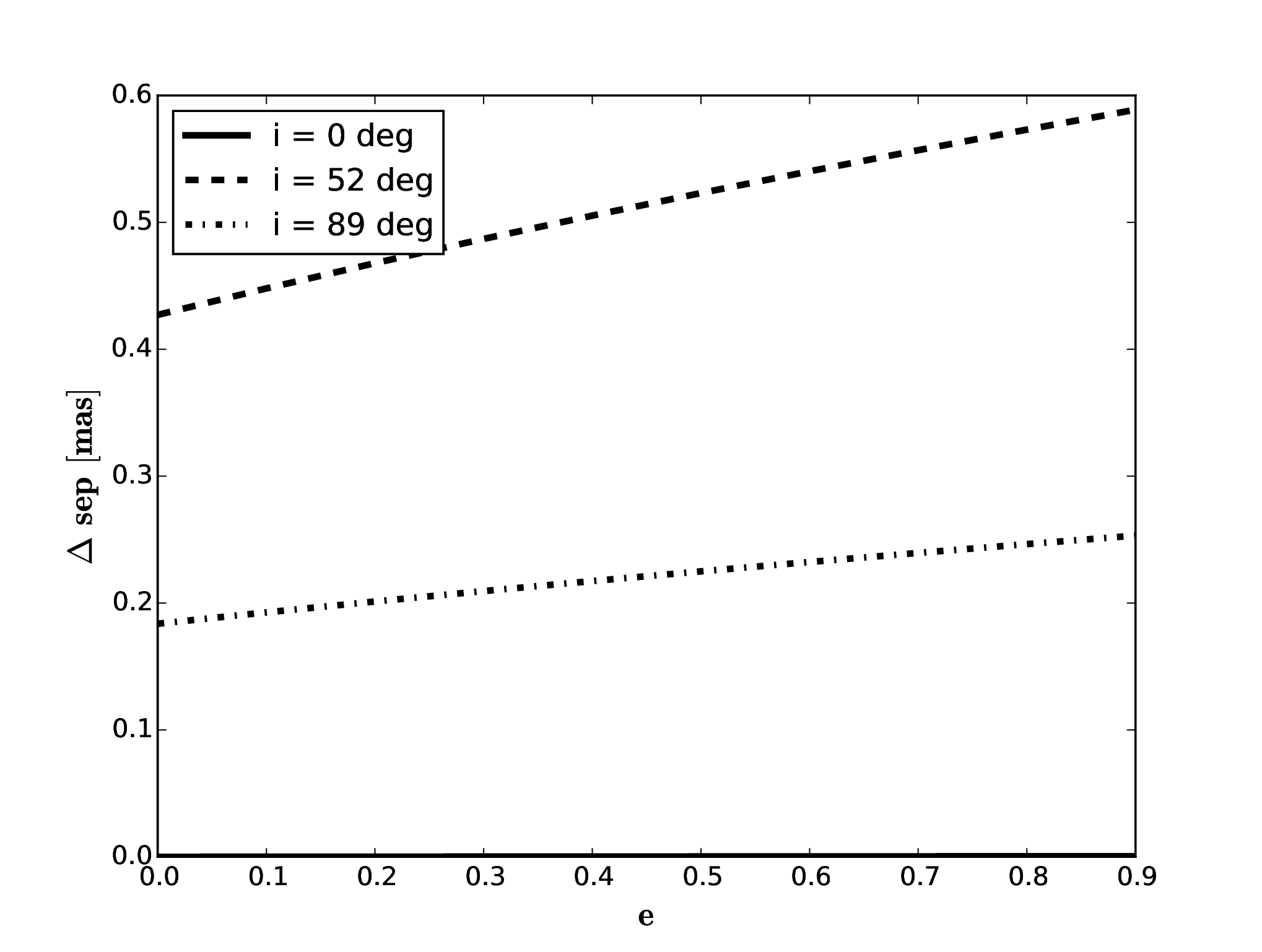}
  \includegraphics[width=0.43\textwidth]{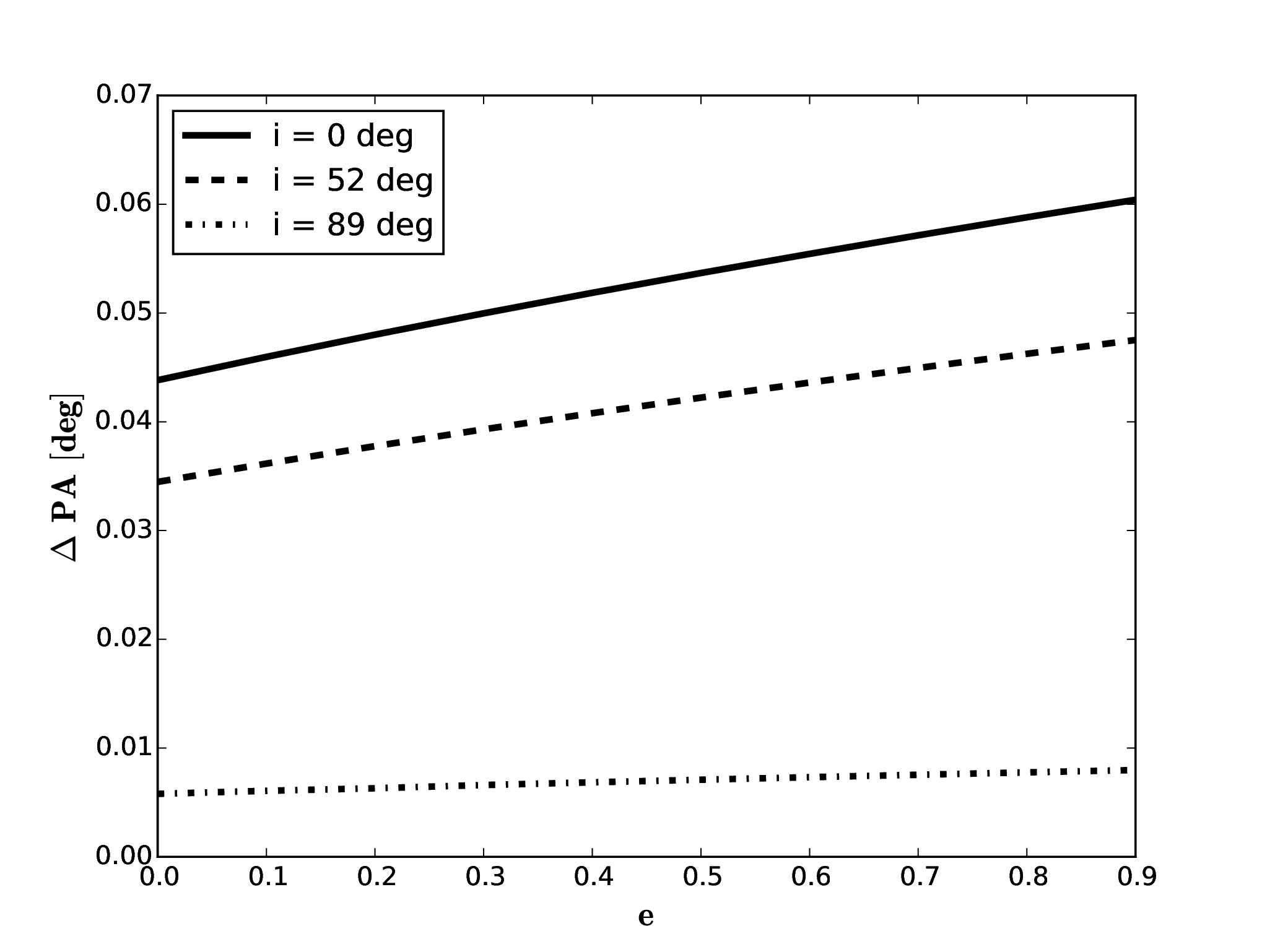}
  \caption{Expected maximum separation (top) and position angle (bottom) change, dependent on inclination and eccentricity of the companion for an epoch difference of three years
  (early in 2016, since the first calibrated epoch was made at
the end of 2012).}
  \label{FigOrbit}
\end{figure}

CVSO 30 is 
currently not suitable for a common proper motion analysis
(Table~\ref{TableCVSOsys}).
Because orbital motion around the host star might be detectable,
we simulated the expected maximum separation (top) and position angle (bottom) change in Fig.~\ref{FigOrbit}, dependent on inclination and eccentricity of the companion for an epoch difference of three years. This corresponds to our first astrometrically calibrated epoch from 2012 to a tentative new observation early in 2016. The dedicated orbital analysis shows that even after two to three years of epoch difference, no significant orbital motion is expected for the wide companion.

\begin{figure}
  \centering
  \includegraphics[width=0.5\textwidth]{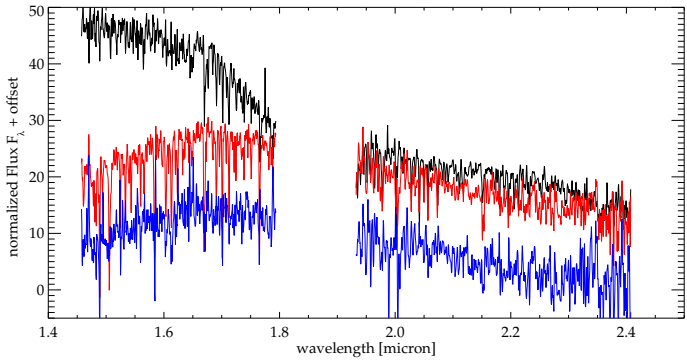}
  \caption{Spectrum of the primary (black) and the companion at the best illuminated pixel as given in the central panel of Fig.~\ref{FigCubes} (red; with OH lines), together with the spectrum after
  subtracting the average spike east and west of the companion (blue), which contributes about 30 \% of the light (beforehand). While the H-band spectrum presents a triangular shape and bluer
  colour, indicating a young sub-stellar companion, the full continuum of the companion is not reliable because different amounts of flux are superimposed by the rotating primary spike, which changes the overall continuum shape because of the different spike removal quality.
  }
  \label{FigSpecAppend}
\end{figure}

A first spectrum of CVSO 30 c at an intermediate reduction step, shown in Fig.~\ref{FigCubes} (central panel), by subtracting an average spectrum of the spike, left and right of the companions psf from the superposition of companion and spike is given in Fig.~\ref{FigSpecAppend}. We find the results before (red spectrum) and after (blue spectrum) spike subtraction, which also removes the still-present OH lines.
In addition, the spectrum of the host star CVSO 30 is shown in black for comparison.

\begin{figure}
  \centering
  \includegraphics[width=0.49\textwidth]{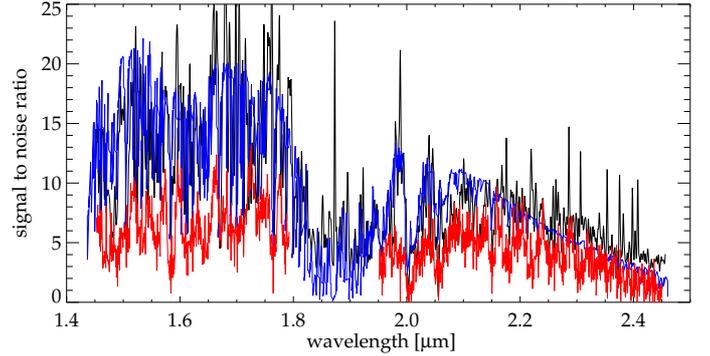}
  \caption{Signal-to-noise ratio (S/N) achieved for the brightest pixel vs.~the background noise in the combined cube (black). For comparison the expected and almost identical S/N is shown, simulated using the exposure time calculator (ETC) of ESO/SINFONI (blue). In red we present the final achieved S/N of the extracted companion spectrum after removing a superimposed spike (Fig.~\ref{FigCubes}), as shown in Fig.~\ref{FigSpec2}.
  }
  \label{FigSpecAppendSN}
\end{figure}

In Fig.~\ref{FigSpecAppendSN} we show the expected signal-to-noise ratio (S/N) for the given conditions and integration times (Tables \ref{TableLog} and \ref{TableCVSOsys}) using ESO's exposure time calculator for SINFONI and the latest available Pickles template spectrum M6 (blue). We derive an almost identical S/N using the flux of the companion after spike removal (Fig.~\ref{FigCubes}) compared to the noise of the background next to the spike (black). However, these S/N estimates are not achieved for our final extracted spectrum and its noise estimate (Fig.~\ref{FigSpec2}) because the spike itself adds slight additional noise, and more importantly, because of the imperfect removal of the spike that dominates the final S/N (red). To take this effect that is most likely caused by imperfect primary star positioning into account, we derived our final noise estimate, given as noise floor in Fig.~\ref{FigSpec2}, as the standard deviation of the neighbouring spectral channels after removing the continuum at the spectral position of interest. This noise was also used for the spectral model fitting (Figs.~\ref{FigSpec2}\,and
\,\ref{FigContSpec}) and the reduced $\chi^2$ estimation for several comparison objects (Table \ref{TableChi}).

We show the colour-magnitude diagram given in Fig.~\ref{FigColMag} in the main document with the identification of all the unlabelled objects in a full version in Fig.~\ref{FigColMagAppend} with the corresponding references in Table~\ref{TableRefColMag} .
The objects seem to follow the prediction of \citet{2008ApJ...689.1327S} quite well, especially around 10 Myr. Only 2M1207 b seems to be far off, possibly because of an edge-on disk
that heavily reddens the object \citep{2007ApJ...657.1064M}. Whether HR 8799 c and d are unusual can hardly be judged because no similar object with a very low luminosity is known at this age.
HR 8799 b is very low in luminosity (Fig.~\ref{FigEvol}), however.
The younger the objects, the higher in luminosity they are at similar spectral type because of 
their larger radius because they are still experiencing gravitational contraction.
The plot (Fig.~\ref{FigColMagAppend}) implies that CVSO 30 c is the first very young ($<$ 10 Myr) L-T transition object.

\begin{figure*}
  \centering
  \includegraphics[width=0.32\textwidth]{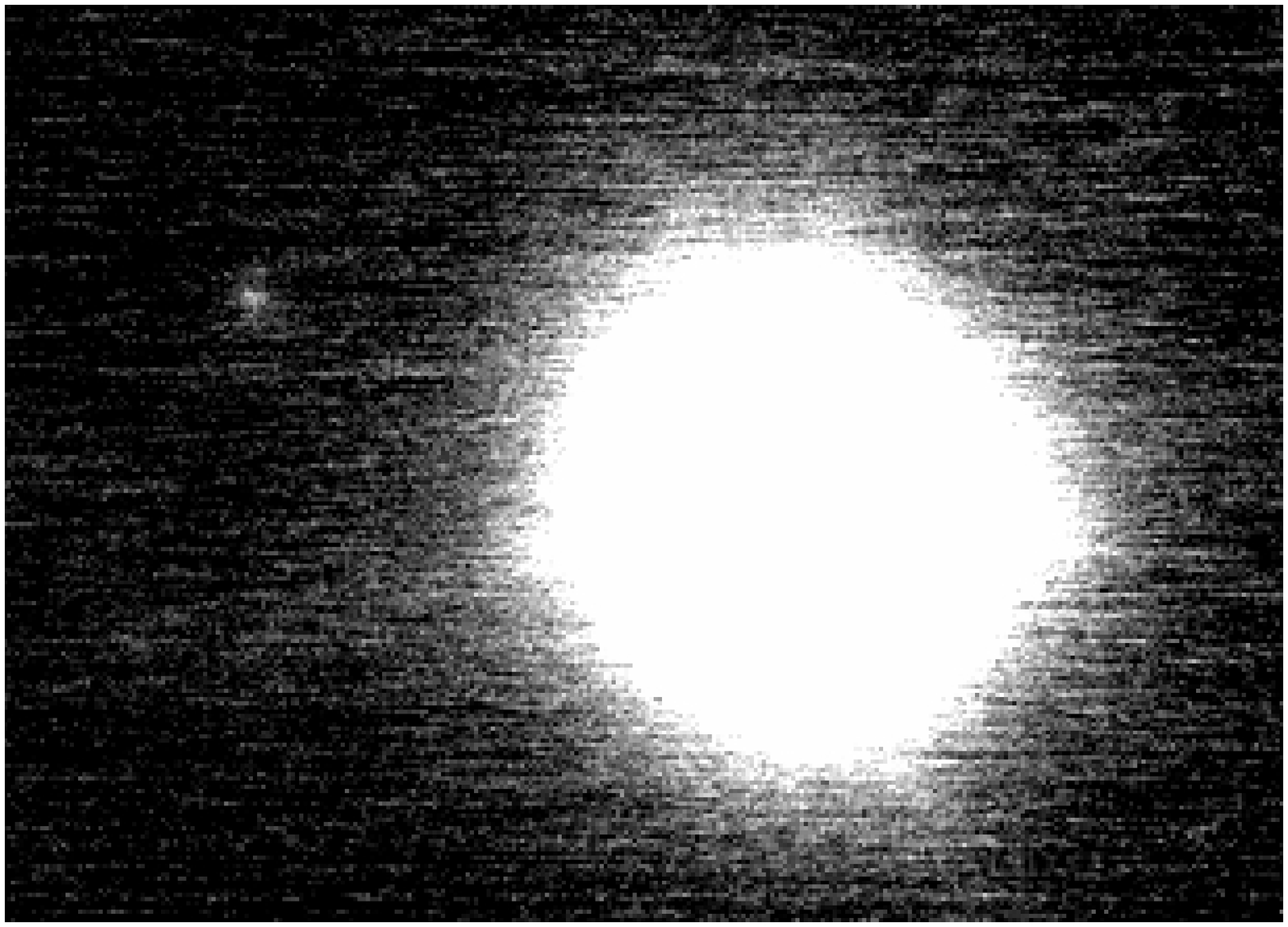}
  \includegraphics[width=0.32\textwidth]{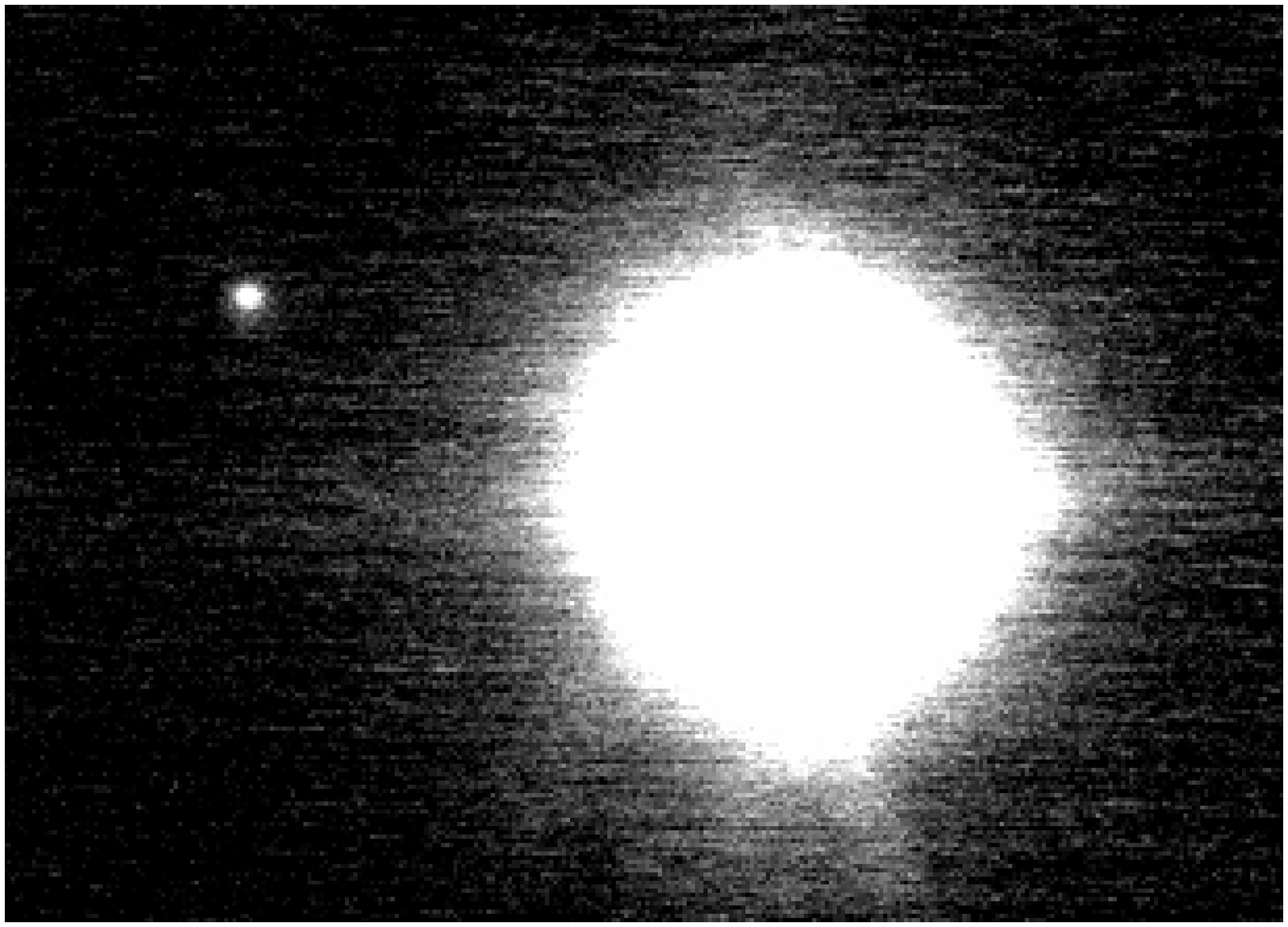}
  \includegraphics[width=0.32\textwidth]{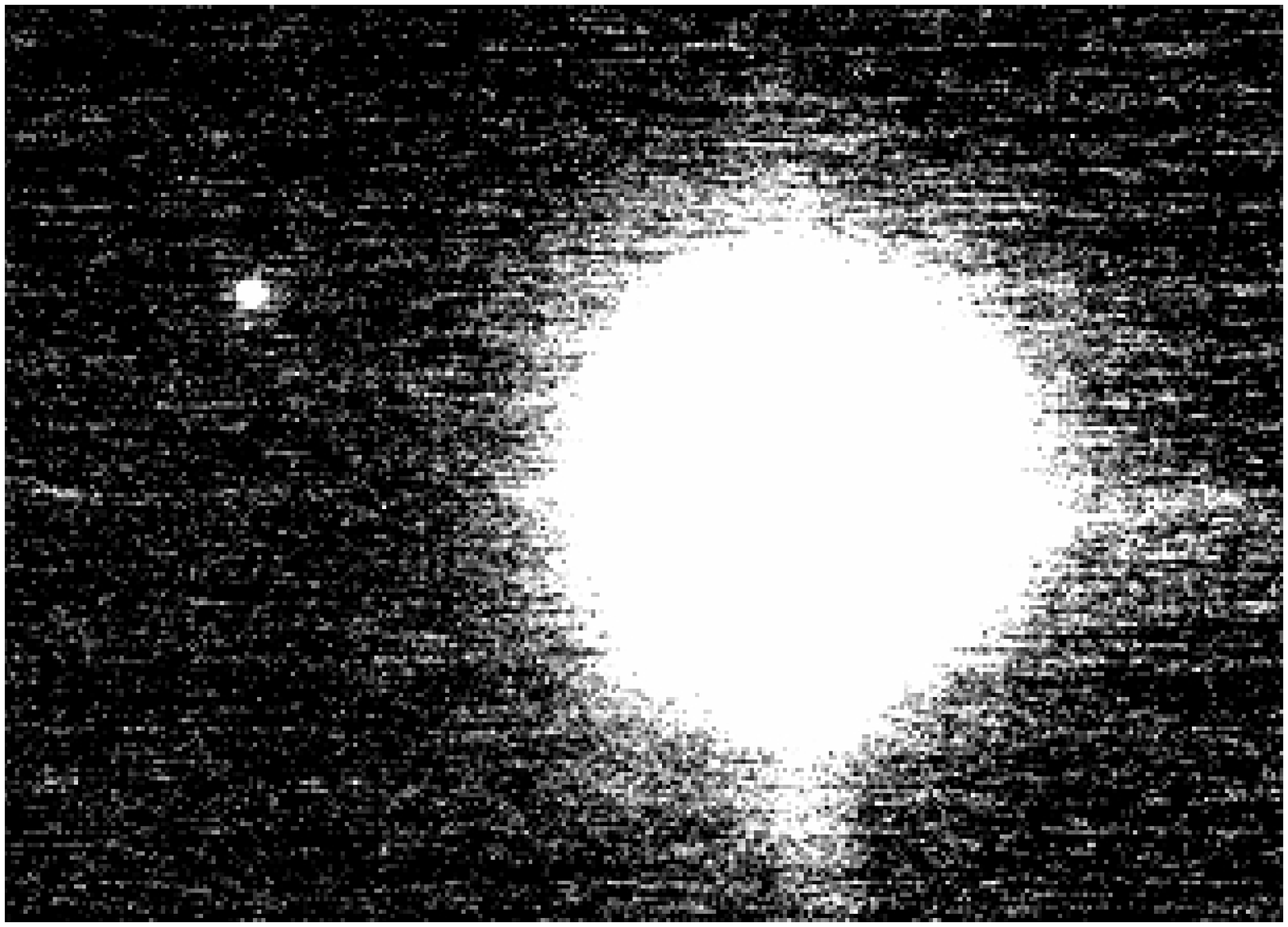}
  \includegraphics[width=0.32\textwidth]{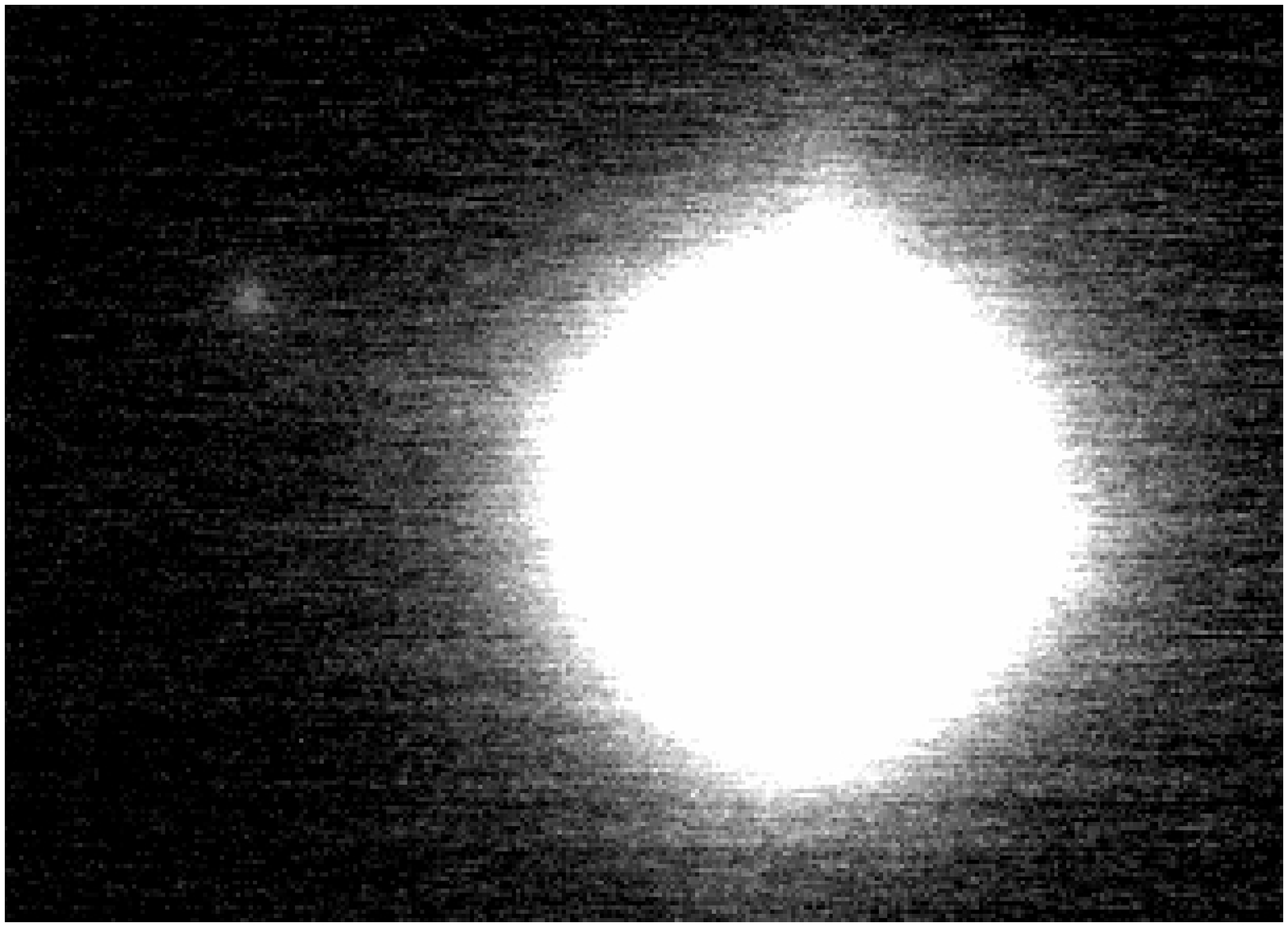}
  \includegraphics[width=0.315\textwidth]{Keckbild4.eps}
  \includegraphics[width=0.32\textwidth]{CVSOFarb3.eps}
  \caption{Direct images of CVSO 30 c. \textit{Top row, left to right:} Quasi-simultaneous VLT NACO J-, H-, and Ks-band data, taken in a sequence and shown in the same percentage of upper cut-off and lower cut-off value 0. \textit{Lower row, left to right:} VLT NACO J band with double exposure time per single image, the same in total, Keck image of data by \cite{2012ApJ...755...42V}, re-reduced. We note that the companion is north-east, not a contaminant south-east, as given in \cite{2012ApJ...755...42V}, and a JHKs colour composite, showing that CVSO 30 c has similar colours as its host star (Fig.~\ref{FigPhot}).}
  \label{FigImageAppend}
\end{figure*}

\begin{figure}
  \centering
  \includegraphics[width=0.5\textwidth]{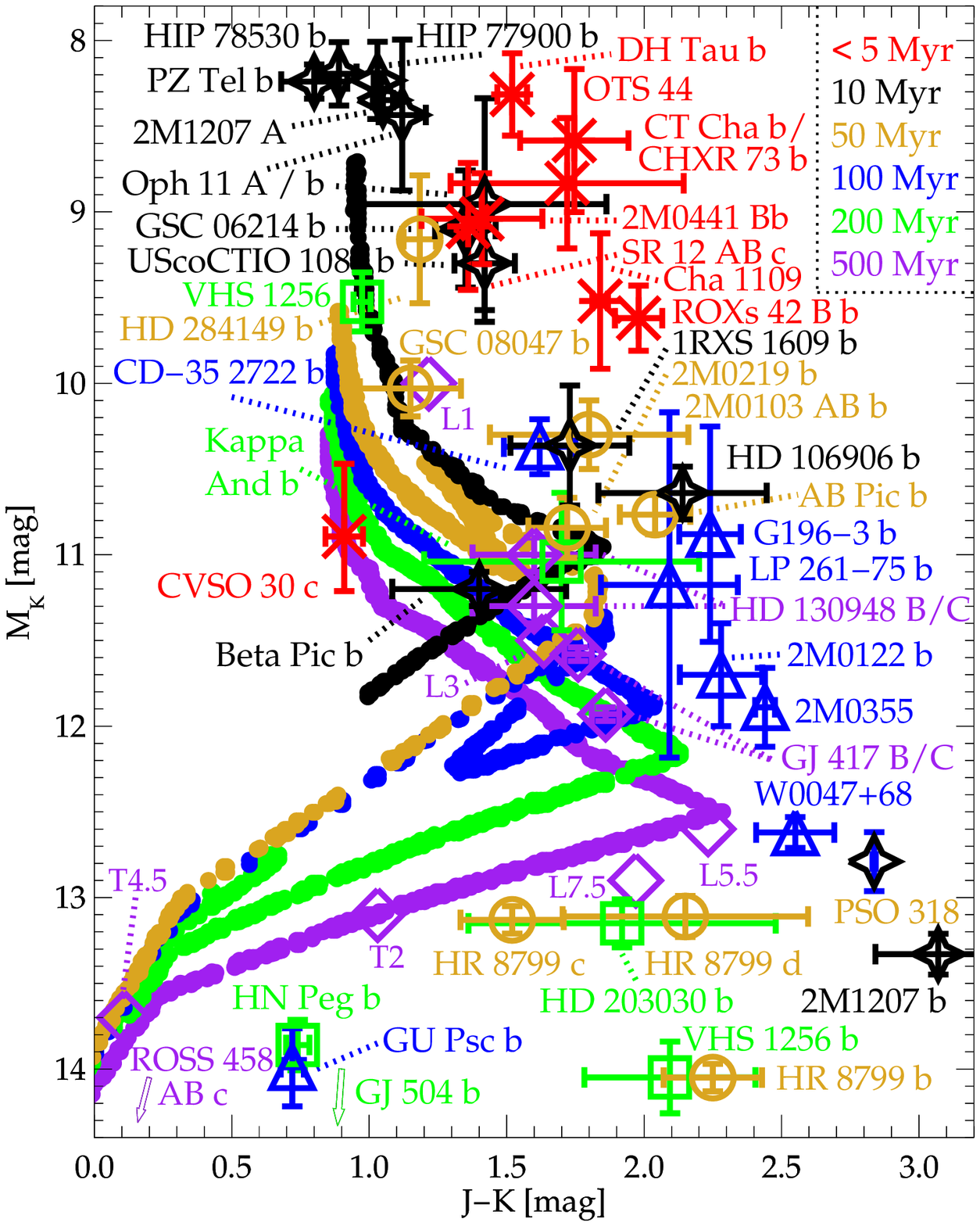}
  \caption{Colour-magnitude diagram of the simulated cluster brown dwarf population from \citet{2008ApJ...689.1327S}. Each sequence corresponds to a different age as given in the legend.
  Superimposed we show the positions of several planet candidates with full identification and CVSO 30 c. 
  See Fig.~\ref{FigColMag} and Table~\ref{TableRefColMag} for further details.
  }
  \label{FigColMagAppend}
\end{figure}

The core-accretion model 
\citep{1969Icar...10..109S,1973ApJ...183.1051G,1996Icar..124...62P},
was also discussed in models for HR 8799 bcde by \citet{2010Natur.468.1048C}, who argued that the inner planet was most likely formed by core accretion, while for the outer planets the gravitational instability of the disk \citep{1978M&P....18....5C,1997Sci...276.1836B}
is the more probable formation scenario. However, HR 8799 is an A- or F-star, and recent numerical simulations \citep{2013A&A...552A.129V} showed that disk fragmentation fails to produce wide-orbit companions around stars with mass $<$\,0.7 M$_{\odot}$, hence this is unfeasible for the $\sim$0.34~--~0.44 M$_{\odot}$ M3 star CVSO 30. In addition, the disk would have to be large enough for in situ formation. The most massive disks around M stars (e.g.~IM Lupi) might be large enough, but in this case, it was shown to possess almost all of its dust within 400 au separation \citep{2009A&A...501..269P}, which is still too small for formation at 662 au. 

If the object has not formed in situ, a very obvious solution would be scattering induced by a stellar flyby close to the system
\citep{2001Icar..150..151A,2015MNRAS.446.2010M}
or with another object of the system. While \citet{2001AJ....122..432R} described this possibility for the formation of brown dwarfs by disintegration of a small multiple system and possibly a cutoff from the formation material reservoir, which might have occurred
for example~for directly imaged
circumbinary planet candidates such as ROSS 458(AB) c \citep{2010ApJ...725.1405B} or SR 12 AB c \citep{2011AJ....141..119K},
the even more obvious possibility would be planet-planet scattering
because an inner planet candidate CVSO 30 b of similar mass is present that might have been scattered inward at the very same scattering event.

A way to distinguish between the formation scenarios would be by higher S/N ratio spectroscopy, as has been done for HR 8799 c \citep{2013Sci...339.1398K}. With both H$_2$O and CO detected, it is possible to estimate the bulk atmospheric carbon-to-oxygen ratio and whether it differs from that of the primary star, which led \citet{2013Sci...339.1393M} to speculate that HR 8799 c formed by core accretion and not by gas instability.

We can place CVSO 30 c best into context by comparing it with the recent M-dwarf survey of \citet{2015ApJS..216....7B},  who found that fewer than 6\% of M dwarfs harbour massive giant planets of 5--13 M$_{\mathrm{Jup}}$ at 10--100 au and that there is currently no statistical evidence for a trend of giant planet frequency with stellar host mass at large separations. We note, however, that CVSO 30 c would probably not have been found at the distance of their targets because it would not have been in the field of view as a result of its large separation of about 662 au.
About 20 of the 49 directly imaged planet candidates at www.exoplanet.eu 
have an M dwarf as host star.

At a projected separation of $\sim$662 au, the system is above the long-term stability limit of $\sim$390 au for an M3 primary star of 0.34 -- 0.44 $M_{\odot}$ (Table \ref{TableCVSOsys}), following the argumentation of \citet{1987ApJ...312..367W} and \citet{2003ApJ...587..407C}. However, as shown in \citet{2005AN....326..701M}, 2M1207 and its companion \citep{2005A&A...438L..25C} are also exceeding this long-term stability limit at about three times the age of CVSO 30.

The currently acquired data are consistent with planet-planet scattering simulations in \citet{2008ApJ...686..621F}, showing that massive planets are more likely to eject one another,
whereas smaller planets are more likely to collide, resulting in stabilised systems, as supported by Kepler satellite and Doppler survey results that find 
predominantly smaller \citep{2009ApJ...693.1084W,2011ApJ...732L..24L} low-density \citep[e.g.~][]{2013ApJ...770..131L} planets in compact close multi-planet systems.

\begin{table}
\caption{Evolutionary plot (Fig.~\ref{FigEvol}) references}
\label{TableRefEvol}
\centering
\begin{scriptsize}
\begin{tabular}{cc|cc}
\hline
Object & reference & Object & reference \\
\hline\hline
GJ 504 b & \citet{2013ApJ...774...11K} & HD 95086 & \citet{2013ApJ...772L..15R}\\
2M1207 & \citet{2004AandA...425L..29C} & b & \citet{2014AandA...565L...4G}\\
 b & \citet{2007ApJ...657.1064M} & HR 8799 & \citet{2008Sci...322.1348M}\\
HR 8799 & \citet{2010Natur.468.1080M} & b, c, d & \citet{2011ApJ...732...61Z}\\
e & \citet{2011ApJ...732...61Z} & & \citet{2010MNRAS.405L..81M}\\ 
 & \citet{2010MNRAS.405L..81M} & $\beta$ Pic b & \citet{2009AandA...493L..21L}\\
1RXS   & \citet{2008ApJ...689L.153L} & & \citet{2014AandA...567L...9B}\\
1609 b & \citet{2012arXiv1201.3537N} & & \citet{2014MNRAS.438L..11B}\\
& \citet{2012ApJ...746..154P} & & \citet{2014MNRAS.445.2169M}\\
CT Cha b & \citet{2008AandA...491..311S} & CHXR 73 & \citet{2006ApJ...649..894L}\\
2M044 & \citet{2010ApJ...714L..84T} & b\\
144 b & & GQ Lup b & \citet{2005AandA...435L..13N}\\
HD & \citet{2013ApJ...766L...1Q} & LkCA15 & \citet{2012ApJ...745....5K}\\
100546b & \citet{2015ApJ...807...64Q} & b, c & \citet{2015Natur.527..342S}\\
ROXs & \citet{2014ApJ...787..104C} & SR 12 & \citet{2011AJ....141..119K}\\
42B b & & AB c\\
DH Tau & \citet{2005ApJ...620..984I} & 2M0103 & \citet{2013AandA...553L...5D}\\
b & \citet{2012arXiv1201.3537N} & AB b\\
AB Pic b & \citet{2005AandA...438L..29C} & HD & \citet{2014ApJ...780L...4B}\\
& \citet{2012arXiv1201.3537N} & 106906 b\\
51 Eri b & \citet{2015Sci...350...64M} & GU Psc b & \citet{2014ApJ...787....5N}\\
& \citet{2015ApJ...813L..11M} & GSC & \citet{2011ApJ...726..113I}\\
USco & \citet{2008ApJ...673L.185B} & 06214 b & \citet{2002AJ....124..404P}\\
CTIO  & \citet{2002AJ....124..404P} & PZ Tel B & \citet{2010AandA...523L...1M}\\
108 b & \citet{2012ApJ...746..154P} &  & \citet{2010ApJ...720L..82B}\\
2M0219 b & \citet{2015ApJ...806..254A} & & \citet{2012MNRAS.420.3587J} \\
\hline
\end{tabular}                                                                                                    
\end{scriptsize}
\end{table}

\begin{table}
\caption{Colour-magnitude plot (Figs.~\ref{FigColMag} and \ref{FigColMagAppend}) references}
\label{TableRefColMag}
\centering
\begin{scriptsize}
\begin{tabular}{cc|cc}
\hline
Object & reference & Object & reference \\
\hline\hline
2M1207 & \citet{2004AandA...425L..29C} & HR 8799 & \citet{2008Sci...322.1348M}\\
A \& b & \citet{2007ApJ...657.1064M} & b, c, d\\
& \citet{2008AandA...477L...1D} & $\beta$ Pic b & \citet{2013AandA...555A.107B}\\
1RXS & \citet{2008ApJ...689L.153L} & ROXs & \citet{2014ApJ...781...20K}\\
1609 b & & 42B b\\
DH Tau & \citet{2005ApJ...620..984I} & SR 12 & \citet{2011AJ....141..119K}\\
b & \citet{2006ApJ...649..894L} & AB c\\
AB Pic & \citet{2005AandA...438L..29C} & 2M0103 & \citet{2013AandA...553L...5D}\\
b & & AB b \\
Ross & \citet{2011MNRAS.414.3590B} & USco & \citet{2008ApJ...673L.185B}\\
458 & & CTIO \\
AB c & & 108 b\\
GSC & \citet{2011ApJ...726..113I} & PZ Tel b & \citet{2010AandA...523L...1M}\\
06214 b & & GJ 504 & \citet{2013ApJ...774...11K}\\
GU Psc & \citet{2014ApJ...787....5N} & b\\ 
b & & 2M0122 & \citet{2013ApJ...774...55B}\\
HD & \citet{2006ApJ...651.1166M} & b\\
203030 & & HD & \citet{2002ApJ...567L.133P}\\
b & & 130948\\
GSC & \citet{2005AandA...430.1027C} & B \& C\\
08047 b & \citet{2014AandA...562A.127B} & 2M0355 & \citet{2013AJ....145....2F}\\
HN Peg & \citet{2007ApJ...654..570L} & CD-35 & \citet{2011ApJ...729..139W}\\
b & & 2722 b\\
$\kappa$ And b & \citet{2013ApJ...763L..32C} & OTS 44 & \citet{2005ApJ...620L..51L}\\
& \citet{2013ApJ...779..153H} & Cha & \citet{2005ApJ...635L..93L}\\
HIP & \citet{2011ApJ...730...42L} & 1109\\
78530 b & & HD & \citet{2014ApJ...791L..40B}\\
Oph 11 & \citet{2006Sci...313.1279J} & 284149\\
A \& b & \citet{2007ApJ...660.1492C} & b\\
LP & \citet{2006PASP..118..671R} & HIP & \citet{2013ApJ...773...63A}\\
261-75 & \citet{2000AJ....120..447K} & 77900 b\\
b & & G196-3 & \citet{1998Sci...282.1309R}\\
GJ 417 & \citet{2001AJ....121.3235K} & b & \\
B \& C & \citet{2014ApJ...790..133D} & HD & \citet{2014ApJ...780L...4B} \\ 
CHXR & \citet{2006ApJ...649..894L} & 106906\\
73 b & & b\\
CT Cha & \citet{2008AandA...491..311S} & W0047 & \citet{2015ApJ...799..203G}\\
b & & +68 \\
VHS & \citet{2015ApJ...804...96G} & 2M0219 & \citet{2015ApJ...806..254A} \\
1256 b & & b \\
2M0441 & \citet{2015ApJ...811L..30B} & PSO & \citet{2013ApJ...777L..20L} \\
Bb & & 318\\
\hline
\end{tabular}                                                                                                    
\end{scriptsize}
\end{table}

\end{appendix}

\end{document}